\documentclass[prd,a4paper,twocolumn,preprintnumbers,nofootinbib,superscriptaddress]{revtex4}

\pdfoutput=1
\usepackage[english]{babel}
\usepackage{amsmath,amssymb,amsfonts, bm,bbm,slashed, subdepth}
\usepackage{graphicx}
\usepackage{hyperref}
\usepackage{enumerate}
\usepackage{setspace}
\usepackage{booktabs, tabularx}
\usepackage{units}
\usepackage{color}
\usepackage{multirow}

\newcommand{\be}{\begin{equation}}
\newcommand{\ee}{\end{equation}}
\newcommand{\ba}{\begin{array}}
\newcommand{\ea}{\end{array}}
\newcommand{\bea}{\begin{eqnarray}}
\newcommand{\eea}{\end{eqnarray}}
\newcommand{\balg}{\begin{align}}
\newcommand{\ealg}{\end{align}}
\newcommand{\bit}{\begin{itemize}}
\newcommand{\eit}{\end{itemize}}

\makeindex

\begin{document}
\preprint{ULB-TH/19-11}

\title{
Minimal self-interacting dark matter models with light mediator
}

\author{Thomas Hambye}
\email{thambye@ulb.ac.be}
\affiliation{Service de Physique Th\'eorique, Universit\'e Libre de Bruxelles, Boulevard du Triomphe, CP225, 1050 Brussels, Belgium}

\author{Laurent Vanderheyden}
\email{lavdheyd@ulb.ac.be}
\affiliation{Service de Physique Th\'eorique, Universit\'e Libre de Bruxelles, Boulevard du Triomphe, CP225, 1050 Brussels, Belgium}

\begin{abstract}

The light mediator scenario of self-interacting dark matter is strongly constrained in many ways.
After summarizing the various constraints, we discuss minimal options and models which allow to nevertheless satisfy all these constraints.
One straightforward possibility emerges if the dark matter and light mediator particles have a temperature sizably smaller than the SM particles. Another 
simple possibility arises if dark matter doesn't annihilate dominantly into a pair of light mediators but into heavier particles. Both possibilities are
discussed with scalar as well as vector boson light mediators. Further possibilities, such as with a hierarchy of quartic scalar couplings, are also identified.

\end{abstract}

\maketitle

\section{Introduction}

In spite of all its many successes, the Cold Dark Matter paradigm cannot 
account for various features that N-body simulations of formation of small scale structure tend to imply.
Features of this kind include in particular the \emph{too-big-to-fail}\cite{BoylanKolchin:2011de, BoylanKolchin:2011dk} and the \emph{core-vs-cusp} problems \cite{Spergel:1999mh,Walker:2011zu,deNaray:2011hy}, as well as the more recently spotted diversity problem \cite{Oman:2015xda}, see also e.g.~Ref.~\cite{Weinberg:2013aya}.

{Self-interacting dark  matter is a plausible solution to these problems \cite{Spergel:1999mh,Wandelt:2000ad,Vogelsberger:2012ku,Rocha:2012jg,Peter:2012jh,Zavala:2012us,Vogelsberger:2014pda,Elbert:2014bma,Kaplinghat:2015aga},  including the diversity one \cite{Kamada:2016euw,Tulin:2017ara}.
Its key ingredient is the hypothesis that dark matter (DM) particles scatter off each other in small-scale structures with a cross section per unit of mass of around $\unit[0.1-10]{cm^2/g}$.
This corresponds to $\unit[10^{12}]{pb}$ for DM masses around $\unit[1]{GeV}$, which is many orders of magnitude above the standard thermal freeze-out cross section of about \unit[1]{pb}. Clearly, if DM  undergoes a thermal freeze-out in the early Universe,
some mechanism should be at work to explain this disparity of cross sections. 

To account for such a disparity, one possibility,  which has been extensively studied in the literature, is to invoke a light mediator enhancing DM self-interactions via non-perturbative effects in small-scale structures~\cite{Feng:2009hw, Buckley:2009in, Loeb:2010gj, Tulin:2013teo}. 
This scenario has the advantage of displaying a DM velocity dependence which allow to accommodate in an easier way the value of $\sigma/m_{DM}$ needed at galactic scales above (or even at dwarf galaxy scale where $v\sim 10$~km/s) with the upper bound from merging clusters (where $v\sim 1000$~km/s), $\sigma/m_{DM} < \unit[0.3]{cm^2/g}$ \cite{Harvey:2015hha,Bondarenko_2018,Clowe:2003tk,Markevitch_2004,Randall_2008}.
However, it has been shown in a series of works that such a scenario is strongly constrained in various ways, and the simplest scenarios one could consider are typically either excluded or marginally allowed \cite{Buckley:2009in,Feng:2009hw,PhysRevD.89.035009,Tulin:2013teo,Nobile_2015,Bernal:2015ova,Bringmann:2016din,Cirelli:2016rnw,Kahlhoefer:2017umn,Hufnagel:2017dgo,Hufnagel:2018bjp,Bernal:2019uqr}.

In the following, after summarizing these various constraints in Section~\ref{sec:constraints}, we will discuss the various minimal ways out one can consider to avoid them. For each option we will discuss the associated phenomenology, with some details for the 2 simplest ways (in Sections  \ref{Tprimewayout}  and \ref{subdomannihwayout}), and more briefly for additional ways (in Sections \ref{sec:pwavemeddecay} to \ref{sec:neutrinowayout}).

\section{Summarizing the constraints holding on the light mediator scenario\label{sec:constraints}}

There is an all jungle of constraints which has been discussed in the literature on the light mediator self-interacting scenario. It is useful to summarize all these constraints in a same paper and same section, together with listing the simple ways out one could in principle consider to avoid each constraint separately. 
To illustrate these constraints we will consider the two usual minimal models of a Dirac fermion DM particle which is charged under a new $U(1)'$ gauge symmetry, so that it couples to the corresponding $\gamma '$,
\begin{equation}
\text{Model}\,\, A_{\gamma '}:\,\,\,  {\cal L} \owns - g_{\gamma '}\overline{\chi} \gamma^\mu \chi \gamma '_\mu +h.c.-\frac{\epsilon}{2} F^Y_{\mu\nu}F'^{\mu\nu}
\label{medcouplV}
\end{equation}
and of a Dirac fermion DM particle which couples to a light scalar through a Yukawa interaction 
\begin{equation}
\text{Model}\,\, A_{\phi}:\,\,\, {\cal L} \owns - y_\phi \phi\overline{\chi} \chi +h.c. -\lambda \phi^\dagger \phi H^\dagger H
\label{medcouplphi}
\end{equation}
In the above we have already written down the possible ways such hidden sector structures could couple to the SM particles, i.e. through the usual kinetic mixing and Higgs portal interactions respectively. For the vector model we assume that the $\gamma'$ acquires a mass from the Stueckelberg mechanism or Brout-Englert-Higgs mechanism.
One could consider a Majorana DM instead of a Dirac DM but we will not consider this possibility here.\footnote{In the scalar mediator case this wouldn't change much the picture with respect to the Dirac case. For a vector mediator instead,
the Sommerfeld effects, which will play an important role below, are expected to be quite different (and more difficult to calculate) for the Majorana case and we will not consider this possibility (although it would be quite interesting to investigate it too). }

Both the models of Eqs.~(\ref{medcouplV}) and (\ref{medcouplphi}) are excluded or marginally allowed by all constraints, see below and in particular Ref.~\cite{Cirelli:2016rnw} for the scalar case and Refs.~\cite{Bringmann:2016din,Hufnagel:2018bjp} for the vector case. In the following, to accommodate the various constraints,  we will either consider these models relaxing some of the assumptions generally made, or consider extensions of these models, or consider other models.
For all these possibilities, we will assume that the DM frameworks fulfill a number of minimal properties beyond which we will not go, i.e.~model building features one can consider
to accommodate these constraints.
These are:
\begin{itemize} 

\item[(0)] \underline{DM production mechanism}:  we will always assume below that the DM relic density is due to a thermal freeze-out of a 2 to 2 scattering process.
Other avenues, not considered here, would have been to consider other DM production mechanisms, such as freezeout based on $n$ ($>2$) to 2 processes \cite{Hochberg:2014dra,1992ApJ...398...43C,Farina:2016llk,Pappadopulo:2016pkp,Bernal:2015ova} or non-freezeout scenarios (i.e.~freezein, reannihilation, ..., see Ref.~\cite{Bernal:2015ova,Hambye:2018dpi}).
For all setups considered below, except the last one of Section \ref{sec:asym}, we will assume that DM is particle-antiparticle symmetric.


\item[(i)] \underline{Dominant or sub-dominant annihilation into light} \underline{mediators}: as said above for all setups we will assume a light mediator to account for the self-interactions constraints. 
In this case DM annihilation into a pair of these mediators is unavoidable (unless DM is asymmetric). If this is the dominant annihilation channel, the corresponding cross section must have the value which follows from thermal decoupling.
If instead there exists another annihilation channel, e.g.~to other particles beyond the SM, which dominates, then it is this one which must have the thermal value and the annihilation rate into light mediators can be smaller. We will allow for such a possibility.

\item[(ii)] \underline{s-wave vs p-wave annihilation}: depending on the nature of the DM and mediator particles the annihilation into a pair of light mediators might proceed in a s-wave, as for Eq.~(\ref{medcouplV}), or p-wave way, as for Eq.~(\ref{medcouplphi}) and this largely affects a number of constraints.

\item[(iii)] \underline{Type and size of the portal interaction}: the self-interaction constraints do not require any portal interaction between the hidden sector (i.e.~DM and light mediator sector) and visible sector. However, a portal might be mandatory to avoid various other constraints. Below we will consider only the 2 usual portals of Eqs.~(\ref{medcouplV}) and (\ref{medcouplphi}). The type and size of the portal affect the direct detection, indirect detection as well as, importantly, the possible decay width of the light mediator into SM particles  (i.e.~to $e^\pm$, photons and neutrinos, given the small value of the mediator mass needed to account for the self-interaction constraints).

\item[(iv)] \underline{Light mediator decay or annihilation into hidden} \underline{sector particles}: whether there are decay and/or annihilation channels of the mediator into other hidden sector particles may affect largely the various constraints too, and we will allow for such a possibility.
 
\item[(v)] \underline{Hidden vs visible temperature ratio}: the temperature of the hidden sector might be the same as the one of the SM sector at time of DM freezeout, or might be different, depending on whether both sectors have thermalized at some point. This is related to the portal issue above because, if the portal interaction is large enough, it will thermalize both sectors, so that $T'/T\simeq1$ at freezeout. Smaller portal interaction
only implies a lower bound on $T'/T$.

\end{itemize}

The (i) to (v) items above constitute the 5 features we will play with to overcome the various constraints. Depending on what we assume for these 5 items, we now discuss the main constraints applying to the light mediator self-interacting DM  paradigm.

\subsection{Relic density constraint}

As said above, DM symmetric scenarios leading to large self-interactions from the multi-exchange of a light mediator between the initial states, unavoidably lead to annihilation of a pair of DM particles into a pair of these light mediators. As the value of the couplings needed to account for the self interactions is sizeable, the corresponding annihilation rate is sizeable and the 
self-interaction constraints can be easily accommodated with the assumption of a freezeout process dominated by this channel.  
Note nevertheless that 
the self interactions constraints can also be accommodated for values of the DM to mediator coupling 
much smaller than the "thermal values".
For example, values of the couplings $g_{\gamma '}$ and $y_{\phi}$, in Eqs.~(\ref{medcouplV}) and (\ref{medcouplphi}), up to $\sim 2$ orders of magnitude smaller than the "thermal values" (leading to annihilation cross section up to $\sim$~8 orders of magnitude below the thermal value),
 turn 
out to be still compatible with the 
self-interactions constraints, see below. 
As already mentioned above, in this case the relic density can be accommodated if the freezeout is dominated by another annihilation channel with annihilation rate equal to the thermal value.
Another possibility to account for the relic density, with a annihilation cross section smaller that the usual thermal freeze-out one arises if both the hidden sector light mediator and DM particles do thermalize (forming a thermal bath with temperature $T'$) but do not with the visible sector (which has temperature $T$), so that $T'\neq T$.
In this case the DM freezeout occurs exclusively in the hidden sector.
This  ``secluded freeze-out'' (see e.g.~Section 3.4 in \cite{Chu:2011be})
implies a DM relic density which scales as $\sim (T'/T)/\langle \sigma v\rangle _{HS}$ (for a s-wave annihilation \cite{Chu:2011be} as well as for a p-wave one, see below), where $\langle \sigma v\rangle_{HS}$ stands for the DM annihilation rate in the hidden sector.
Thus, for $T'/T<1$, the relic density constraint needs $\langle \sigma v\rangle_{HS}$ smaller than the thermal value by a factor of $\sim T'/T$.
This is possible as long as  the portal between the hidden and SM sectors is feeble enough to not heat up the hidden sector from the SM thermal reservoir to a temperature $T'$ higher than the one assumed. 

\subsection{Non overclosure of the Universe by the light mediator}

In most minimal models, such as in models of Eqs.~(\ref{medcouplV}) and (\ref{medcouplphi}), the light mediator decouples chemically from the thermal bath at same temperature as DM, when DM freezes-out.
Thus, since typically the light mediator must be lighter than the DM particles by a factor larger than 20 (for self-interactions), the light mediator decouples while it is still relativistic. In this case, for a real scalar boson, the mediator to photon number density ratio, $n_{med}/n_\gamma$, is predicted to be as large as $1/2$ at decoupling, whereas for a vector boson it is predicted to be $3/2$. 
Thus, if it is absolutely stable, and unless it is extremely light (below few eV), the light mediator hot relic will overclose the Universe.
Consequently, one needs a mechanism to reduce its number density after DM freeze-out.  
Below we will consider both the possibilities of a light mediator decay and of a light mediator annihilation into lighter particles.
In both cases it is highly desirable to have a portal interaction between the hidden sector  and the SM visible sector, so that the decay products of these decay or annihilation are ultimately SM particles.
There is nevertheless one very simple and viable way out which doesn't require any decay or annihilation of the relativistic light mediator after DM freeze-out (and thus doesn't require a portal): if the temperature of the hidden sector
is sufficiently low with respect to the one of the visible sector to suppress the light mediator relic density accordingly, see below. 

\subsection{Small scale structure constraints} 

Simulations show that the \textit{too-big-to-fail} and \textit{core-vs-cusp} problems could be alleviated if the self-interaction cross section divided by the DM mass, lies within the range $0.1$ $\hbox{cm}^2/g \lesssim \sigma_T/m_{DM} \lesssim 10$ $\hbox{cm}^2/g$. Nevertheless, the non-observation of an offset between the mass distribution of DM and galaxies in the Bullet Cluster constrains such self-interacting cross section, concretely $\sigma_T/m_{DM} < 1.25$ $\hbox{cm}^2/\hbox{g}$ at 68$\%$ CL \cite{Clowe:2003tk,Markevitch_2004,Randall_2008}. Similarly, recent observations of cluster mergers lead to the constraint $\sigma_T/m_{DM} < 0.47$ $\hbox{cm}^2/\hbox{g}$ at 95\% CL \cite{Harvey:2015hha} (or even smaller with $\sigma_T/m_{DM} < 0.3$ $\hbox{cm}^2/\hbox{g}$ in \cite{Bondarenko_2018}).

Along the light mediator scenario, such large cross section is obtained from the non-perturbative effects arising from the multi-exchange of the light mediator between the two highly non-relativistic DM initial state particles. 
As a reference we show here in Fig. \ref{fig:SI_swave_Oh0} what is the value of $m_{DM}$ we need as a function of $m_{med}$ for the scalar and vector mediator models of Eqs.~(\ref{medcouplV}) and (\ref{medcouplphi}), fixing the $y_\phi$ and $g_{\gamma '}$ couplings from requiring successful freeze-out from DM annihilation into a pair of light mediators and assuming $T'=T$ at the DM freezeout time. 
For the (p-wave) scalar mediator case, one distinguishes from top to bottom the classical, resonant and Born regimes. The resonant regime arises from the fact that the associated potential is attractive.
The (s-wave) vector mediator case leads to similar values, even though the potential in this case has both an attractive and a repulsive piece.
The assumptions one could make on the size and type of the portal, as well as on the decay of the mediator, have a secondary impact on the self-interactions. The other two simple model building features one could play with, see above, do nevertheless clearly matter for the self-interactions:

\begin{itemize}
\item \underline{Dominant or sub-dominant annihilation into light} \underline{mediators}: 
as already mentioned above self-interaction constraints can be accommodated for a DM to light mediator coupling much smaller than its "thermal value". Quantitatively this is shown 
in  
Fig. \ref{fig:SI_swave_Oh2}, which gives  the masses one needs assuming that 
 the annihilation cross section is a factor $10^{2}, 10^{4}, 10^6$ smaller than the one needed to account for the observed relic density. In this case, the resonant regime regions shrinks and the classical and Born region works for a somewhat narrower range of $m_{DM}$  and slightly smaller mediator masses. Once again, if $T'/T=1$ at the freezeout time, this requires another (faster) annihilation channel to account for this freezeout.
  
 \item \underline{Hidden vs visible temperature ratio}: as already mentioned above too, if $T'/T<1$ the annihilation rate accounting for the freezeout must be smaller by a factor of $\sim T'/T$ (for the case of a s-wave annihilation as well as for a p-wave annihilation). Thus, in this case the annihilation into a pair of light mediators could be responsible for the freezeout with a smaller value than the usual freezeout value, so that Fig.~\ref{fig:SI_swave_Oh2} is relevant for this case too.
 
 \end{itemize}


\begin{figure}[h!]
\centering
\begin{minipage}{0.23\textwidth}
  \centering
\includegraphics[scale=0.35]{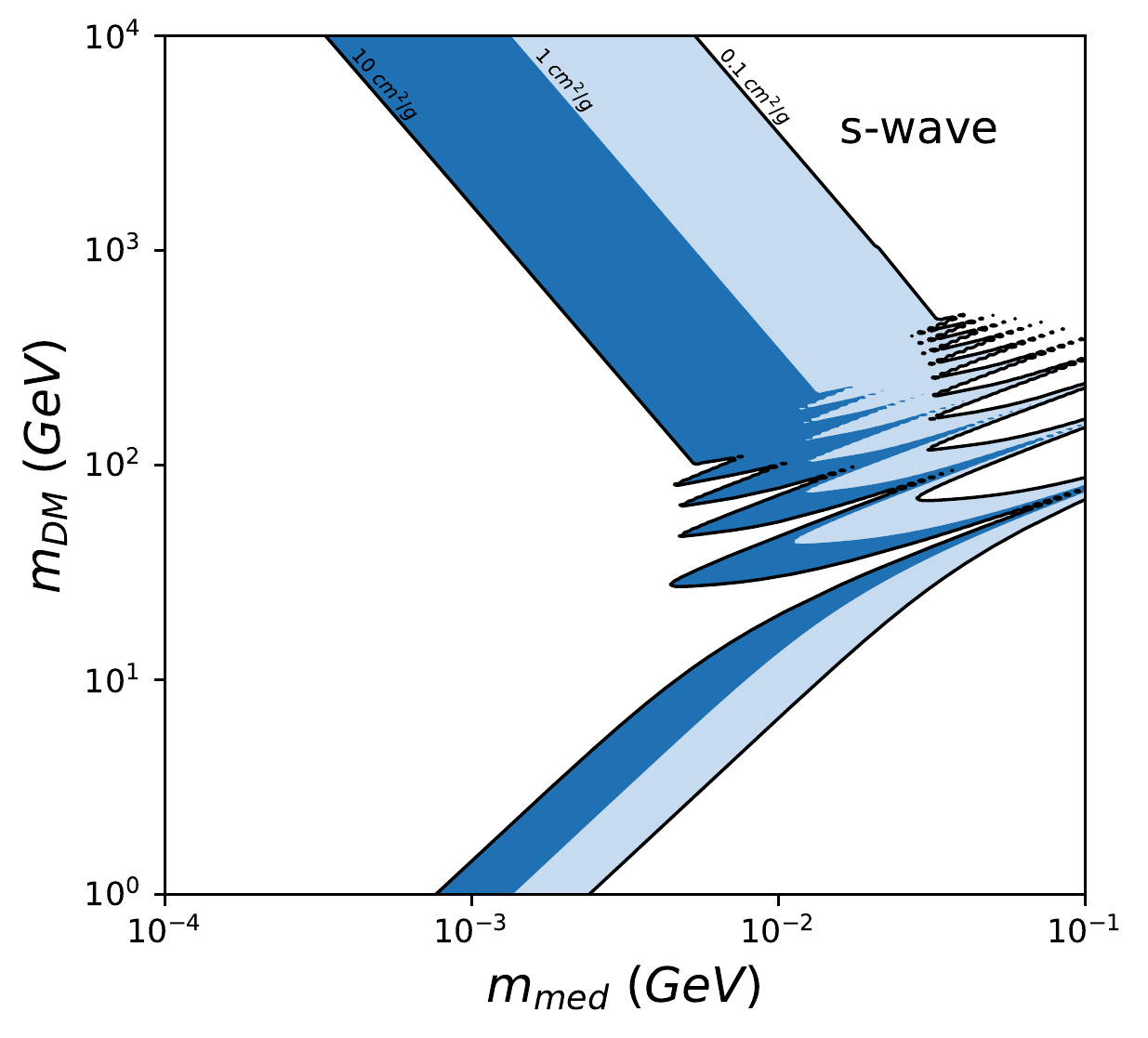}
\end{minipage}
\begin{minipage}{0.23\textwidth}
  \centering
\includegraphics[scale=0.35]{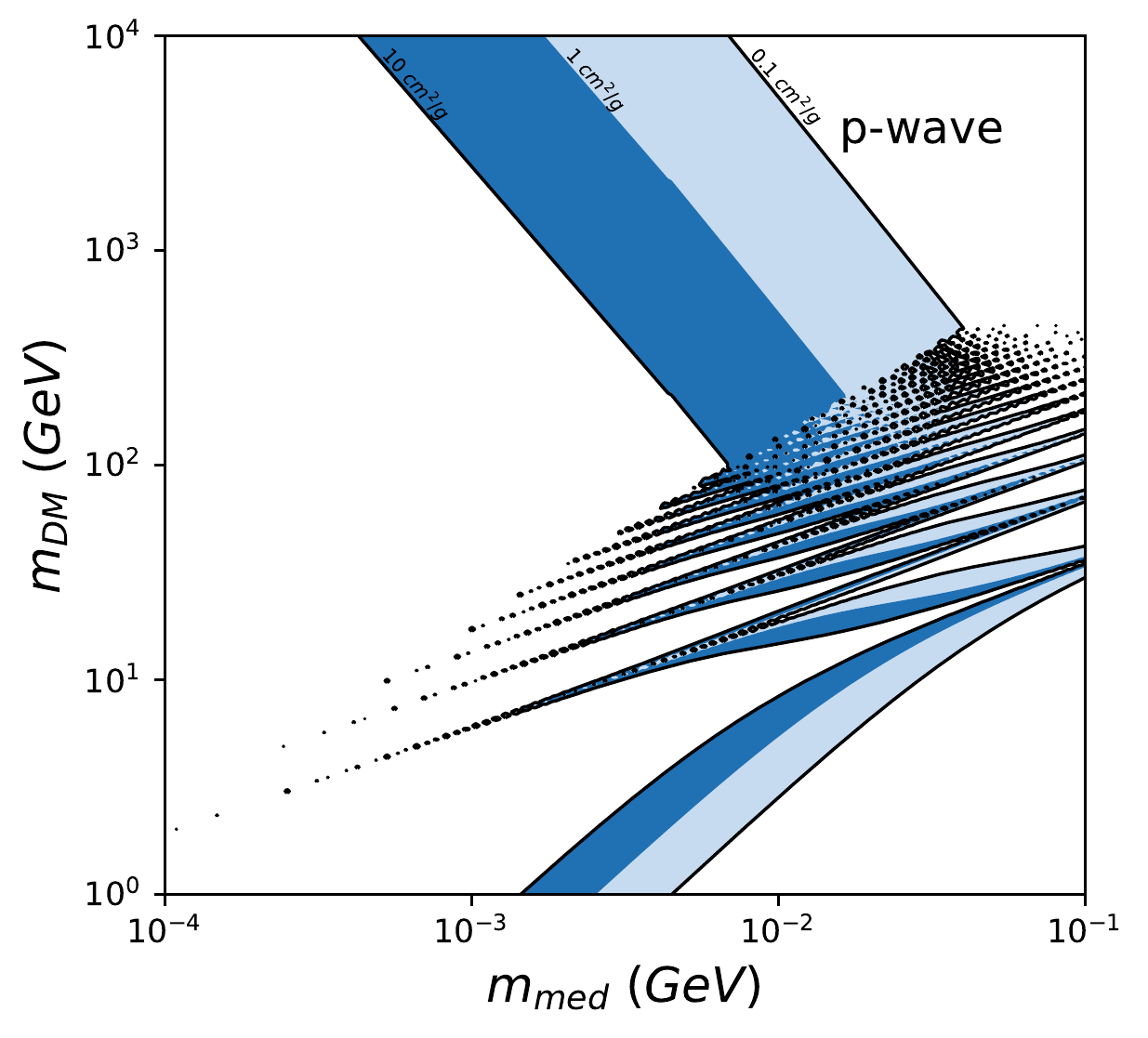}
\end{minipage}
\caption{Self interaction constraints. Left: Dirac DM with vector mediator (i.e. s-wave annihilation case). Right: Dirac DM with scalar mediator (i.e.~p-wave annihilation case). Note that there is no significative difference if we take into account the Sommerfeld enhancement at freeze-out time, what we didn't do here.} \label{fig:SI_swave_pwave}
\label{fig:SI_swave_Oh0}
\end{figure}

\begin{figure}[h!]
\centering
\begin{minipage}{0.24\textwidth}
  \centering
\includegraphics[scale=0.35]{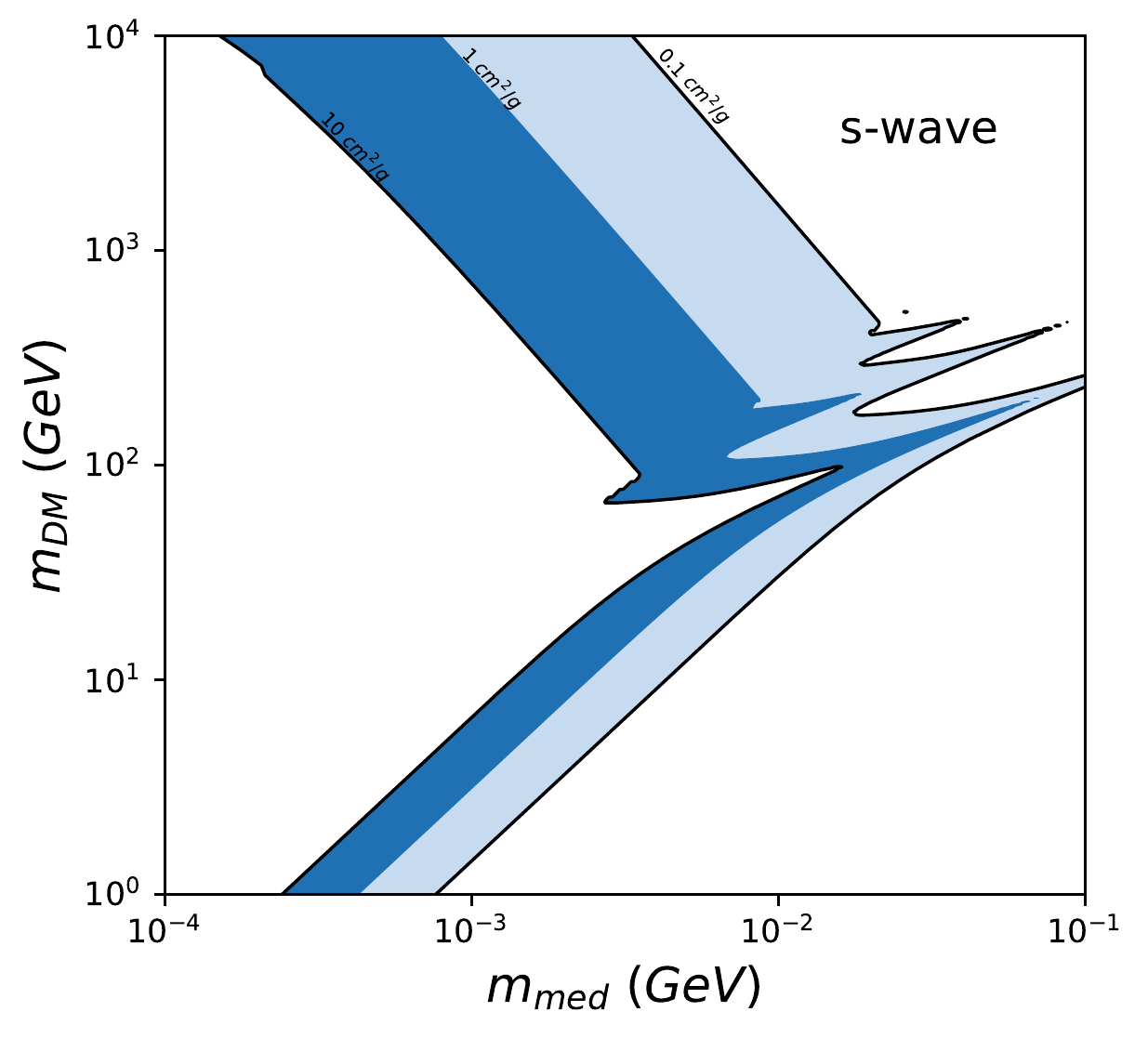}
\end{minipage}
\begin{minipage}{0.23\textwidth}
  \centering
\includegraphics[scale=0.35]{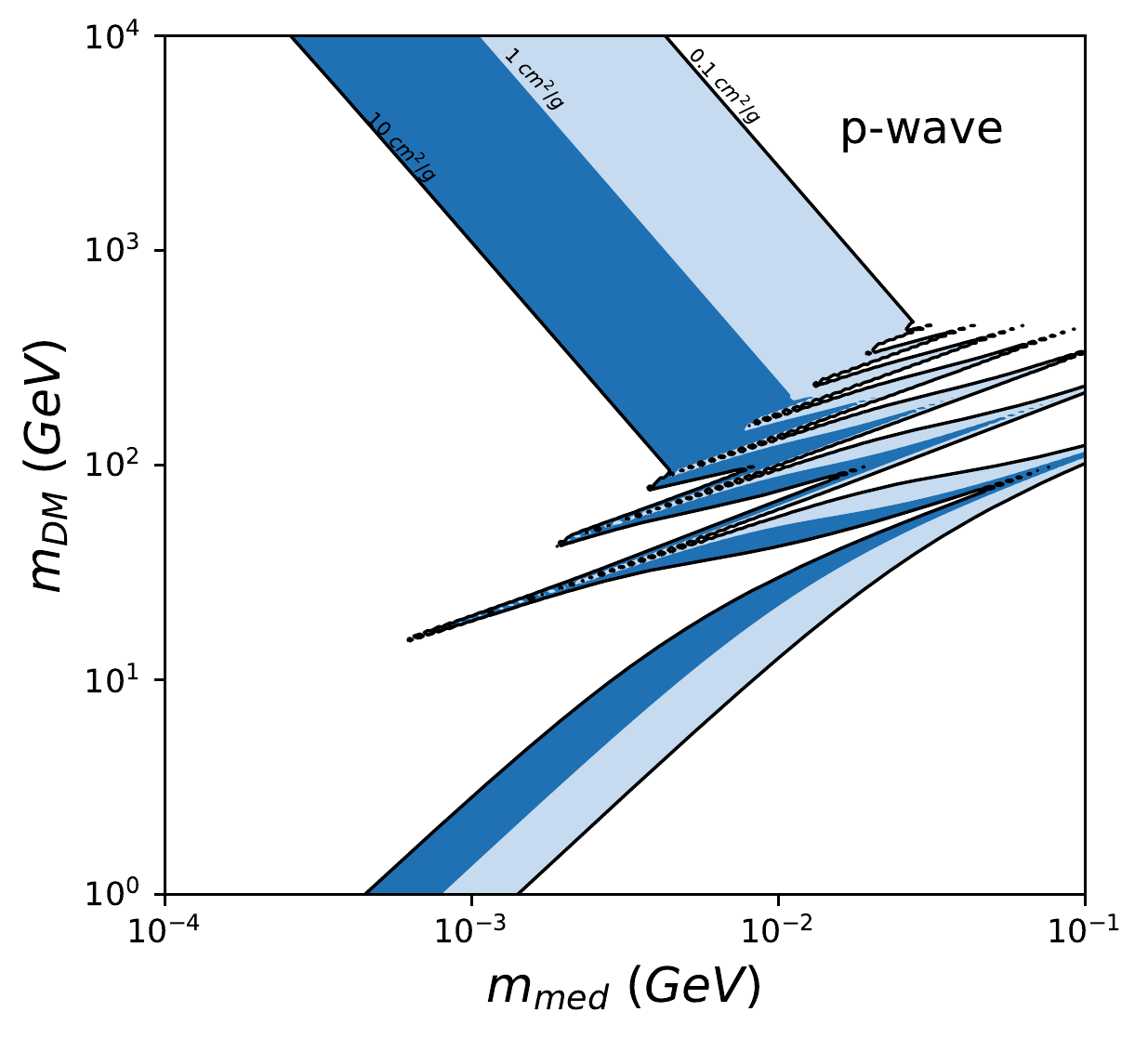}
\end{minipage}\\
\begin{minipage}{0.24\textwidth}
  \centering
\includegraphics[scale=0.35]{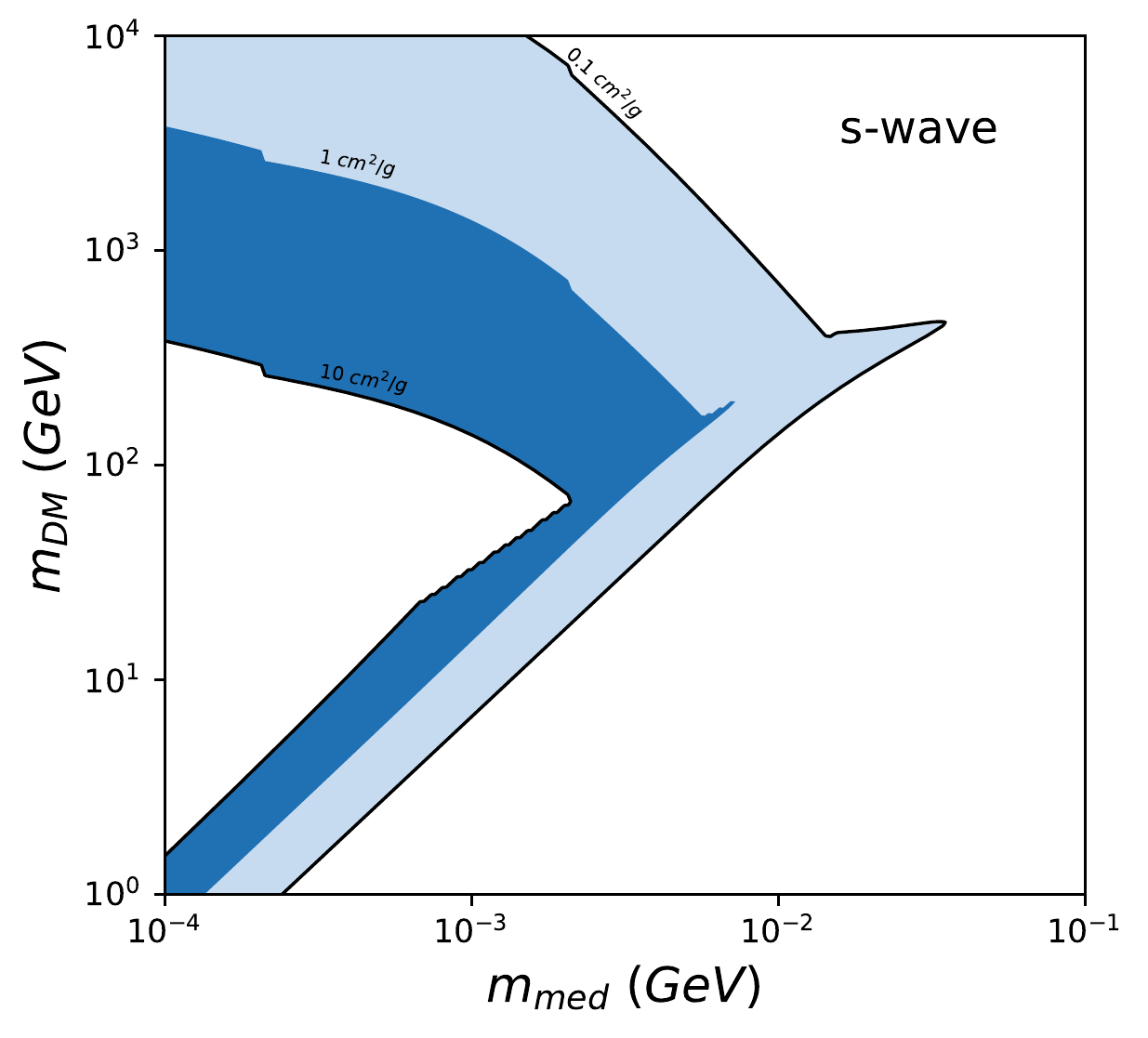}
\end{minipage}
\begin{minipage}{0.23\textwidth}
  \centering
\includegraphics[scale=0.35]{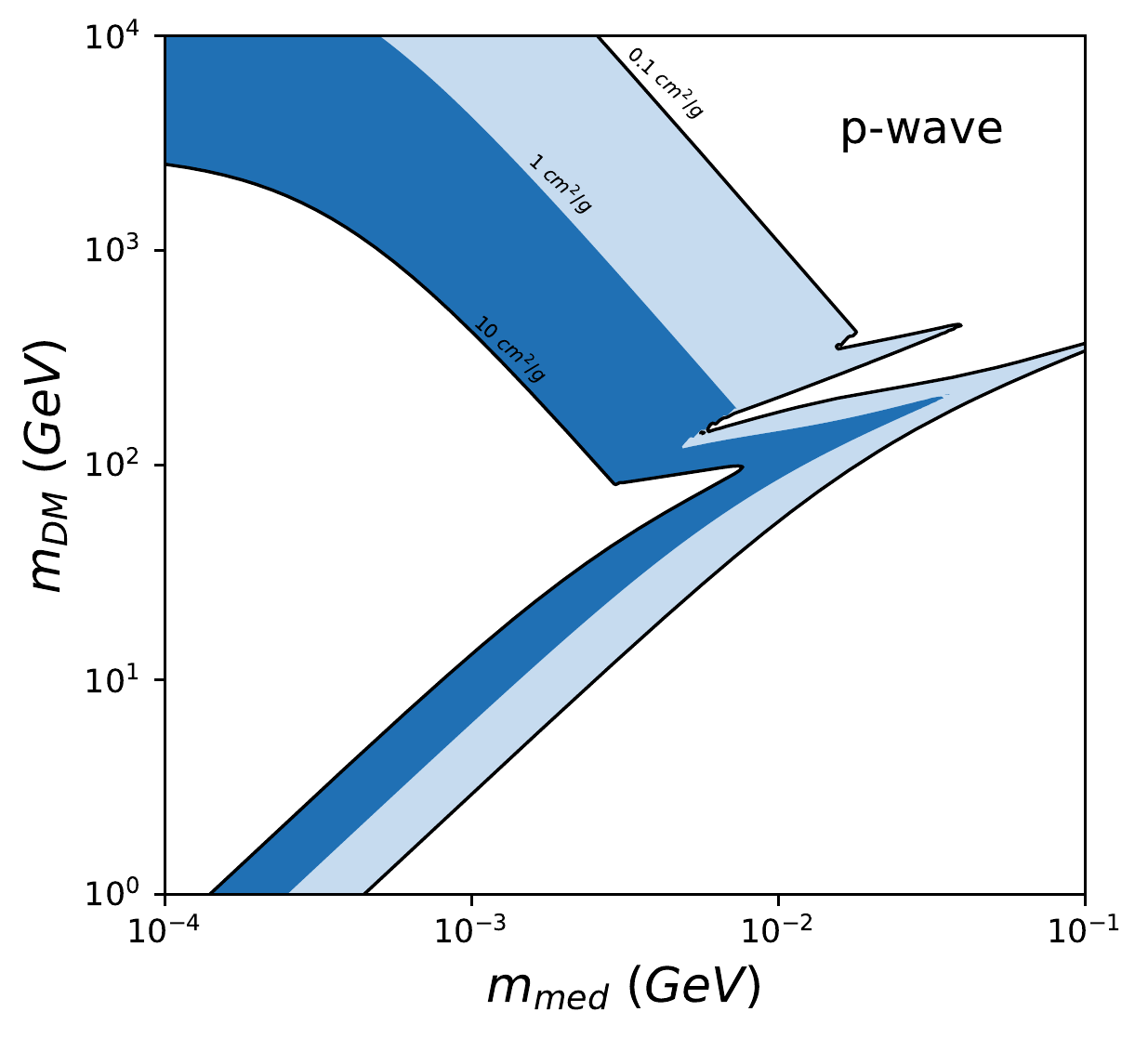}
\end{minipage}\\
\begin{minipage}{0.24\textwidth}
  \centering
\includegraphics[scale=0.35]{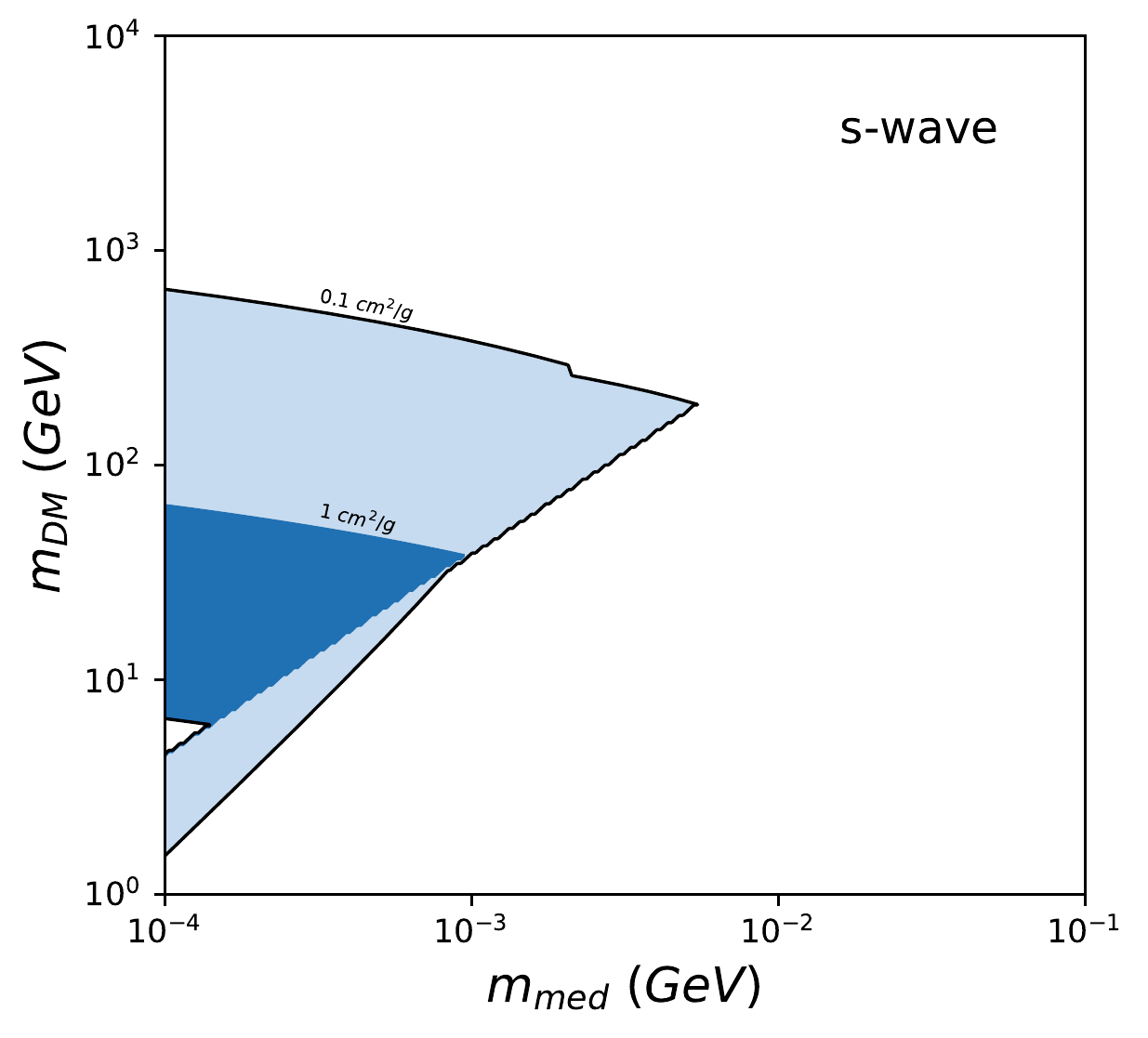}
\end{minipage}
\begin{minipage}{0.23\textwidth}
  \centering
\includegraphics[scale=0.35]{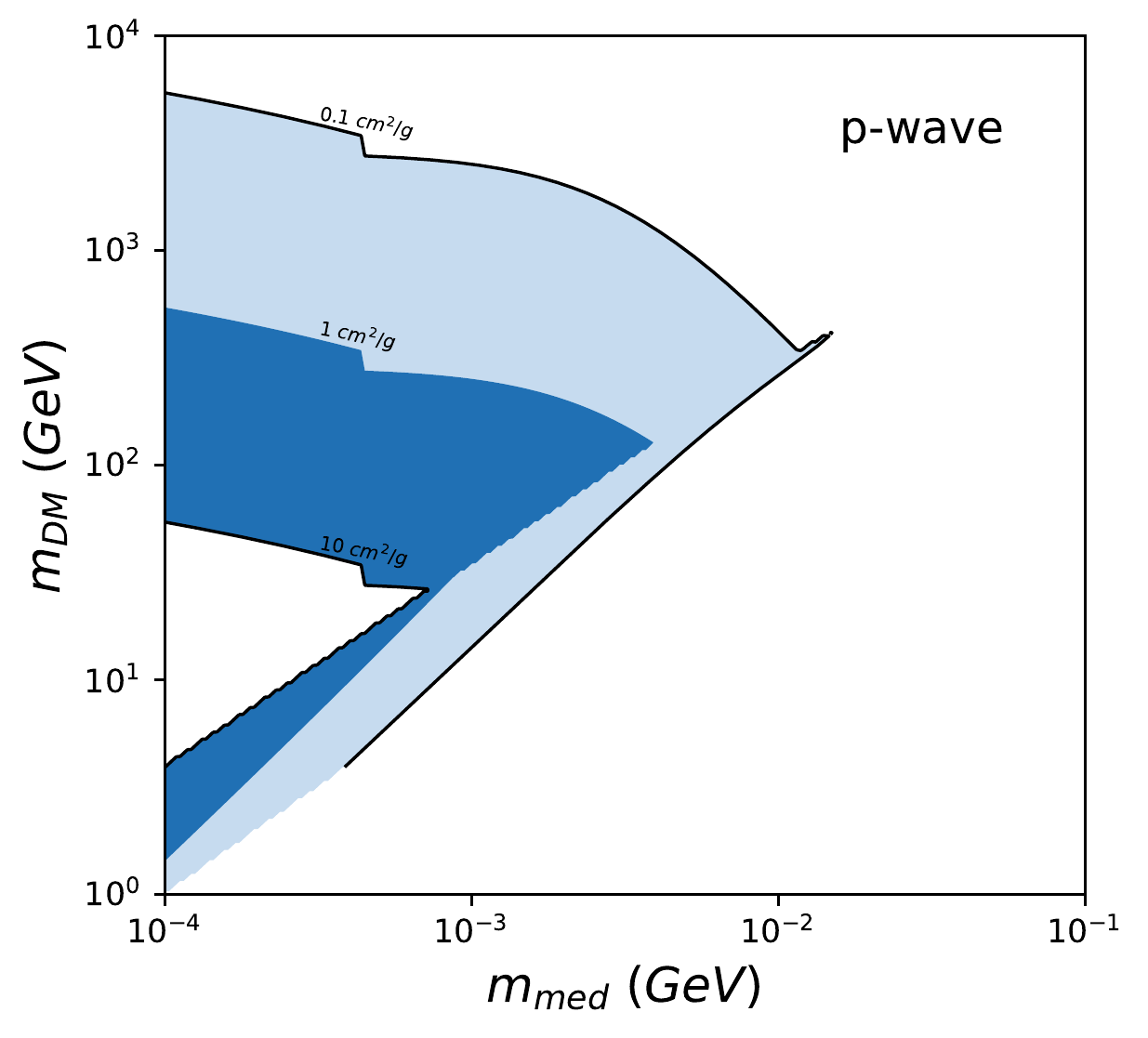}
\end{minipage}\\
\caption{Left (Right): Self interaction constraints, from top to bottom, for a DM annihilation into a pair of light vectors (scalars) mediator a factor $10^{2,4,6}$ smaller than the thermal value at freezeout, which (if there were no other annihilation channel) would lead to  $\Omega_{\chi}h^{2}= (1.2 \cdot 10^{2}, \,1.4 \cdot 10^4, \,1.6 \cdot 10^{6})\times\Omega_{0}h^{2}$ (from the fact that $\Omega_{\chi}h^2 \propto \log (\langle \sigma v \rangle...)/\langle \sigma v\rangle$). Note that there is no significative difference if we take the Sommerfeld enhancement at freeze-out into account, what we didn't do here.} \label{fig:SI_swave_Oh2}
\end{figure}

\subsection{CMB constraints} 

There is an all series of CMB constraints applying on the light mediator scenario. The main constraints are both on the DM annihilation and on the decay of the light mediator \cite{Bernal:2015ova,Bringmann:2016din}. 

\subsubsection{CMB constraint on DM annihilation rate}

The DM annihilation cross section is bounded from above by the CMB if its annihilation produces photons. This constraint obviously applies to the DM annihilation 
into a pair of light mediators if, through the portal interaction, the light mediator decays into any SM particles except neutrinos. The constraint on the annihilation cross section at the time of recombination  (i.e.~at redshift $z\sim1100$) is
\begin{equation}
\langle \sigma v\rangle_{rec}\lesssim N_\chi \cdot 4 \times 10^{-25} \hbox{ cm}^3 \hbox{s}^{-1} \Big(\frac{f_{eff}}{0.1}\Big)^{-1}\Big(\frac{m_\chi}{100\hbox{ GeV}}\Big)\,,
\label{cmbconstraint}
\end{equation}
where $f_{eff}$ is related to the fraction of the released energy ending up in photons or electrons, with $f_{eff}\gtrsim 0.1$ for any SM final states except neutrinos (see e.g.~\cite{Slatyer:2015jla}), and where $N_\chi=1,2$ for Majorana and Dirac dark matter respectively. The implications of this constraint have been analyzed at length in \cite{Bringmann:2016din,Cirelli:2016rnw}. Fixing the annihilation cross section into a pair of light mediators to the thermal value, and assuming a decay of the light mediators into SM particles other than neutrinos, a s-wave annihilation scenario is excluded due to the fact 
that in this case at the recombination time the cross section is largely boosted by the Sommerfeld mediator multi-exchange effect (see Fig.~1 of \cite{Bringmann:2016din}).
The boost is especially large at this time because DM at recombination time is highly non-relativistic ($v/c\lesssim 10^{-7}$). 
This implies in particular that the vector model of Eq.~(\ref{medcouplV}) is excluded for $m_{med}>2 m_e$, because in this case the light vector boson essentially decays into charged SM leptons, see Ref.~\cite{Bringmann:2016din} for a detailed discussion.
Looking at the various ways to avoid this CMB constraint along the "model-building features" above, one finds the following possible ways out:

\begin{itemize} 

\item {\it The annihilation into mediators proceeds in p-wave}. 
In this case
the annihilation turns out to be much less boosted at recombination time than in the s-wave case and is not excluded by the CMB constraint, see Fig.~1 of \cite{Bringmann:2016din}.
This constitutes the most straigthforward way out. This singles out a scalar mediator with DM a Majorana or Dirac fermion, as for Eq.~(\ref{medcouplphi}).

\item {\it The mediator decays dominantly into neutrinos}. If the light mediator decay only into SM particles via a portal interaction, the CMB constraint can be avoided
for s-wave annihilation if the decay proceeds into neutrinos.
As the light mediator must be a SM singlet, decay of neutrinos in simple models means also similar decay rate to the associated charged leptons (which do produce photons), which excludes the scenario unless 
$m_{med}$ is below the corresponding mass threshold. The neutrino option is therefore possible only if $m_{med}<2m_e$ \cite{Bringmann:2016din} or if the decay proceed only to muon and/or tau neutrinos (with $m_{med}<2m_{\mu,\tau}$). 
For the vector model of Eq.~(\ref{medcouplV}), none of these options is possible because in this model the vector boson couples equally to all flavors and because for $m_{med}<2m_e$ the dominant decay is not into neutrinos but into 3$\gamma$. The last option will be nevertheless discussed below for a model with a light $Z'$ coupling only to muon and tau flavors, model $E_{\mu-\tau}$ in Section~\ref{sec:neutrinowayout}.

\item {\it The mediator has a very large lifetime}.  If the mediator has a lifetime much larger than the age of the Universe at recombination time,
the CMB constraint of Eq.~(\ref{medcouplV}) disappears. However, if the lifetime is larger than the age of the Universe today, one has nevertheless to make sure that the light mediator doesn't overclose the Universe, see above. Actually this is what happens in the vector model of Eq.~(\ref{medcouplV}) for $m_{med}\lesssim 10$ keV.\footnote{Below the electron threshold, the light mediator lifetime is given by $\tau _{3\gamma} \simeq 2\times 10^{3} \left(10^{-4}/\alpha '\right)\left(100~\text{keV}/m_{\gamma '}\right)^{9}$ sec \cite{Pospelov:2008jk}.} In this case the dominant decay proceeds into 3$\gamma$ and leads to a lifetime so large that the vector boson is effectively stable. As a result, together with the fact that for $m_{med}\lesssim 2 m_e$ one also produces $\gamma$-rays beyond the extragalactic $\gamma$-ray background, there is no allowed window for this model at all \cite{Bringmann:2016din,Cirelli:2016rnw}.

\item {\it The mediator decays dominantly into lighter hidden sector particles}: if the decay of the light mediator dominantly proceeds into non SM particles, i.e.~into lighter hidden sector particles,
then the constraints on the light mediator are traded for constraints on the fate of these lighter particles. If it is absolutely stable or if its mass is below the $e^\pm$ threshold, so that it decays dominantly into neutrinos, the CMB constraints could be avoided. These options allow a mass mediator above the $2m_e$ threshold, even if the annihilation is of the s-wave type. We will not look any further at this possibility but it could be worth exploring it.
  
\item {\it DM annihilation rate into light mediators smaller than usual thermal value}: considering a cross section into a pair of mediators below the thermal value obviously helps. An upper bound on the annihilation into light mediators cross section over the thermal value can be expressed using the upper bound coming from CMB, Eq.~(\ref{cmbconstraint}),

\begin{eqnarray}
&& \frac{\left\langle\sigma v\right\rangle}{\left\langle\sigma v\right\rangle _{th}} \leqslant 15.7  N_{\chi} \frac{S_{fo}}{S_{rec}}\left(\frac{f_{eff}}{0.1}\right)^{-1}\left(\frac{m_{\chi}}{100\text{~GeV}}\right)
\label{upperboundsigmas}
\end{eqnarray}

with $S_{fo}$ and $S_{rec}$ the Sommerfeld factors at freeze-out and at the recombination era respectively. In Fig.~\ref{fig:CMBsvratio} we give the value of the upper bound of Eq.~(\ref{upperboundsigmas}), as a function of $m_{DM}$ for $m_{med}=0.3,3,30,300$~MeV.
This plot shows that for many values of $m_{DM}$ and $m_{med}$, one doesn't need to reduce the cross section by a large factor with respect to the thermal value to accommodate the CMB constraint. For example, for $m_{DM}=1$~TeV and $m_{med}=3$~MeV we obtain in this way that the annihilation cross section at freezeout must be about $120$ times smaller than the thermal value.\footnote{Actually, as Fig.~\ref{fig:CMBsvratio} shows, there are instances which does not exclude the thermal value (for $m_{DM}$ equal to a few tens of GeV and $m_{med}$ larger than a few tens of MeV) but these values are not compatible with the self-interactions constraints for the vector model of Eq.~(\ref{medcouplV}), as a comparison with 
Fig.~\ref{fig:SI_swave_pwave} shows (but interesting to point out because possibly such instances could work for other models).}

\begin{center}
\begin{figure}[h!]
\includegraphics[scale=0.8]{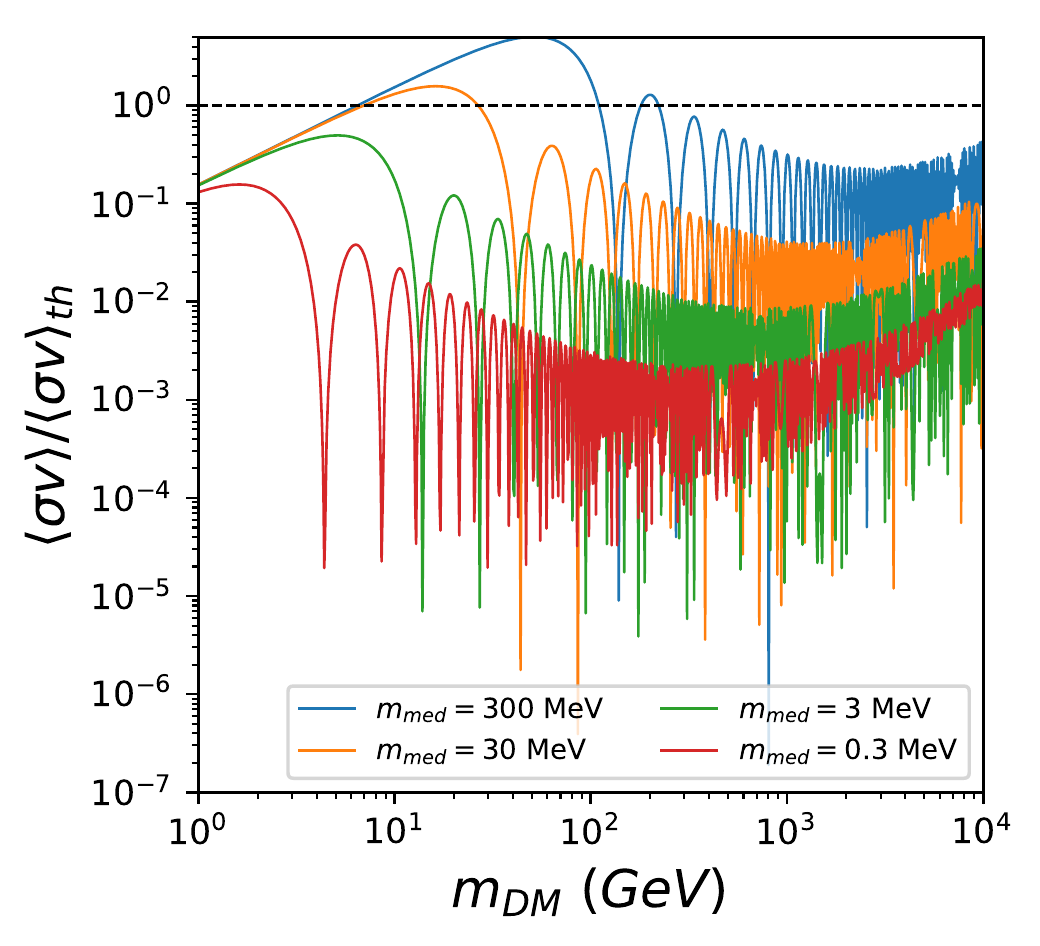}
\caption{CMB upper bound on the s-wave annihilation rate into light mediators cross section, normalized to the thermal value. The dashed horizontal line corresponds to a cross section with the thermal value.}\label{fig:CMBsvratio}
\end{figure}
\end{center}

As stressed above, values of  $\frac{\left\langle\sigma v\right\rangle}{\left\langle\sigma v\right\rangle _{th}}$ down to $\sim 10^{-8}$ turns out to be compatible with self-interaction constraints. As also stressed above, if $T'/T=1$ at freezeout time, this implies another (faster) annihilation channel for the freezeout. Such annihilation channel must also satisfy Eq.~(\ref{cmbconstraint}), and is boosted in the same way if it proceeds in s-wave. Thus, the constraints on this annihilation channel are the same as above,  either the annihilation must be p-wave or this light particle must be stable or must have a decay channel which do not produce photons.
This suggests the following simple scenario: subdominant s-wave annihilation of DM into a pair of light mediators, with dominant  p-wave annihilation into an extra (e.g.~heavier) particle.  This simple way out will be at the basis of the $B_{\gamma '}$ and $B_\phi$ models in Section \ref{subdomannihwayout} below.
  
\item {\it $T'/T\ll 1$}:  as already said above too, if $T'/T<1$, the relic density constraint requires a smaller DM annihilation cross section than for $T'/T=1$. 
Thus, in this way too, one could satisfy the constraint of Fig.~\ref{fig:CMBsvratio}, even if the annihilation  into a pair of light mediators is of the s-wave type and dominant.
This will be the first way out we will consider in more detail below, models $A_{\gamma '}$ and $A_{\phi}$ in Section \ref{Tprimewayout} below. Note that, if $T'/T<1$, the DM velocity at recombination is lower than if $T'/T=1$. But this doesn't change the Sommerfeld boost factor, because already for $T'/T=1$ one lies in the regime where this boost is saturated and doesn't vary with $v$ anymore, see Fig.~1 of \cite{Bringmann:2016din}.
\end{itemize}


\subsubsection{CMB constraints on mediator decay}


If the light mediator has a few MeV masses, it can decay only to $e^\pm$, neutrinos and photons.
If it does mainly to electromagnetic channels, $e^+e^-$ or $\gamma \gamma$, and has a lifetime larger than the age of the Universe at recombination, $\tau_{med}\gtrsim 10^{12}$~sec, the energy injection it implies can 
largely change the anisotropy structure of the CMB. This would have been already observed unless the light mediator abundancy is low. For a mediator lifetime equal to $10^{13,15,17,19,21,23,25}$~sec this gives an upper bound on the amount of mediator particle (there would be today if there were no decay).
This upper bound is \cite{Slatyer:2012yq,Poulin:2016anj} 
$\Omega_{med}h^2\lesssim  10^{-11.7,-10.9,-9,-7,-5,-3,-1}$ respectively. 
CMB distortion constraints on decay into $e^+e^-$ or $\gamma \gamma$ are also relevant for lifetime shorter than the age of the Universe at recombination, basically implying that for 
$\Omega_{med} h^2 =10^{0,-1,-2,-3,-4,-5}$
, the lifetime should be shorter than 
$\sim10^{5.7,6.2,6.8,8.2,10.2,12.2}$~sec respectively, see \cite{Poulin:2016anj}  and refs therein.
Note again that since a MeV light mediator must be a SM singlet, production of $\nu\bar{\nu}$ basically comes (in simple frameworks) with similar production of $e^+ e^-$ as soon as $m_{med}> 2 m_e$. 

For a  lifetime larger  than the age of the Universe and $m_{med}<2 m_e$, there are also constraints coming from X-ray observation, which in a conservative way requires $\tau_{med}\gtrsim 10^{28}\,\hbox{sec}\times \Omega_{med}h^2/0.12$ \cite{Essig:2013goa,Boddy:2015efa,Riemer-Sorensen:2015kqa}, and which applies basically to any scenario where the mediator decay would produce SM particles (apart from neutrinos).

\subsubsection{CMB constraint on $N_{eff}$}

Last but not least, the CMB put stringent constraints on the modification of the Hubble constant at recombination time, i.e.~on the number of relativistic degrees of freedom at this time. The latter is in general expressed in terms of the effective number of neutrinos, $N_{eff}$. The best constraint on the minimum of effective neutrinos, from Planck satellite, gives $2.66<N_{eff}< 3.33$ at 2$\sigma$ level \cite{Aghanim:2018eyx}. For the light mediator this translates into constraints on its lifetime and abundancy at the time of neutrino decoupling. Light mediator particles which decay after neutrino decoupling reheats the photons but not the neutrinos. A detailed study of this effect has been recently done in Ref.~\cite{Hufnagel:2018bjp}. A bound applies only when a sizeable fraction of the mediators decay after neutrino decoupling, when $T<T^{\nu}_{dec} \simeq 1$~MeV, or equivalently when $t>t_{dec}^\nu\simeq 7$~sec.
This means that a bound applies only when $t_{dec}=\tau_{med}T(t_{dec})/m_{med}\gtrsim t_{dec}^\nu$ where $T(t_{dec})/m_{med}$ is approximately the Lorentz boost factor which applies when $m_{med}\lesssim T^{\nu}_{dec} $.

Assuming for instance that the light mediator has $m_{med}=10$~MeV and that it decouples chemically from the thermal bath at $T=T_{dec}=10$~GeV, this constraint excludes scenarios with a mediator lifetime equal to $10^{1,2,3,4}$~sec if 
$n_{med}/n_\gamma\gtrsim 10^{-0.3,-0.7,-1.2,-1.7}$ \cite{Hufnagel:2018bjp}, which corresponds to 
$\Omega_{med}h^2\gtrsim 10^{5.6,5.2,4.7,4.2}$
 respectively. Here we have used the conversion formula
\begin{eqnarray}
\Omega_{med}h^{2}&\simeq & 8\times 10^{4}\left(\frac{g_{\star}^{eff}}{g_{\star}^{S}}\right)\left(\frac{m_{med}}{\text{MeV}}\right)\left(\frac{n_{med}}{n_{\gamma}}\right)\left(\frac{T'}{T}\right)^{3},\nonumber\\
\end{eqnarray}
where all quantities are meant to be taken at $T=T_{dec}$.
For relativistically decoupling light mediators this $N_{eff}$ constraint is fully relevant.
From the results of Ref.~\cite{Hufnagel:2018bjp}, one finds that  the $N_{eff}$ constraint is compatible with such relativistic decoupling values of the light mediator number densities, if the light mediator lifetime is above a lower bound which we give for various values of $m_{med}$ in Table~\ref{table:Neff_Az} for the $A_{\gamma '}$ model (top) and for the $A_\phi$ model (bottom). All these lifetimes have been obtained in Ref.~\cite{Hufnagel:2018bjp} assuming here too that  $T_{dec}=10$~GeV. For another decoupling temperature, the bound will be moderately affected by a factor of the relativistic degrees of freedom contributing to the entropy, $g_s(10\hbox{ GeV})/g_s(T_{dec})$ due to decoupling of relativistic species between both temperatures. 

\begin{table*}[t]
\begin{tabular}{|c|c|c|c|c|c|c|c|c|c|c|}
\hline 
$m_{\gamma '}$~(MeV)            & 0.01    & 0.03    & 0.1     & 0.3   & 1    & 3    & 10    & 30    & 100   & 300   \\ \hline
$N_{eff}$            & $10^{6.3}$       & $10^{5.3}$      & $10^{4.3}$     & $10^{3.3}$     & $10^{2.2}$      & $10^{1.1}$      & $10^{0.1}$     & $10^{-0.4}$     & $10^{-0.7}$    & $10^{-0.9}$     \\
Photodis./Entropy inj.         & $10^{5.9}$       & $10^{4.9}$      & $10^{3.9}$     & $10^{3.1}$     & $10^{2.4}$      & $10^{2.0}$      & $10^{1.0}$     & $10^{0.0}$     & $10^{-0.4}$    & $10^{-0.5}$     \\ \hline
\hline 
$m_{\phi}$~(MeV)            & 0.01    & 0.03    & 0.1     & 0.3    & 1    & 3    & 10    & 30    & 100   & 300   \\ \hline
$N_{eff}$            & $10^{7.3}$       & $10^{6.3}$      & $10^{5.3}$     & $10^{4.3}$     & $10^{3.2}$      & $10^{2.1}$      & $10^{1.1}$     & $10^{0.1}$     & $10^{-0.5}$    & $10^{-0.7}$     \\
Photodis./Entropy inj.         & $10^{6.8}$       & $10^{5.9}$      & $10^{4.8}$     & $10^{3.9}$     & $10^{3.0}$      & $10^{2.4}$      & $10^{2.0}$     & $10^{1.0}$     & $10^{-0.1}$    & $10^{-0.4}$     \\ \hline
\end{tabular}
\caption{Upper bound on the light mediator lifetime (in seconds) from CMB $N_{eff}$, BBN photodisintegration and BBN Hubble constant/entropy injection constraints, assuming a relativistic decoupling of the light mediator 
for the $A_{\gamma '}$ model (top) and the $A_{\phi}$ model (bottom). A value of $T'/T=1$ has been assumed at DM freezeout time.}\label{table:Neff_Az}
\end{table*}

\subsection{BBN constraints} 

As well known, new light degrees of freedom may easily affect the Big Bang Nucleosynthesis process. Here we will distinguish 2 main ways: a) {photodisintegration of light nuclei} during and after the BBN process from decay of the light particle and, b)  modification of light nuclei abundances from modification of the Hubble constant and entropy injection by the light mediator.

\subsubsection{Photodisintegration BBN constrains} 

Photons from the decay product of the light mediator can dissociate the light nuclei formed during BBN. This will be in particular the case if the light mediator decays into $\gamma\gamma$ or $e^+ e^-$ and if the photons produced in this way have an 
energy above the photodisintegration thresholds of $^2H$,$^3H$ and $^4He$, $E_\gamma^{^2H}=2.22$~MeV, $E_\gamma^{^3H}=6.92$~MeV and $E_\gamma^{^4He}=28.3$~MeV.
The constraints on the decay lifetime into $e^+e^-$ can be found in Fig.~9 of \cite{Hufnagel:2018bjp} (see also \cite{Jedamzik:2006xz,Forestell:2018txr}) for a light mediator between keV and GeV. 
If the light mediator has a mass below twice the $^2H$ threshold, $m_{med}<4.4$~MeV, there are no bounds.
Due to the Deuterium bottleneck, the light nuclei predominantly form only when $t\geq180$~sec. Thus, there is not much of a constraint on the light mediator for shorter lifetimes (or actually even up to $\sim 10^{3.5}$~sec).
For instance, for $m_{med}=10$ MeV and if the lifetime is equal to $10^4$~sec, the photodisintegration constraint requires $\Omega_{med}h^{2}<10^{4.9}$. For lifetimes equal to $10^6$~sec and $10^8$~sec
one needs $\Omega_{med}h^{2}<10^{3.8}$ and $\Omega_{med}h^{2}<10^{3.1}$ respectively.

\subsubsection{BBN constraint from modification of the Hubble constant and entropy injection}

As long as the mediator has not decayed it modifies the Hubble constant. Subsequently its decays inject entropy into the medium.
Both effects modify the relation between time and temperature, which leads to modification of the light nuclei abundances.
From the observational constraints on these abundances this gives an upper bound on the light mediator number (prior to decay) as a function of its lifetime. 
Similarly to the $N_{eff}$ bound, this bound applies when the number of light mediators is large prior to the decay, as is the case when it decouples relativistically.
The entropy injection constraint is relevant only when a sizable fraction of the mediators decays after the BBN process has started, when $t\gtrsim t_{BBN}\simeq 180$~sec, or equivalently when $T \lesssim T_{BBN} \simeq 0.07$~MeV.
This means that it can be relevant when $t_{dec}=\tau_{med}T(t_{dec})/m_{med}\gtrsim t_{BBN}$, where $T(t_{dec})/m_{med}$ is the relativistic Lorentz boost factor which applies when $m_{med}\lesssim T^{BBN} $.
The Hubble constant constraint instead can be relevant for smaller lifetimes, i.e.~$t_{dec}$ of order a second or more.
This can be particularly relevant if the mass of the mediator is in the multi-MeV range or more, because, in this case, when it becomes non-relativistic, its contribution to the Hubble constant increases faster than the one from relativistic species (unless it decays or annihilates), see \cite{Scherrer:1987rr,Menestrina:2011mz,Hufnagel:2017dgo,Hufnagel:2018bjp,Forestell:2018txr}.
Using the results of Ref.~\cite{Hufnagel:2018bjp}, in Table~\ref{table:Neff_Az} we give the upper bounds on the light mediator lifetime that require the Hubble constant/entropy injection constraints, together with the previous photodisintegration constraint, assuming that it decouples relativistically at $T_{dec}=10$~GeV. Note that for a mediator which decouples relativistically, the Hubble constant and entropy injection constraints turn out to be always more stringent than the photodisintegration one.

If the mediator is stable, the entropy injection constraint doesn't apply but the modification of the Hubble constant constraint can still be relevant.
If the mediator yield, $Y_{med}=n_{DM}/s$, has already by the BBN epoch the value it has today, the non-oveclosure constraint implies that the mediator number density is very suppressed at this time and the Hubble constant constraint is totally irrelevant. If this is not the case, for instance if the light mediator annihilates and undergoes a Boltzmann suppression around or after the BBN epoch (i.e.~when $m_{med}\lesssim 10$~MeV), the Hubble constant constraint can be relevant. An estimate of the constraint applying in this case can be obtained by requiring that the number of effective additional neutrino degrees of freedom the mediator implies at $T\sim 1$-$10$~MeV doesn't exceed the one allowed by BBN (see also  \cite{Duerr:2018mbd,Hufnagel:2017dgo}):
\begin{equation}
\Delta N_{eff}^{med}< 0.31 \hspace{0.5cm}\text{(95\% C.L.)}
\label{DeltaNeff}
\end{equation}
taken from \cite{Patrignani:2016xqp,Hufnagel:2017dgo}.
If the mediator has a mass well below MeV it is still relativistic and non Boltzmann suppressed by the BBN time and $\Delta N_{eff}^{med}\simeq \frac{4g_{med}}{7}\left(\frac{g_{\star}^{S}\left(10\text{~MeV}\right)}{g_{\star}^{S}\left(T_{dec}\right)}\right)^{4/3}$. For $m_{med}\gtrsim 10$~MeV the mediator yield is already Boltzmann suppressed and $\Delta N_{eff}^{med}$ is suppressed accordingly. For intermediate masses, to assume that the mediator yield doesn't vary during this epoch, as assumed in Eq.~(\ref{DeltaNeff}),  is not much realistic and a proper determination of the constraint must be performed, see \cite{Hufnagel:2017dgo}.


\subsubsection{Summing up the constraints on light mediator decay and number density}

From the above one concludes that, on top of the CMB constraint on $N_{eff}$, both the BBN photodisintegration and Hubble constant/entropy injection constraints are also relevant for what concerns the abundancy and lifetime of the mediator. 
These 3 constraints together basically exclude all simple scenarios unless one of the following main options are fulfilled:
\begin{itemize}
\item The light mediator lifetime is below the values given in Table~\ref{table:Neff_Az}  \cite{Hufnagel:2018bjp}.  In this way we avoid all these constraints even if the light mediator decouples relativistically and if afterwards the only process it undergoes is to decay largely into electromagnetically interacting SM particles. 
This will be the way out we will use for models $B_{\gamma '}$ and $B_{\phi}$ in Section \ref{subdomannihwayout}.
\item There is a mechanism to suppress the light mediator population below the value resulting from its relativistic decoupling at DM freezeout. 
 The reduction of the number density required by the $N_{eff}$ and entropy injection constraints are relatively mild (2-3 orders of magnitude), whereas the one required by the photo-disintegration for large lifetime and $m_{DM}>4.4$~MeV is larger (up to $\sim$ 7-12 orders of magnitude).
 In order to reduce this number density the simplest options are: 
 \begin{itemize} 
  \item a) that the $T'/T$ ratio is smaller than unity so that its number is reduced by a $(T'/T)^3$ factor. If the mass is below 4.4 MeV one just needs $T'/T\lesssim 1/5$ to reduce the mediator abundancy by the $\sim$ 2-3 orders of magnitude quoted above.
For a larger mass, whatever the lifetime is, one can always reduce the abundance in this way so that  
$\Omega_{med}h^2\lesssim 10^{-4,-5}$ (by today if it was not decaying). This requires 
$T'/T\lesssim 10^{-2,-3}$. This will be the simple way out we will use for models $A_{\gamma'}$ and $A_{\phi}$ in Section \ref{Tprimewayout} below.

 \item b) that the light mediator population is reduced after decoupling with DM through an extra annihilation process (so that it decouples non-relativistically) or through a decay to an extra particle (with constraints holding then on this extra particle). This will be the (more complicated) way out we will consider for model C in Section \ref{sec:pwavemeddecay} below.
Let us repeat here that such a reduction of the number density is necessary even if the mediator is stable, due to the non overclosure constraint above. 
As explained above, in this case the constraint from a modification of the Hubble constant can also be relevant, Eq.~(\ref{DeltaNeff}).

 \end{itemize}

 \end{itemize}

All these constraints are on the mediator abundancy and lifetime. Whether the DM annihilation into light mediators is dominant or subdominant or whether it is s-wave or p-wave, is of  less importance here.

\subsection{Supernovae constraints}
Dark matter models with a MeV-scale mediator can also be constrained by observations of supernovae collapse, and more specifically SN1987A \cite{Kazanas:2014mca}. The physics of supernovae is subject to systematic uncertainties and such constraints are less robust that those based on cosmological production and decay of light mediator. In particular, for the models of Eqs. \ref{medcouplV} and \ref{medcouplphi} and for $\sim$MeV mediator, one can derive constraints on the portal parameter $\epsilon\notin [10^{-10},10^{-6}]$ \cite{Kazanas:2014mca,Chang:2016ntp,Mahoney:2017jqk,Chang:2018rso} and $\lambda _{\phi}\notin [10^{-5},10^{-3}]\times (v_{\phi}/\text{GeV})$ \cite{Cline:2013gha,Krnjaic:2015mbs,Chang:2017ynj} respectively. These constraints can be relevant in some cases, in particular for the scalar mediator case \cite{Hufnagel:2018bjp} but are in most cases less stringent than the other constraints, so that they do not change much the overall picture. In the following all the results we will present will be always for parameters which fulfill those constraints.
\subsection{Direct detection constraints} 

This constraint is all about the type and size of portal one assumes. If the light mediator couples sizably to nucleons through a portal, the DM scattering on nucleon is largely enhanced by the lightness of the mediator. The t-channel mediator propagator is proportional to $1/(2m_N E_R+m^2_{med})$, with $m_N$ and $E_R$ the mass and recoil energy of the nucleon. Since the recoil energy threshold of direct detection experiments is of order few keV, this means that the lightest is the mediator the larger is the signal, down to value of $m_{med}\sim 30$~MeV where the denominator saturates. With respect to usual frameworks where the mediator would lie at the electroweak scale or above, the enhancement of the cross section is huge. This strongly bounds from above the combination of couplings involved in this scattering. For the vector and scalar boson mediators of Eqs.~(\ref{medcouplV}) and (\ref{medcouplphi}) 
the direct detection constraint is on $\kappa _{KM}\equiv\sqrt{\alpha'/\alpha}\cdot \epsilon$  and $\kappa _{HP}\equiv y_{\phi}\sin\left(2\theta\right)/2$, with $\alpha'=g_{\gamma '}^2/4\pi$ and $\tan2\theta = \lambda v_{\phi}v_{h}/\left(m_{h}^{2}-m_{\phi}^{2}\right)$.

\begin{center}
\begin{figure}[h!]
\includegraphics[scale=0.70]{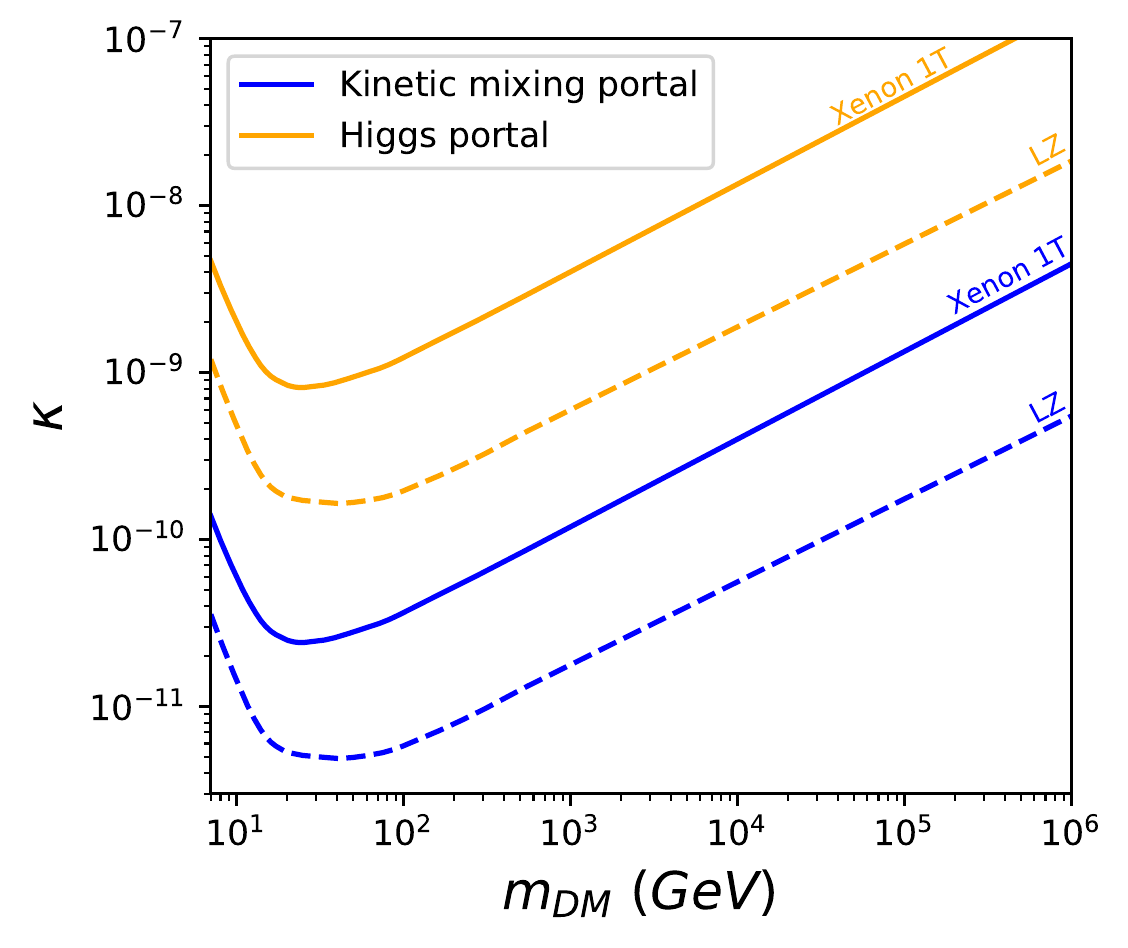}
\caption{Direct detection constraints for an interaction mediated by a light mediator (i.e. $m_{med}\lesssim 30$~MeV). The blue (orange) solid line is the current constraints from Xenon 1T experiment \cite{Aprile:2018dbl}, while the dashed blue (orange) line shows prospects for LZ experiment after 1000 days of exposure in the kinetic mixing (Higgs) portal model, from \cite{Hambye:2018dpi}.}\label{fig:DD_Constraints}
\end{figure}
\end{center}

For a light mediator below $\sim 30$~MeV and $m_{DM}$ above $\sim$~GeV, as typically required by self-interactions constraints, the direct detection constraint doesn't depend on the mass of the mediator, and typically requires $\kappa_{HP} \lesssim  10^{-9}$ and $\kappa_{KM}\lesssim 3\times 10^{-10}$ for a Higgs portal or a kinetic mixing portal respectively. This is shown in Fig. \ref{fig:DD_Constraints} where limits on $\kappa _{KM}$ have been taken from \cite{Hambye:2018dpi} and limits on $\kappa _{HP}$ have been obtained from the one on $\kappa _{KM}$ from a proper recasting of the couplings\footnote{The substitution to be made is $\kappa _{HP}\rightarrow \kappa_{KM}\times\left(\frac{4\pi\alpha Z}{y_{hnn}A}\right)$, with $Z$ and $A$ the number of proton and of nucleon in a Xenon nucleus. $y_{hnn}$ is the effective coupling between the Higgs boson and nucleon, $y_{hnn}\simeq 1.2\times 10^{-3}$ \cite{Cline:2013gha}.}. To consider a value of $m_{med}\sim100$ MeV hardly relaxes this bound. 

As said above, a tiny portal is no problem at all for fulfilling the self-interaction constraints, as Figs. \ref{fig:SI_swave_Oh0} and \ref{fig:SI_swave_Oh2} are obtained assuming a negligible effect of the portal on the self-interactions, but such a tiny portal is easily in contradiction with the BBN and CMB constraints which, as discussed above, require in many setups a larger portal in order that the mediator decays fast enough into SM particles. 
For the scalar model of Eq.~(\ref{medcouplphi}) the tension between the BBN constraints and the direct detection constraint, together with self-interactions constraints, have been shown \cite{Hufnagel:2018bjp} to exclude the model, except for a very small region of the parameter space, which gets even tinier when one adds the Supernovae constraints. The still allowed region requires $m_{med}\sim 1.1$ MeV and $m_{DM}\sim 0.5$~GeV \cite{Hufnagel:2018bjp}. The smallness of this still allowed parameter space region
 stems from the fact that in this case the decay width is suppressed by the small value of the electron Yukawa coupling, whereas direct detection proceeds through the Higgs boson to nucleon coupling which is not so suppressed. For the vector model of Eq~(\ref{medcouplV}) this tension is not so strong, as a result of the fact that the vector boson decays is not more suppressed than the vector boson to nucleon coupling; both couplings involve one power of $\epsilon$ and nothing else. However as explained above, this latter case is excluded by the CMB constraint of Eq.~(\ref{cmbconstraint}) because it proceeds in a s-wave way.

To weaken the tension just explained, one possibility is to weaken the BBN constraints, along the various ways listed above.  The other possibility is to weaken the direct detection constraint.
There are various simple options to do so:
\begin{itemize}


\item {\it Frameworks where the DM to mediator coupling is reduced.} In scenarios where this coupling is reduced, i.e.~in models where the annihilation rate into light mediators is below the thermal value, because subdominant or because $T'/T< 1$, not only the  BBN/CMB constraints are reduced (see above) but also the direct detection rate. The tension just explained is consequently reduced from both the BBN/CMB side and from the direct detection side. This increases the region allowed by these constraints.

\item{\it Pseudoscalar and axial interactions}. If the scalar light mediator couples in a pseudoscalar way to SM quarks and/or to DM, the spin independent direct detection rate is largely suppressed. Note that a pure pseudoscalar interaction of the scalar to DM does not lead to any relevant self-interactions \cite{Kahlhoefer:2017umn,Agrawal:2020lea}. A CP-violating mixture of scalar and pseudoscalar interactions (with the mediator to quark  couplings essentially of the pseudoscalar type, as induced e.g.~from some UV physics) can nevertheless relax the tension between the constraints in special regions of the parameter space, see \cite{Kahlhoefer:2017umn}. Similarly axial interactions for a light vector boson mediator doesn't give much self interactions \cite{Agrawal:2020lea} (i.e.~Sommerfeld enhancement)  and leads to a suppressed spin-independent direct detection rate. We will not consider further these possibilities.

\item{\it The DM to nucleon interaction is spin dependent and/or velocity suppressed.} If DM couples to a light vector boson in an axial way or to a light scalar boson in a pseudoscalar way, the DM to nucleon interaction will be either spin dependent or suppressed by the small DM velocity, see e.g.~\cite{deSimone:2014pda}. 

Such kind of couplings lead to specific 
Sommerfeld enhancement behaviors for the self-interactions as well as for the DM annihilation. It would be worth to explore further these possibilities, which we will not do in this work.   

\item {\it Small DM to mediator mass ratio.} In this case the direct detection signal is not boosted anymore and the self-interactions constraints can still be fulfilled in the Born regime. A very minimal model of this kind fulfilling all the constraints can be found in Ref.~\cite{Chu:2016pew}. In this case one has no more velocity dependence of the self-interactions, which is not excluded but not ideal to accommodate all self-interactions constraints at different velocities. We will not consider any further this possibility here, as it is not anymore a ``light mediator'' scenario.

\item {\it Inelastic DM.} The direct detection constraint can be relaxed if DM doesn't couple elastically to the light mediator but inelastically \cite{Blennow:2016gde}. For instance 
if DM has 2 components, $\psi_1$ and $\psi_2$ with masses $m_{\psi_2}>m_{\psi_1}$ and $(m_{\psi_2}-m_{\psi_1})/(m_{\psi_2}+m_{\psi_1})\ll 1$, the $\psi_1$-$\psi_2$-$med$ interaction can lead to self interactions with a suppressed direct detection rate. This can hold if the heavier component is subleading enough to suppress the $\psi_2+N\rightarrow \psi_1+N$ rate and if the mass splitting is large enough to supress kinematically the $\psi_1+N\rightarrow \psi_2+N$ rate. We will not consider any further this possibility.
\end{itemize}


\subsection{Indirect detection constraints}

A mediator much lighter than the DM particle not only boosts the self-interactions, the direct detection and the effects on the CMB. It also boosts the annihilation today into lighter particles. It is well-known that the Sommerfeld effect for s-wave annihilation boosts indirect detection rates, as a result of the small dark matter particle velocity today. For a p-wave annihilation, indirect detection signals are in general considered as hopeless because suppressed by 2 powers of the velocity $v$. Nevertheless, in presence of a light mediator, the Sommerfeld effect can compensate for this suppression, a property which has been hardly considered (see \cite{Das:2016ced} for an example of non self-interacting model). The Sommerfeld enhancement factor in the p-wave case goes like $1/v^3$ , giving an overall $1/v$ dependence which is similar to the one arising in the Sommerfeld enhanced s-wave case. This arises in the same way as for the Sommerfeld boost enhancing the cross section at the CMB time, except that for the CMB the velocity is so small that the scaling in $v$ doesn't go anymore in $1/v$ as for indirect detection but scales as $v^0$ and $v$ for s-wave and p-wave respectively \cite{Bringmann:2016din}.

\begin{center}
\begin{figure}[h!]
\includegraphics[scale=0.70]{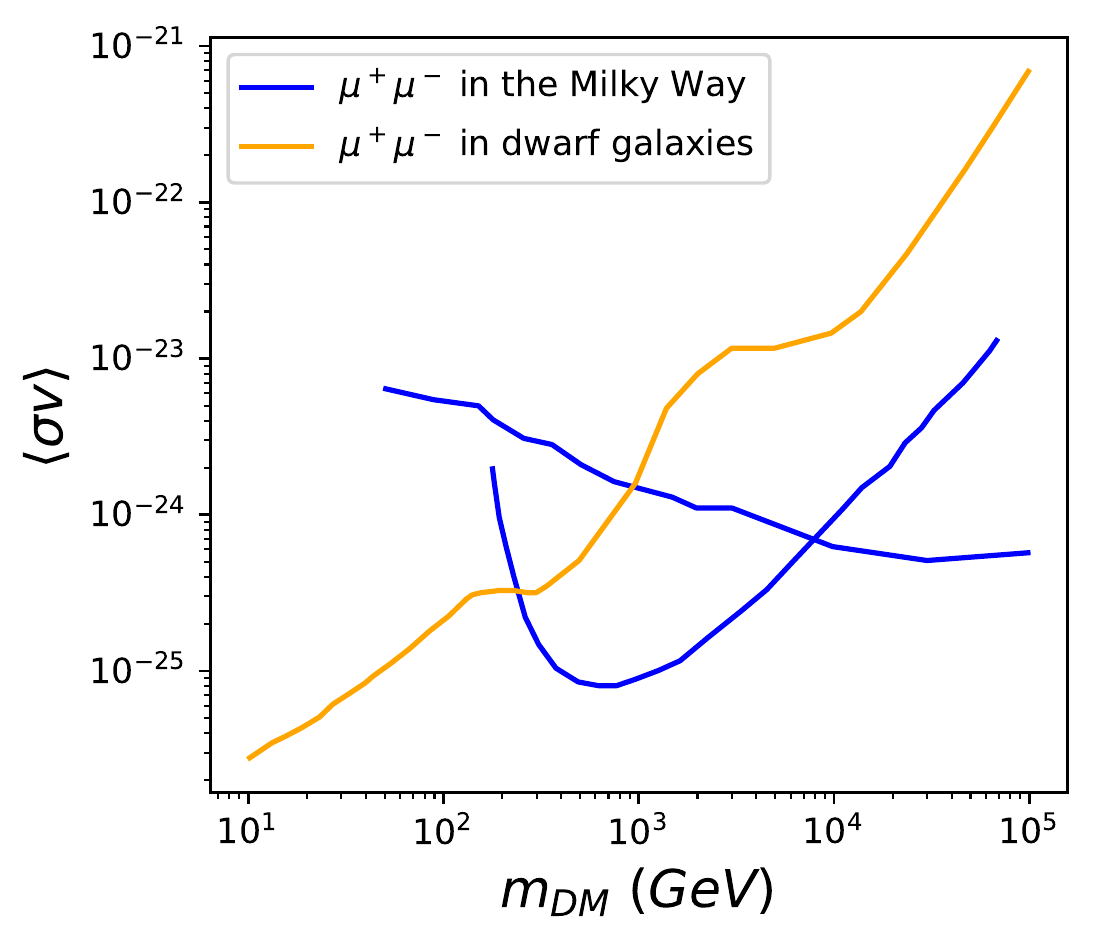}
\caption{Indirect detection constraints on the DM annihilation cross section. Blue solid lines show current constraints for annihilation in the Milky Way and have been taken from \cite{Ackermann_2012,Abramowski_2013,Ackermann_2015,Abdallah_2016} and \cite{Albert:2016emp}. Orange solid line shows current constraints for annihilation in dwarf galaxies \citep{2016}. All these constraints assume a NFW profile}\label{fig:ID_Constraints}
\end{figure}
\end{center}

Below we will show that for one of the scenarios considered, the p-wave DM annihilation can lead to an observable indirect detection signal, see section \ref{sec:pwavemeddecay}.

\subsection{Direct limits on the size of the portals}

For the light scalar mediator case the Higgs portal interaction, Eq.~(\ref{medcouplphi}), induces an invisible decay channel for the Higgs boson, $H\rightarrow \phi\phi$. The current LHC bound is $\text{BR}\left(h\rightarrow \text{inv.}\right) < 0.19$ (95\% C.L.) \cite{Sirunyan:2018owy}. This can be translated into an upper bound on the invisible decay width: $\Gamma _{Inv}<0.96$~MeV. Another constraint on the invisible decay width comes from the observed Higgs signal strength $\mu\equiv \left[\sigma_{h}\text{BR}\left(h\rightarrow \text{SM}\right)\right]_{\text{exp}}/\left[\sigma_{h}\text{BR}\left(h\rightarrow \text{SM}\right)\right]_{\text{SM}} <0.89$ (95\% C.L.) \cite{Khachatryan:2016vau} where $\sigma_{h}$ is the Higgs boson production cross section. This translates in the bound $\Gamma _{Inv}<0.50$~MeV. Given the fact that
\begin{eqnarray}
\Gamma _{H\rightarrow \phi\phi}&\simeq & 0.50 \left(\frac{\lambda _{\phi H}}{0.01}\right)^{2}\text{ MeV}\,,
\end{eqnarray}
this gives: $\lambda _{\phi H}<0.01$.

Other constraints from colliders and beam dump experiments also exist from searching for meson decays involving the mediator in the final state, see e.g. Fig. 3 of \cite{Bondarenko:2019vrb}. For $m_{\phi}\lesssim 100$ MeV, they require $\sin\theta < 3\times 10^{-4}$.
\\

For the kinetic mixing portal, there are many constraints applying on the $\epsilon$ parameter, on top of the Supernovae constraints already mentioned above, see~e.g.~\cite{Berezhiani:2000gw,Goodsell:2009xc,Jaeckel:2010ni,Berezhiani:2008gi,McDermott:2010pa,Bernal:2019uqr}.

\section{Minimal ways out}

From the long list of constraints above it is clear that the simplest light mediator models one can consider, models  $A_{\gamma '}$} and $A_{\phi}$ of Eqs~(\ref{medcouplV}) and (\ref{medcouplphi}), are either excluded ($A_{\gamma '}$ ) or very marginally allowed ($A_{\phi}$). 
The discussion above nevertheless suggests specific ways out. In this Section we will present a number of simple scenarios/models which, based on these ways out, can easily work well.


We start by showing that nothing but the $A_{\gamma '}$ and $A_{\phi}$ models, which do not involve any extra particles beside the DM and light mediator ones, actually are in agreement with all constraints if we simply relax the assumption that $T'/T\simeq1$ when DM decouples.
Subsequently we will consider models which do assume extra light particles, to which DM can annihilate dominantly (models $B_{\gamma '}$ and $B_{\phi}$) or to which the light mediator can decay or annihilate, models C and D. Some more comments on the neutrino option (which does not require extra light particles) will be given in subsection E.

\subsection{The simplest option: DM annihilation into light mediator in a hidden sector with $T'<T$ \label{Tprimewayout}}

As said above, the self-interaction constraints do not require any connector between the DM/light mediator sector and the SM sector. Thus, it is perfectly conceivable that the DM/light mediator populations form a thermalized hidden sector which has never thermalized with the SM one. This possibility goes well along the fact that 
a MeV mediator must have anyway somewhat reduced interactions with the SM particles to have not been observed already. In particular,  the experimental constraints applying on the size of the kinetic mixing portal, which we mentioned above~\cite{Berezhiani:2000gw,Goodsell:2009xc,Jaeckel:2010ni,Berezhiani:2008gi,McDermott:2010pa,Bernal:2019uqr}, require this portal to be feeble, especially for a $\gamma'$ below the MeV scale. Thus, let us assume that $T'/T$ is sizably smaller than unity. This option doesn't require any new particle beside the DM and light mediator ones.
In the following we will show that nothing but the two truly minimal $A_{\gamma '}$ and $A_{\phi}$ models of Eqs.~(\ref{medcouplV}) and (\ref{medcouplphi}) works well as soon as we allow for $T'/T$ to be sizably below unity.\footnote{It was already suggested long ago \cite{Chu:2016pew} that
to consider $T'/T<1$ will relax the direct detection and BBN constraint and could work, for instance with a kinetic mixing portal, even if for the model considered in this reference (i.e. spin-1 hidden vector DM model with Higgs portal) the (too?) stringent BBN 
constraints applied there excluded this scenario. A value of $T'/T$ slightly below unity has been considered in \cite{Foot:2014uba,Foot:2016wvj}, along another general ("dissipative") framework.}

Before discussing how these 2 models behave when $T'/T<1$, it is interesting to stress that if DM thermalizes within a hidden sector that has temperature $T'$, the DM relic density constraint sets  a model independent lower bound on the value of $T'/T$. This is to be anticipated because, if $T'/T$ is too low, there will be simply too few DM particles. The maximum number of relic DM particles one can have today cannot exceed the number there were when these particles were relativistic and is saturated if DM decouples relativistically. In this case the number of particles at the temperature DM is decoupling is
\begin{equation}
n_{DM}= \frac{\zeta(3)}{\pi^2} g^{eff}_{DM} T'^3.
\end{equation}
Thus, it is suppressed by a $(T'/T)^3$ factor. Imposing that it gives a relic density at least as large as the observed one, this gives the lower bound
\begin{equation}
\frac{T'}{T}\geq 2.46\times 10^{-4}\times \left(\frac{100\text{~GeV}}{m_{DM}}\right)^{1/3}\times  \left( \frac{g^{S}_{\star}(T_{dec})}{g^{eff}_{DM}(T'_{dec})} \right)^{1/3},\label{eq:TpTlowerbound}
\end{equation}
with $T_{dec}$ and $T'_{dec}$ the value of $T$ and $T'$ when DM decouples.
This lower bound is basically model independent. Although it is difficult to conceive that such a model independent ``$T'/T$ floor'' has not been already derived, we didn't find any paper presenting it.

Next, it is useful to derive the lower bound which holds on the hidden sector interaction by requiring that the hidden sector has thermalized before DM decouples, so that a hidden sector $T'$ temperature can be defined. For models $A_{\gamma '}$ and $A_{\phi}$ this bound can be approximately obtained by requiring that the $\Gamma/H |_{T'=m_{DM}}>1$ inequality holds. Naming $\alpha '\equiv g_{\gamma '}^2/4\pi$ and $\alpha_\phi\equiv y^2_\phi/4\pi$, this requires
\begin{eqnarray}
\alpha'&\gtrsim& 5.98\times 10^{-9}\frac{T }{T'}\left(\frac{\sqrt{g_{\star}^{eff}(T )}}{g_{DM}^{eff}(T')}\right)^{1/2}\left(\frac{m_{DM}}{100\text{~GeV}}\right)^{1/2}\label{eq:alphapth} \\
\alpha _{\phi}&\gtrsim& 3.95\times 10^{-9}\frac{T }{T'}\left(\frac{\sqrt{g_{\star}^{eff}(T )}}{g_{DM}^{eff}(T')}\right)^{1/2}\left(\frac{m_{DM}}{100\text{~GeV}}\right)^{1/2}\label{eq:alphaphith}
\end{eqnarray}

\begin{figure}[h!]
\centering
\begin{minipage}{0.235\textwidth}
  \centering
\includegraphics[scale=0.4]{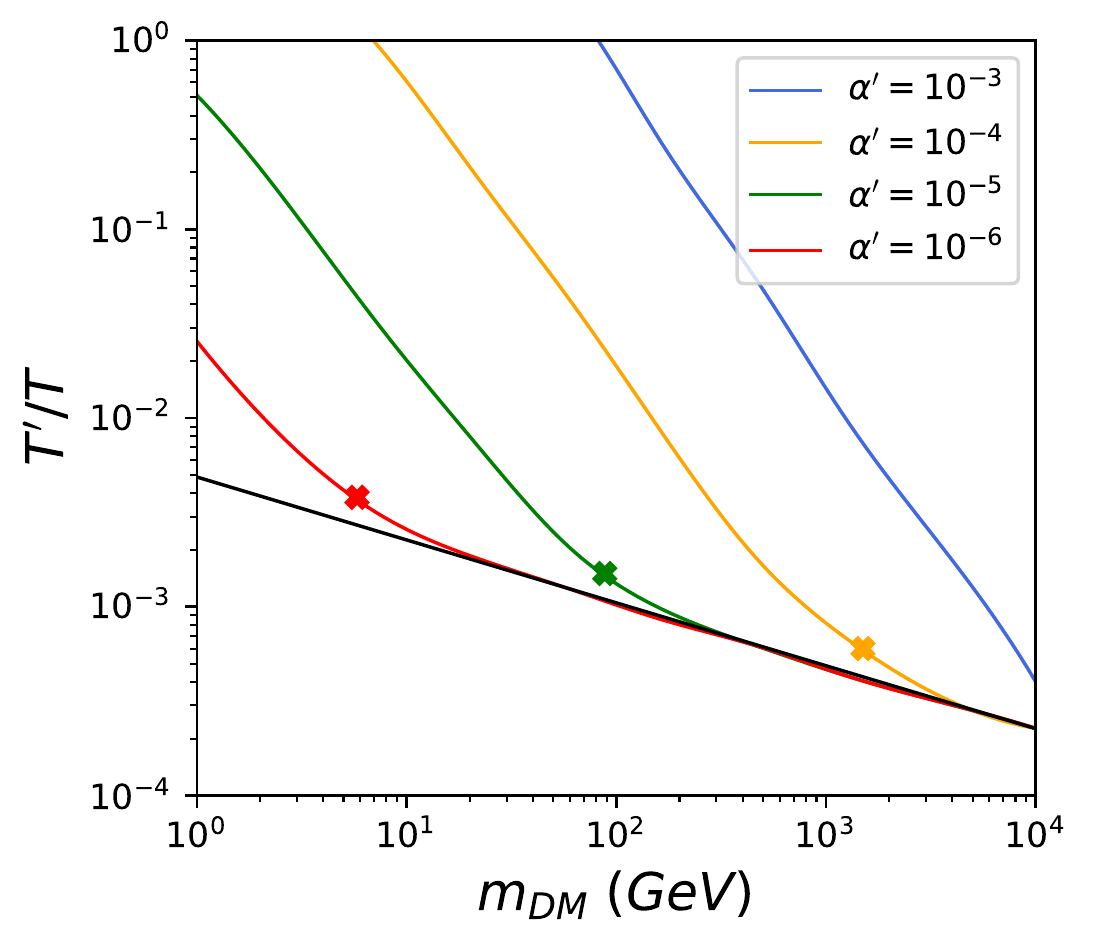}
\end{minipage}
\begin{minipage}{0.23\textwidth}
  \centering
\includegraphics[scale=0.4]{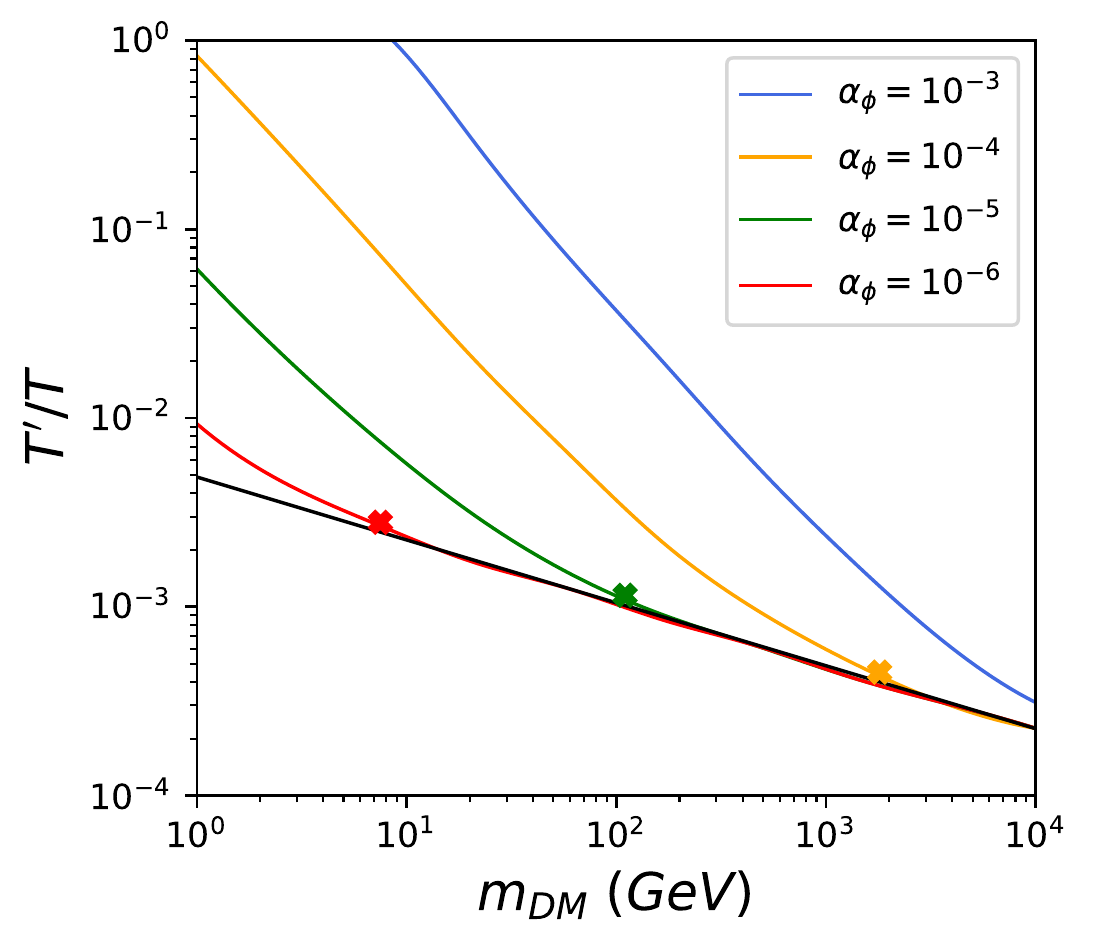}
\end{minipage}
\caption{As a function of the DM mass, value of $(T'/T)_{T=T_{dec}}$  one needs in order to reproduce the observed relic abundance of DM for a given set of couplings $\alpha '=10^{-3,-4,-5,-6}$ (left) and $\alpha _{\phi}=10^{-3,-4,-5,-6}$ (right), for the $A_{\gamma '}$ and $A_{\phi}$ models respectively, and assuming no connection to the SM. The black solid line shows the model independent ``$T'/T$ floor''  lower bound on $T'/T$ obtained in Eq. (\ref{eq:TpTlowerbound}). The crosses show the value of $m_{DM}$  above which Eqs.~(\ref{eq:alphapth}) and (\ref{eq:alphaphith}) are no longer satisfied, i.e.~above which $\Gamma/H |_{T'=m_{DM}}<1$.}
\label{fig:TpT_epsilon}
\end{figure}

The left and right panels of Fig.~\ref{fig:TpT_epsilon} gives, as a function of $m_{DM}$, the value of $(T'/T)_{T=T_{dec}}$ which leads to the observed relic density, for four example values of the dark gauge coupling, $\alpha '= 10^{-3,-4,-5,-6}$, and four example values of the Yukawa coupling, $\alpha_\phi=10^{-3,-4,-5,-6}$, respectively.


\begin{figure}[h!]
\centering
\begin{minipage}{0.23\textwidth}
  \centering
\includegraphics[scale=0.40]{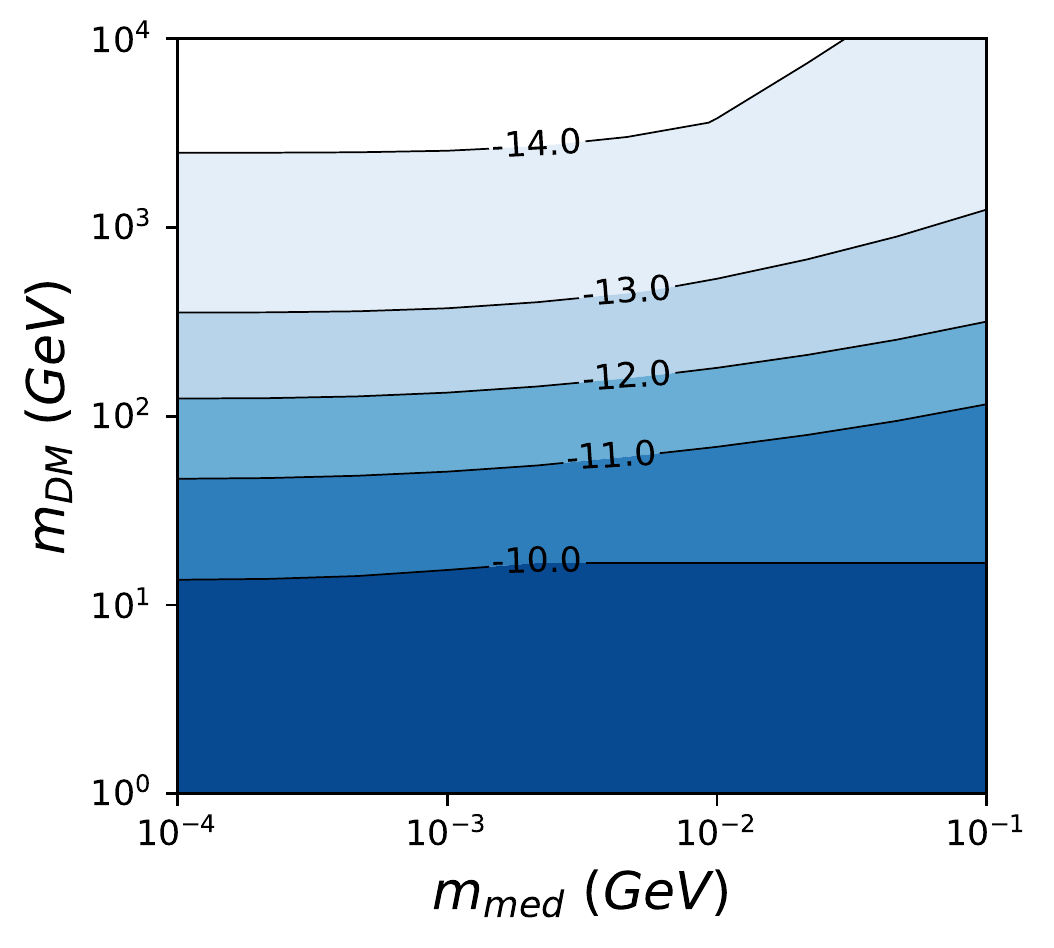}
\end{minipage}
\begin{minipage}{0.23\textwidth}
  \centering
\includegraphics[scale=0.40]{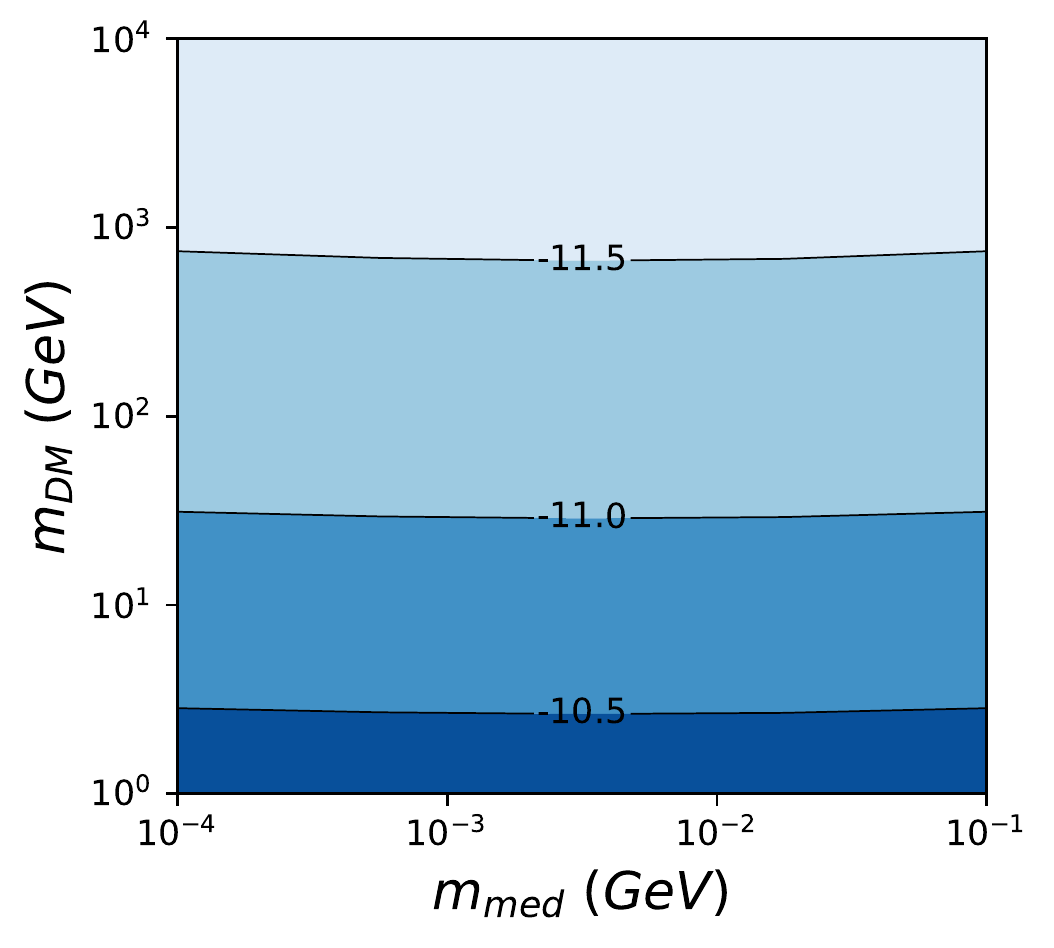}
\end{minipage}
\caption{For  $\alpha '=10^{-4}$ (left) and $\alpha_\phi=10^{-4}$ (right), maximal values of the kinetic mixing $\log_{10}\left[\epsilon\right]$ (left) and Higgs portal coupling $\log_{10}\left[\lambda_\phi\right]$ (right) for which
the number of DM and light mediator particles created through the portal never exceeds the number one has without portal assuming that $(T'/T)_{T=T_{dec}}$ is as given in Fig.~\ref{fig:TpT_epsilon}.}\label{fig:portalmax}
\end{figure}

Various features of Fig.~\ref{fig:TpT_epsilon} can be understood in the following way. For a given value of $\alpha'$ (or $\alpha_\phi$), the larger $m_{DM}$ is, the smaller $T'/T$ must be, as a result of the fact that $\Omega_{DM}$ scales as $\sim (T'/T)/\langle \sigma v\rangle$. Thus, the larger $m_{DM}$ is, the smaller must be the Boltzmann suppression after DM becomes non-relativistic. Thus, for fixed value of $\alpha'$ (or $\alpha_{\phi}$), when one increases  $m_{DM}$, at some point one asymptotically reaches the 
``$T'/T$ floor'' of Eq.~(\ref{eq:TpTlowerbound}) (shown by black solid line in Fig.~\ref{fig:TpT_epsilon}). This floor corresponds to a decoupling when the DM number density has still its relativistic value, i.e. no Boltzmann suppression.
For each value of $\alpha'$ (or $\alpha_{\phi}$) we have also indicated with a cross the value of $m_{DM}$, which we call $m^*_{DM}$, for which the approximate hidden sector thermalization condition holds, $\Gamma/H |_{T'=m_{DM}}=1$. Thus, along each line, 
for $m_{DM}<m^*_{DM}$ the hidden sector thermalizes, whereas for $m_{DM}>m^*_{DM}$ it does not thermalize. This is to be expected because, for fixed hidden sector coupling, the larger $m_{DM}$, the smaller $\Gamma/H |_{T'=m_{DM}}$. Thus, for approximately $m_{DM}>m^*_{DM}$ the line makes sense only if one assumes that
the hidden sector has thermalized before from another interaction in the UV. That the $m^*_{DM}$ crosses lie close to the blue line $T'/T$ floor is not surprising. If the approximate thermalization condition, $\Gamma/H|_{T'=m_{DM}}>1$, holds,
one expects that DM will decouple with a number density which is already Boltzmann suppressed, so that in Fig.~\ref{fig:TpT_epsilon} the crosses lie above the ``$T'/T$ floor''. Conversely, if this condition doesn't hold
one approximately does not expect any sizable Boltzmann suppression so that one lies close to the black ``$T'/T$ floor'' line.

The results of Fig.~\ref{fig:TpT_epsilon} have been obtained assuming no portal interaction at all. Switching on this interaction, these results are basically unchanged as long as the portal is small enough not to heat the hidden sector to a temperature $T'$ beyond the one assumed above.  Fig.~\ref{fig:portalmax} gives as a function of $m_{DM}$ and $m_{med}$ the upper bound on the portal parameter one gets imposing that the portal never lead to numbers of DM and light mediator particles which are larger than the ones one has without portal, taking for $T'/T$ the value given in Fig.~\ref{fig:TpT_epsilon}. Here too we fixed the dark sector interaction to the values $\alpha ' =10^{-4}$ and $\alpha _{\phi}=10^{-4}$.
Although there are viable regions of parameters where the portal exceeds these values (along reannihilation and secluded freeze-out ways where the hidden sector temperature $T'$ is dominantly due to the energy transferred through the portal \cite{Chu:2011be,Hambye:2019dwd}), here for simplicity we will calculate the relic density assuming that the portal has a negligible effect on it (which means our results will be strictly valid only for portal values below the ones reported in Fig.~(\ref{fig:portalmax})).\footnote{Related to Fig.~\ref{fig:TpT_epsilon} too, let us stress that the $T'/T$ floor in this figure is interesting in itself for any model, independently of the self-interaction issue. It gives the value of $T'/T$
for which DM can be a hot relic (but still cold DM subsequently since $m_{DM}\gg\hbox{keV}$) and still leads to the observed relic density (i.e.~without the need of any Boltzmann suppression), despite of the fact that $m_{DM}\gg\hbox{eV}$.} 


Let's now discuss the various constraints one by one in this $T'/T<1$ scenario for both models:
\begin{itemize}
\item {\it CMB}: the CMB constraint on the DM annihilation cross section, Eq.~(\ref{cmbconstraint}), is avoided in the scalar $A_{\phi}$ model because the annihilation is p-wave.
For $T'/T=1$, at DM decoupling, this constraint excludes the vector $A_{\gamma '}$ model for $m_{med}>2 m_e$ (see above), but now it is relaxed by the fact that if $T'/T<1$, the annihilation cross section leading to the observed relic density is reduced by a factor of $\sim T'/T$. As a result, one can easily avoid this constraint, see Fig.~\ref{fig:CMBsvratio}. Fig.~\ref{fig:CMB_BBN} shows the values of $m_{DM}$ and $m_{med}$ which, with $T'/T$ as given in Figs.~\ref{fig:TpT_epsilon}, allow to satisfy this constraint (for $\alpha'$ (left) and $\alpha _{\phi}$ (right) equal to $10^{-4}$). This CMB constraint is also trivially satisfied if there is no  portal at all (or a very tiny one) so that the light mediator has a lifetime sizably larger than the age of the Universe at recombination time.


\begin{figure}[h!]
\centering
\begin{minipage}{0.23\textwidth}
  \centering
\includegraphics[scale=0.40]{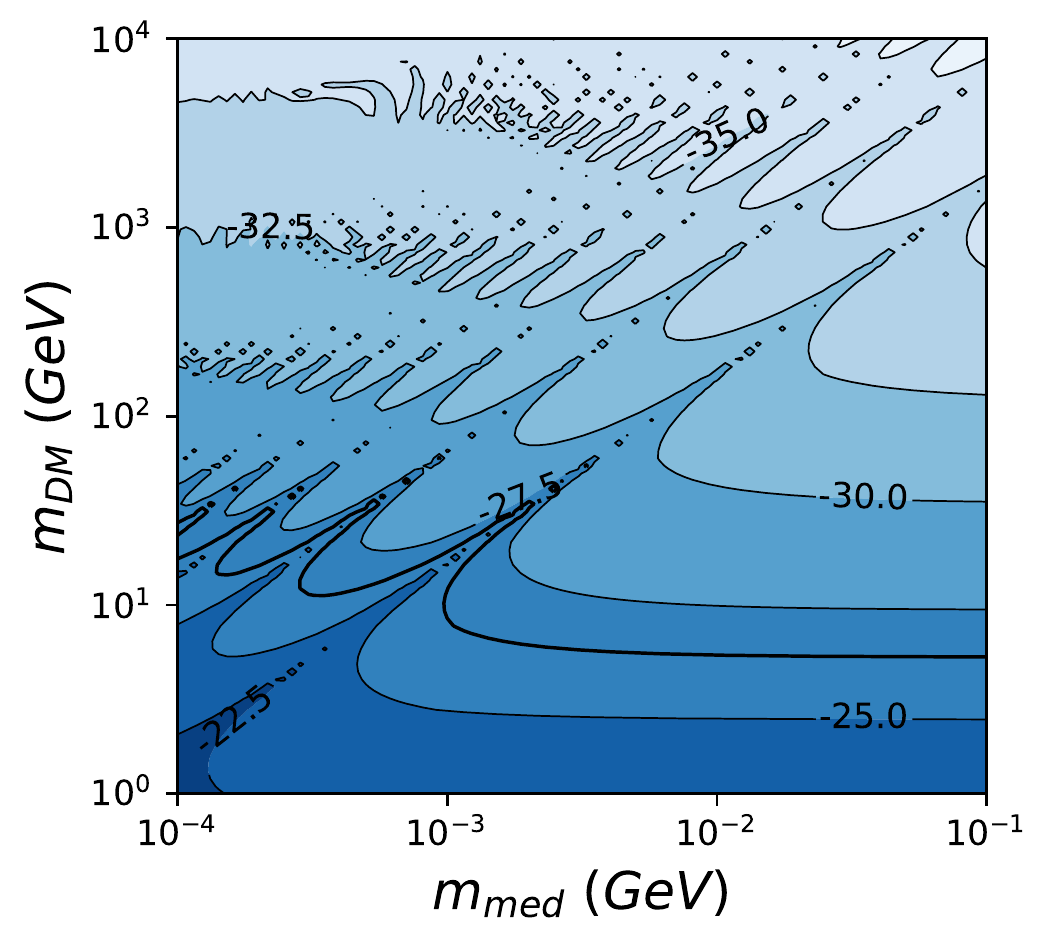}
\end{minipage}
\begin{minipage}{0.23\textwidth}
  \centering
\includegraphics[scale=0.40]{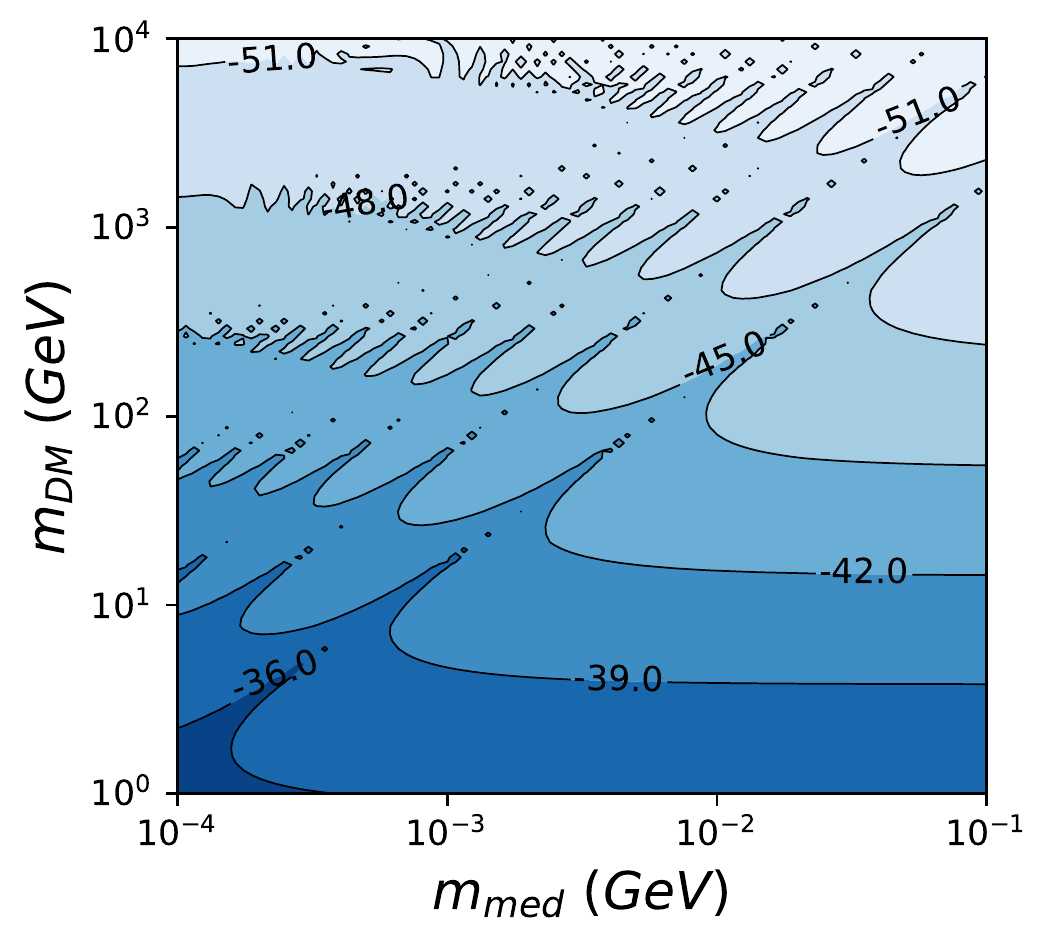}
\end{minipage}
\caption{$\log_{10}\left[\left\langle \sigma v\right\rangle_{rec}/m_{DM}\right]$ at the CMB epoch as a function of DM and mediator masses for the vector $A_{\gamma '}$ model with $\alpha '= 10^{-4}$ (left) and for the scalar $A_{\phi}$ model with $\alpha _{\phi}=10^{-4}$ (right). The solid black line gives $\left\langle \sigma v\right\rangle_{rec}/m_{DM} = 4\times 10^{-27}$ $\text{cm}^{3}\text{s}^{-1}\text{GeV}^{-1}$ which is the upper bound value resulting from Eq.~(\ref{cmbconstraint}). The region below this line is excluded.}\label{fig:CMB_BBN}
\end{figure}


The CMB constraints on the light mediator decay are also relaxed  when $T'/T<1$, as the number of mediators left when they decouple from the DM particles is a factor $(T'/T)^3$ smaller than for $T'/T=1$. As a result, with portal values bounded from above by the values given in Fig.~\ref{fig:portalmax}, one can check that the $\gamma '$ lifetime obtained can be easily fast enough to avoid this bound. 
It shows that, for example for $m_{med}=10$~MeV (1.5~MeV), the CMB constraint on the decay is satisfied provided $m_{DM}\gtrsim 117$~GeV (74~GeV). Fig.~\ref{fig:TpT_epsilon} shows that this is obtained thanks to the fact that in this case $T'/T$ is smaller than $0.02$ ($0.04$).
This shows how easy it is to avoid this CMB constraint as soon as $T'/T$ is sizably smaller than unity.

Finally, as soon as one considers values of $T'/T$ below unity, the $N_{eff}$ CMB constraint quickly gets irrelevant, again because the number density of light mediators is relaxed by a factor of $(T'/T)^3$. Moreover, 
if the light mediator still decays relativistically after neutrino decoupling, each decay injects less energy after neutrino decoupling than in the $T'/T\sim1$ case and its decay is less boosted, so that there are less decays after neutrino decoupling.
This weakens the constraint further.
Thus, from the upper bound obtained on the light mediator number density as a function of its lifetime for $T'/T=1$, one can get conservative upper bounds on this lifetime holding for the $T'/T<1$ case, simply by multiplying 
the light mediator number density by a $(T'/T)^3$ factor.
Assuming the light mediator decouples relativistically, as it is the case for model $A_{\gamma '}$ and $A_{\phi}$, Table \ref{table:Neff} gives the upper bound on the lifetime this leads to for various values of $m_{med}$ and $T'/T$. These can be compared with the first line of the same table, where $T'/T=1$.

\begin{table}[h!]
\fontsize{7.5}{12}
\begin{tabular}{|c|c|c|c|c|c|c|c|c|c|c|c|}
\hline
\multicolumn{2}{|c|}{\multirow{2}{*}{$\tau_{med}$ (sec)}} & \multicolumn{4}{c|}{$m_{\gamma '}$ (MeV)} & \multicolumn{4}{c|}{$m_{\phi}$ (MeV)} \\ \cline{3-10} 
\multicolumn{2}{|c|}{}                                    & 3     & 10     & 30    & 100    & 3     & 10     & 30    & 100   \\ \hline
\multirow{4}{*}{$\frac{T'}{T}$}          & 1              & $10^{1.1}$       & $10^{0.1}$      & $10^{-0.4}$     & $10^{-0.7}$      & $10^{2.1}$       & $10^{1.1}$      & $10^{0.1}$     & $10^{-0.5}$     \\ \cline{2-10} 
                                         & 0.1            & $10^{7.3}$       & $10^{6.3}$      & $10^{5.3}$     & $10^{4.3}$      & $10^{8.3}$       & $10^{7.3}$      & $10^{6.3}$     & $10^{5.3}$     \\ \cline{2-10} 
                                         & 0.01           & $>$       & $>$      & $>$     & $>$      & $>$     & $>$     & $>$    & $>$     \\ \cline{2-10} 
                                         & 0.001          & $>$       & $>$      & $>$     & $>$      & $>$     & $>$     & $>$    & $>$     \\ \hline
\end{tabular}
\normalsize
\caption{For various values of $T'/T$ taken at DM freeze-out, upper bound on the light mediator lifetime from $N_{eff}$ at CMB, assuming a relativistic decoupling for the $A_{\gamma '}$ model (left) and the $A_{\phi}$ model (right). The $T'/T=1$ line is as in Table \ref{table:Neff_Az} and "$>$" indicates an upper bound greater than $10^{8}$ sec.}\label{table:Neff}
\end{table}

Note also that if the light mediator is effectively stable, as it is the case if there is no portal or for the vector $A_{\gamma '}$ model for $m_{med}<2 m_e$ (see above), the only relevant constraints is the non-overclosure one (or more exactly that its relic density is smaller than the DM observed one) and the modification of the Hubble constant. The $(T'/T)^3$ suppression of the mediator number density allows to easily
fulfill these constraints.

\item {\it BBN}: if the mediator is effectively stable there is no relevant BBN constraint. In particular, the modification of the Hubble constant the mediator implies at the BBN epoch, Eq.~(\ref{DeltaNeff}), is negligible because the number of mediator is suppressed by a $(T'/T)^3$ factor.
If instead the portal doesn't vanish, the Hubble constant/entropy injection BBN constraint, as well as photodisintegration constraints (for $m_{med}> 4.4$~MeV) are also relaxed by the $(T'/T)^3$ suppression of the light mediator number density. 
Applying a conservative $(T'/T)^3$ suppression to the light mediator number density, Table \ref{table:BBN} gives the upper bounds on the light mediator lifetime one gets from these 2 constraints for various values of $m_{med}$ and $T'/T$. 
Again one sees how quickly these bounds get relaxed as soon as $T'/T$ is below one.

\begin{table}[h!]
\fontsize{7.5}{12}
\begin{tabular}{|c|c|c|c|c|c|c|c|c|c|c|c|}
\hline
\multicolumn{2}{|c|}{\multirow{2}{*}{$\tau_{med}$ (sec)}} & \multicolumn{4}{c|}{$m_{\gamma '}$ (MeV)} & \multicolumn{4}{c|}{$m_{\phi}$ (MeV)} \\ \cline{3-10} 
\multicolumn{2}{|c|}{}                                    & 3     & 10     & 30    & 100    & 3     & 10     & 30    & 100   \\ \hline
\multirow{4}{*}{$\frac{T'}{T}$}          & 1              & $10^{2.0}$       & $10^{1.0}$      & $10^{0.0}$     & $10^{-0.4}$      & $10^{2.4}$       & $10^{2.0}$      & $10^{1.0}$     & $10^{-0.1}$     \\ \cline{2-10} 
                                         & 0.1            & $10^{6.9}$       & $10^{4.1}$      & $10^{3.9}$     & $10^{3.8}$      & $10^{7.8}$       & $10^{4.2}$      & $10^{4.0}$     & $10^{3.9}$     \\ \cline{2-10} 
                                         & 0.01           & $>$       & $>$      & $10^{4.7}$     & $10^{4.5}$      & $>$     & $10^{7.6}$     & $10^{4.9}$    & $10^{4.7}$     \\ \cline{2-10} 
                                         & 0.001          & $>$       & $>$      & $>$     & $10^{7.0}$      & $>$     & $>$     & $>$    & $>$     \\ \hline
\end{tabular}
\normalsize
\caption{For various values of $T'/T$ taken at DM freeze-out, upper bound on the light mediator lifetime from photodisintegration and Hubble constant/entropy injection during BBN, assuming a relativistic decoupling for the $A_{\gamma '}$ model (left) and the $A_{\phi}$ model (right). The $T'/T=1$ line is as in Table \ref{table:Neff_Az} and  "$>$" indicates an upper bound greater than $10^{8}$ sec.}\label{table:BBN}
\end{table}

\item {\it Direct detection}: For the entire DM and mediator masses range we consider in Figs. \ref{fig:portalmax}, the connector $\kappa$ is below or almost below current direct detection constraints given in Fig.~\ref{fig:DD_Constraints} ($\kappa _{KM}\lesssim 10^{-10.5}$ and $\kappa _{HP}\lesssim 10^{-13}\left(\frac{v_{\phi}}{\text{GeV}}\right)$). But, again, if $T'/T<1$, the DM to mediator coupling needed to get the observed relic density is even smaller than the one one needs in the $T'/T=1$ case. As a result, the direct detection constraint can be easily further satisfied.

\item {\it Indirect detection}: Here too the signal is reduced as a result of the fact that the DM to mediator coupling is reduced.

\end{itemize}

\begin{figure}[h!]
\centering
\begin{minipage}{0.24\textwidth}
  \centering
\includegraphics[scale=0.42]{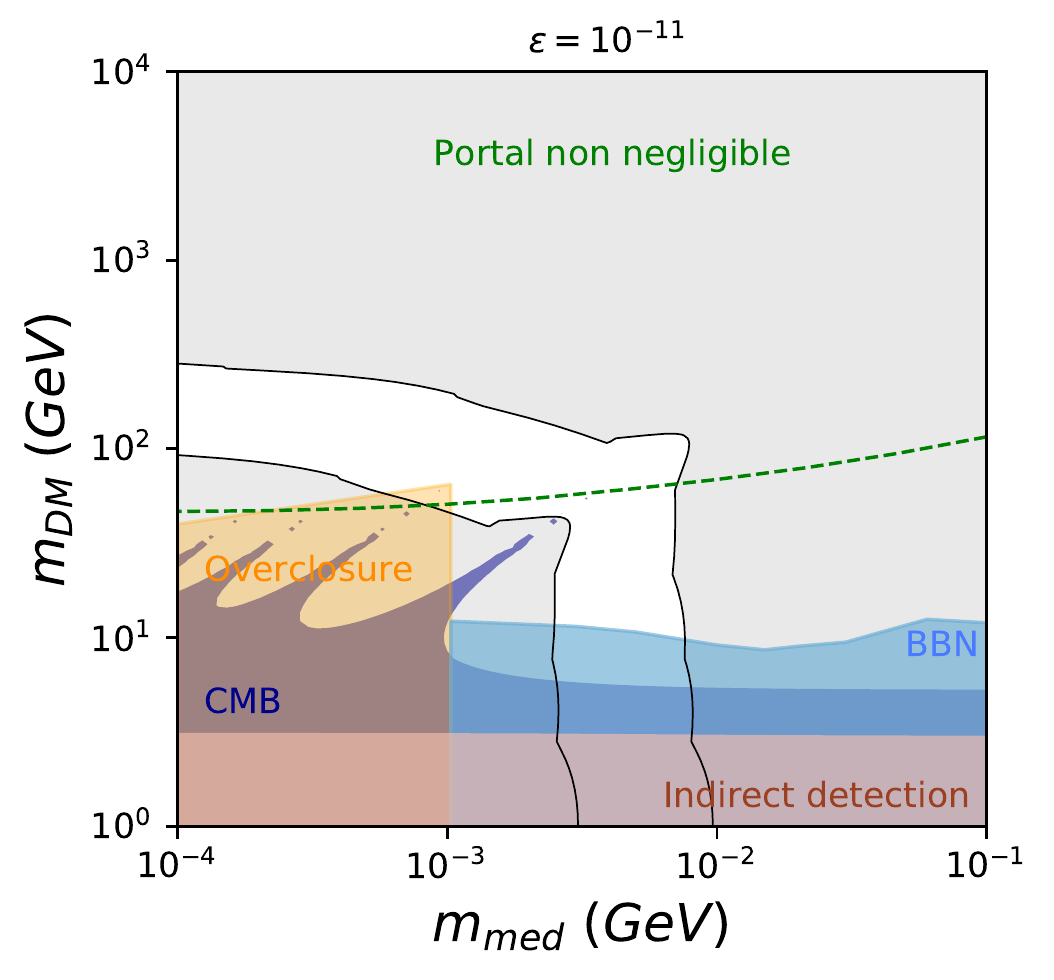}
\end{minipage}
\begin{minipage}{0.23\textwidth}
  \centering
\includegraphics[scale=0.42]{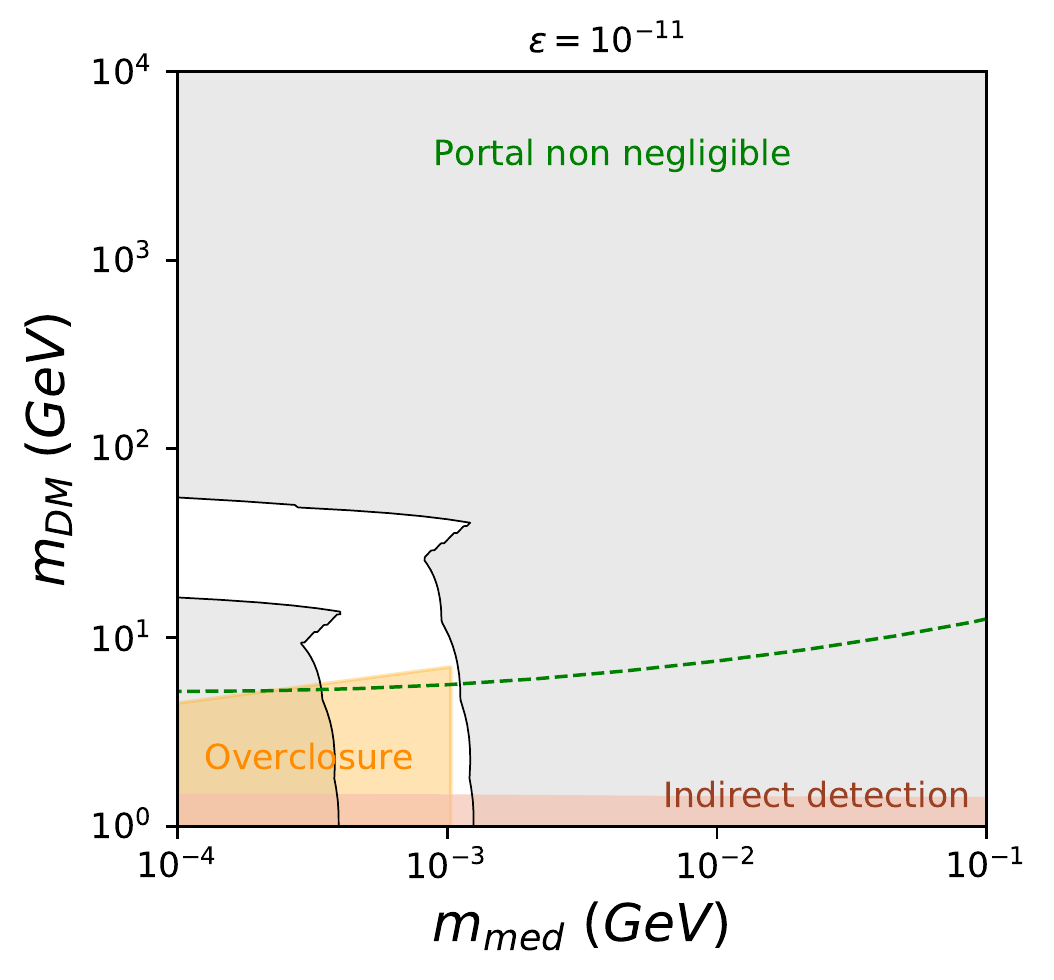}
\end{minipage}\\
\begin{minipage}{0.24\textwidth}
  \centering
\includegraphics[scale=0.42]{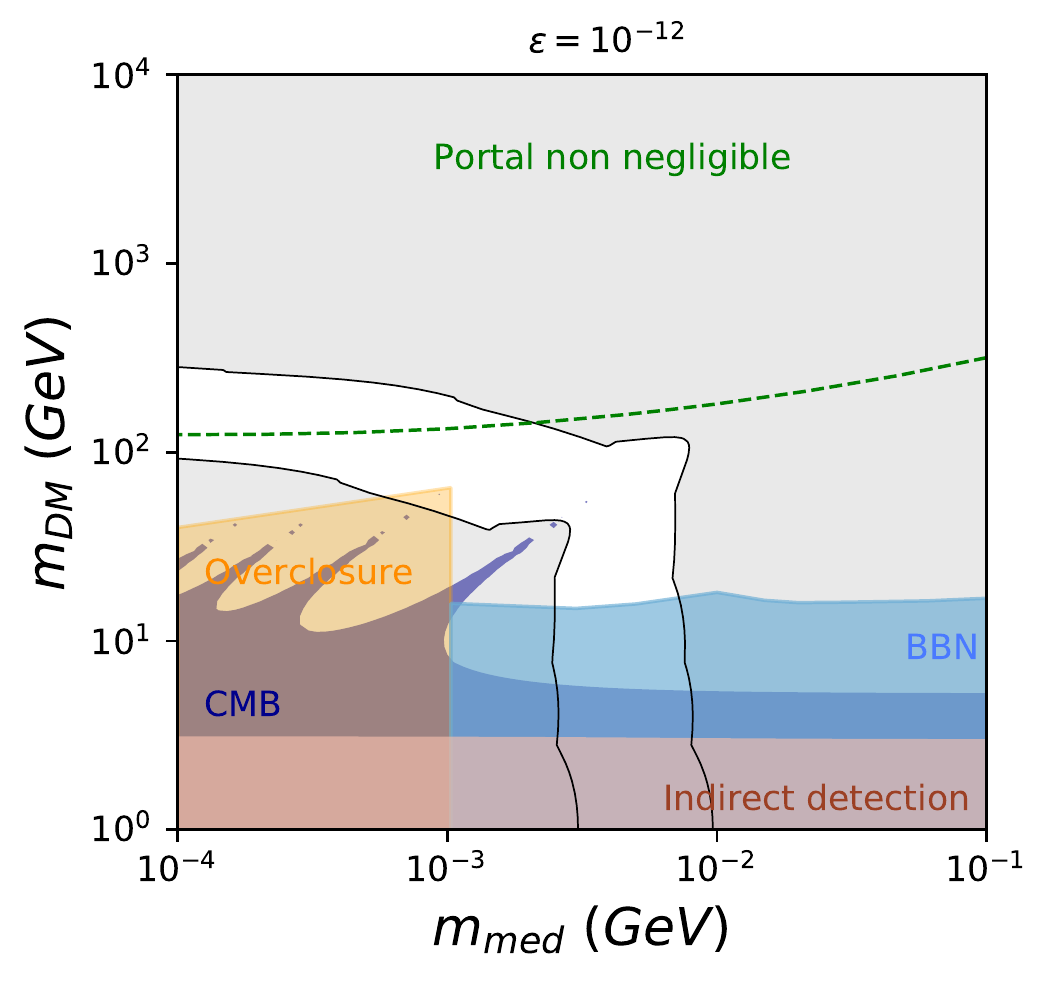}
\end{minipage}
\begin{minipage}{0.23\textwidth}
  \centering
\includegraphics[scale=0.42]{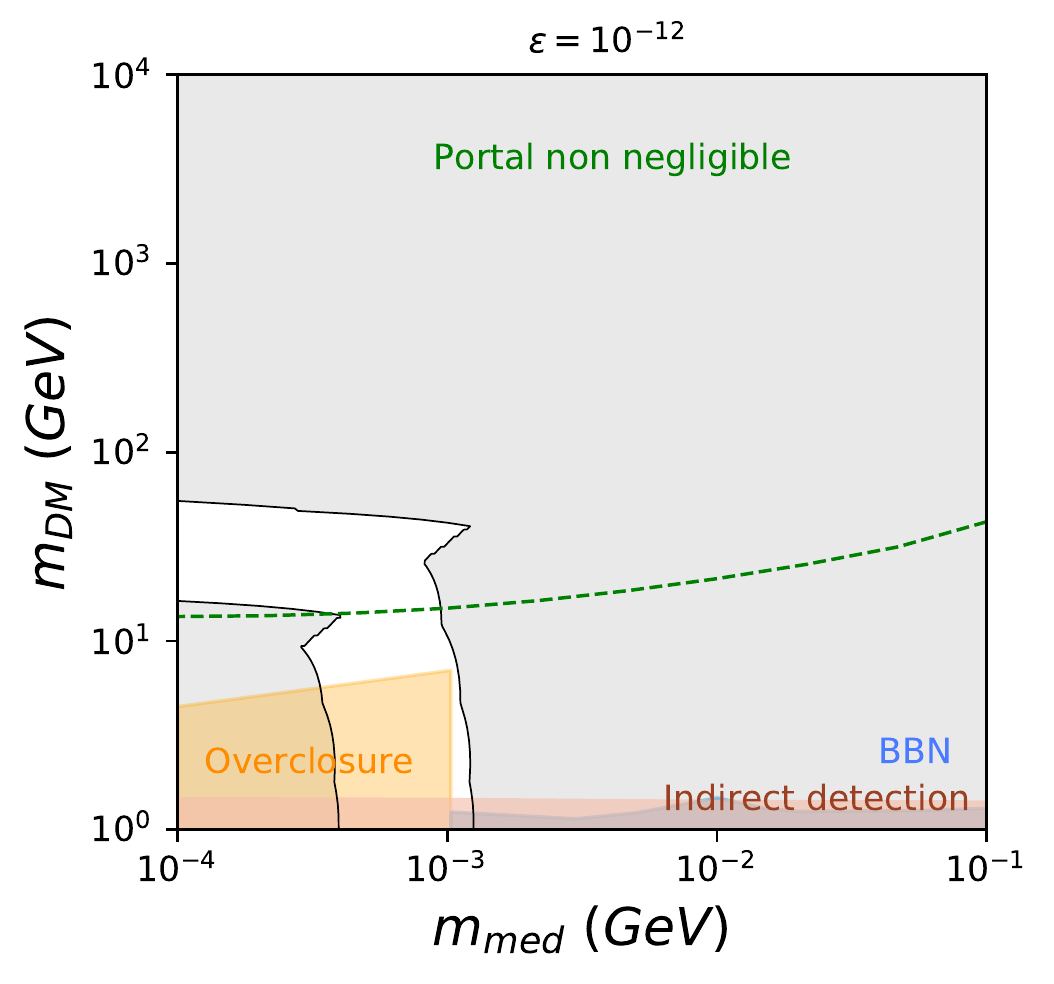}
\end{minipage}\\
\begin{minipage}{0.24\textwidth}
  \centering
\includegraphics[scale=0.42]{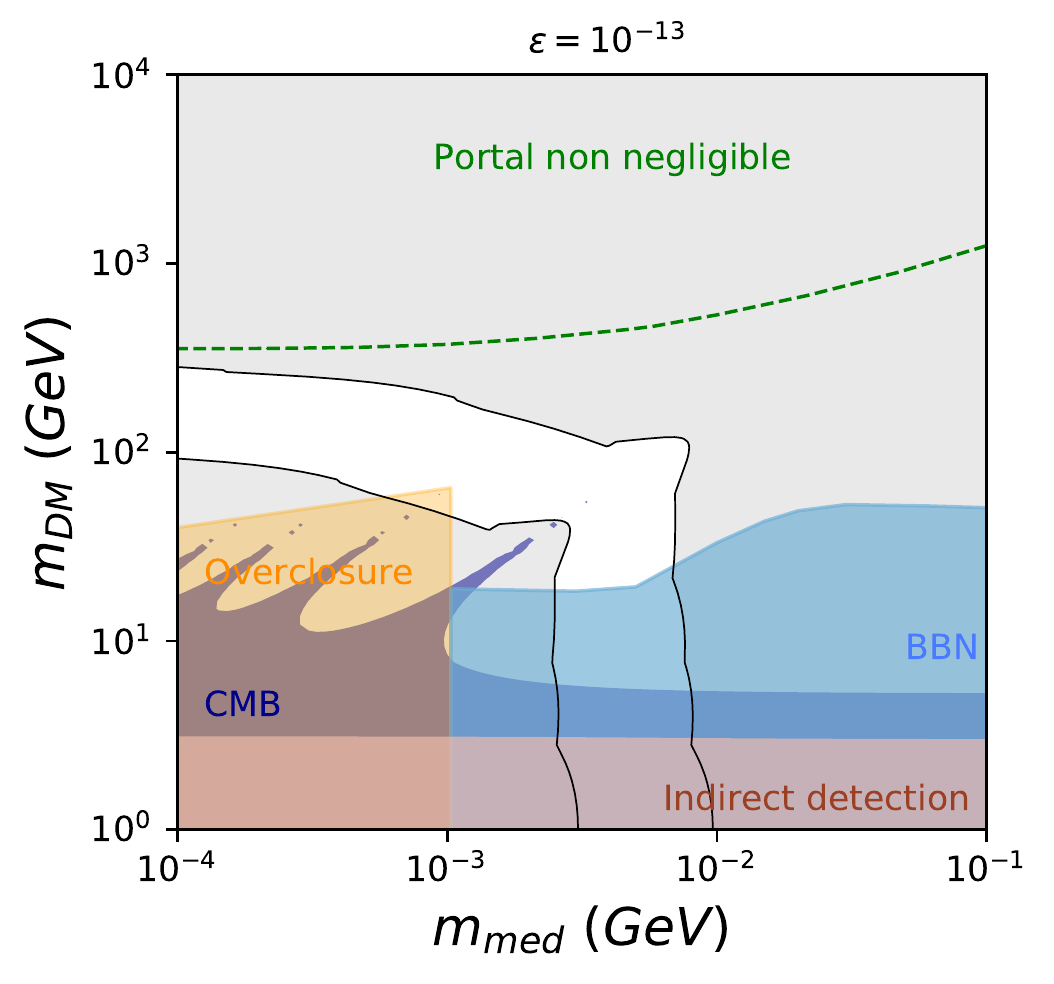}
\end{minipage}
\begin{minipage}{0.23\textwidth}
  \centering
\includegraphics[scale=0.42]{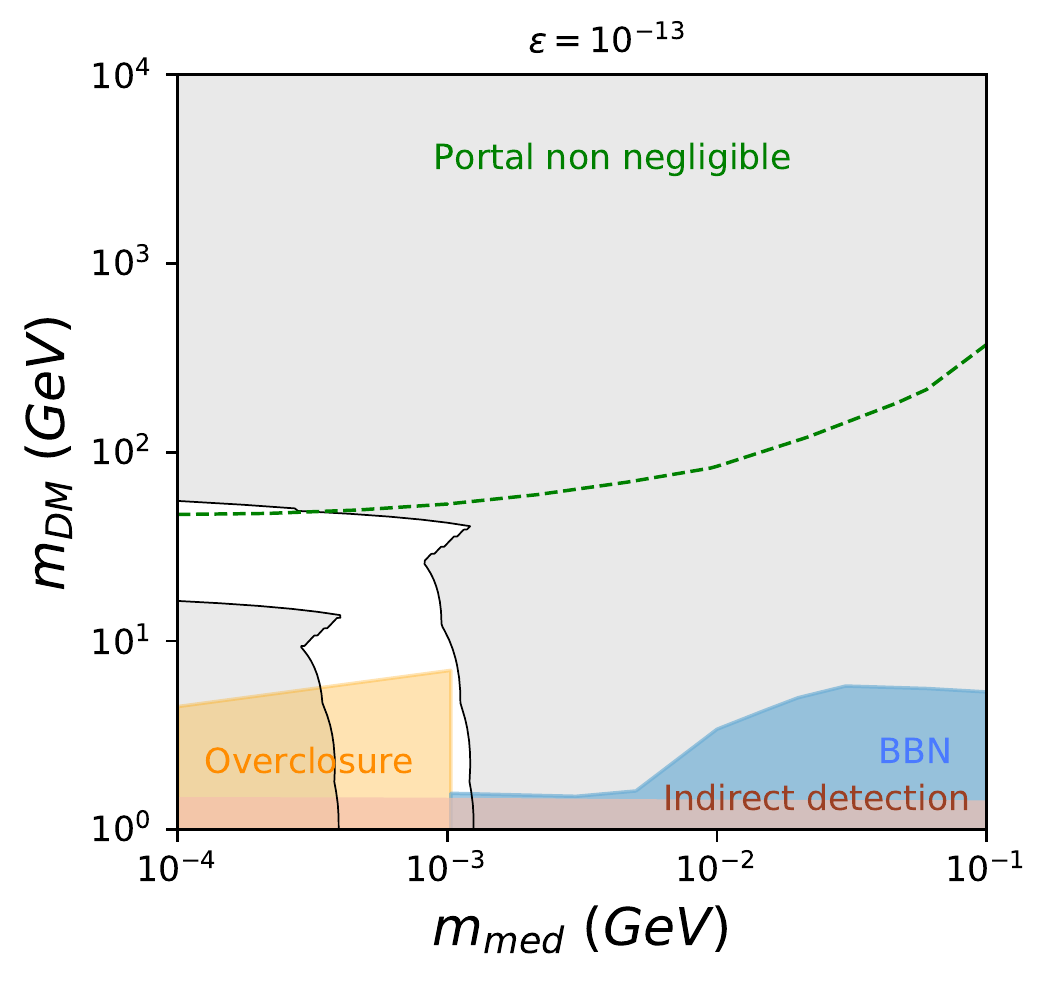}
\end{minipage}\\
\caption{Constraints from CMB, BBN, self-interaction, indirect and direct detection all together for the vector $A_{\gamma '}$ model with $\alpha '= 10^{-4}$  (left) and $\alpha '= 10^{-5}$ (right), for three different values of $\epsilon$.}\label{fig:CMB_SI_alphapV}
\end{figure}

In Fig.~\ref{fig:CMB_SI_alphapV} we show in the $m_{med}$-$m_{DM}$ plane what are the various constraints just discussed for the vector $A_{\gamma '}$ model, for $\alpha '=10^{-4}$ (left) and $\alpha '=10^{-5}$ (right), and for 3 values of $\epsilon$.
In this plot the BBN excluded region combines both photodisintegration and Hubble constant/entropy injection BBN constraints, as well as the $N_{eff}$ constraint. These constraints have been obtained applying the conservative $(T'/T)^3$ suppression of the light mediator number density in the bounds existing for $T'/T=1$ \cite{Hufnagel:2018bjp}. We also show on the same plot the indirect detection constraints as well as the region which fulfills the self-interactions constraints. Note that the all parameter space is allowed by direct detection experiments.
For these plots the value of $T'/T$ is set by the relic density constraint, assuming, as said above, that the portal has no effect on it. The dashed green line gives the value of $m_{DM}$ below which this holds, i.e.~below which $\epsilon$ lies below the values given in Fig.~\ref{fig:portalmax}. Thus, all the white region below this line in Fig.~\ref{fig:CMB_SI_alphapV} is clearly allowed, whereas above this line the white region is expected to be still largely allowed but to show that explicitly would require to calculate the relic density including the effect of the portal (along a reannihilation or secluded production regime \cite{Chu:2011be}, which is beyond the scope of this work). Note that for $\alpha'=10^{-5}$ the annihilation rate is smaller than for $\alpha'=10^{-4}$, so that the DM number density is less Boltzmann suppressed at freeze-out. Thus, in this case, one needs a smaller value of $T'/T$, see Fig.~\ref{fig:TpT_epsilon}, and the mediator density is more suppressed. As a result, the associated BBN and CMB constraints are further relaxed,
see Fig.~\ref{fig:CMB_SI_alphapV}.

\begin{figure}[h!]
\centering
\begin{minipage}{0.24\textwidth}
  \centering
\includegraphics[scale=0.42]{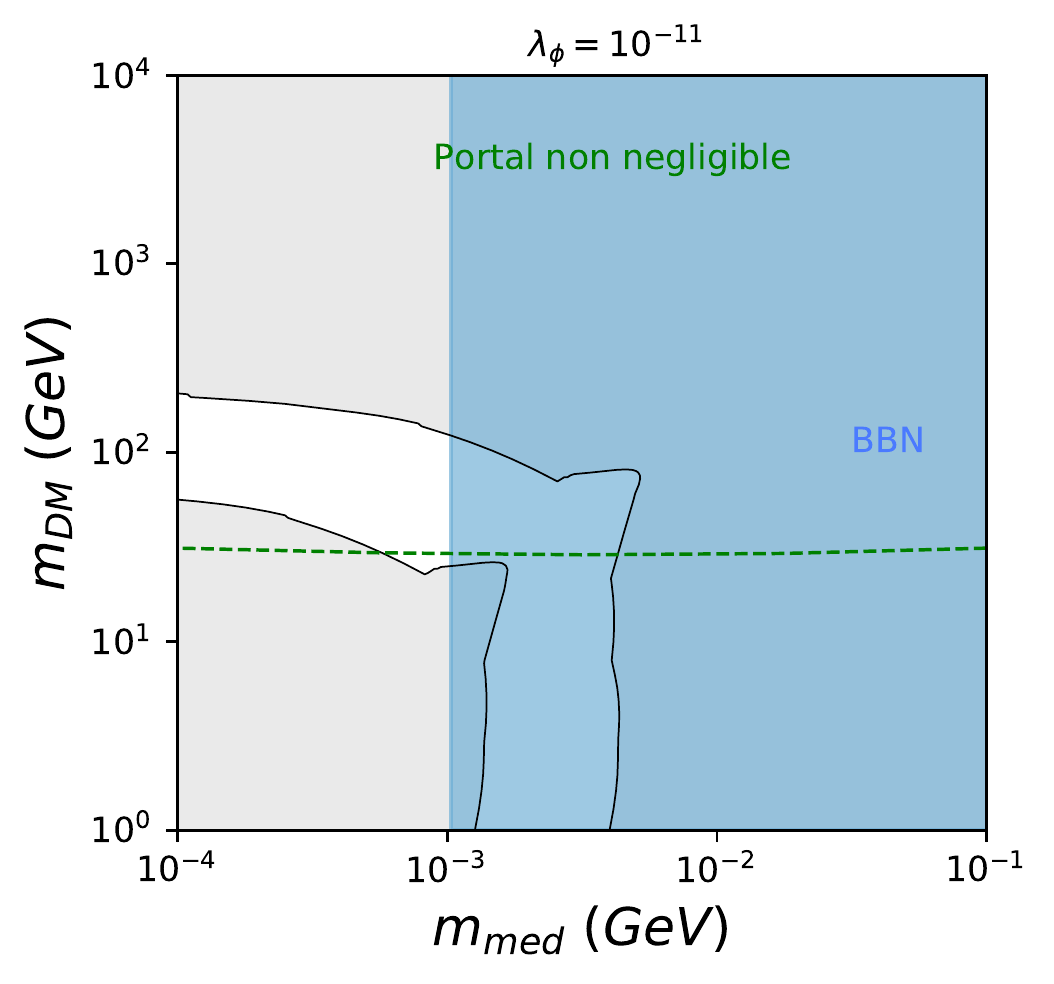}
\end{minipage}
\begin{minipage}{0.23\textwidth}
  \centering
\includegraphics[scale=0.42]{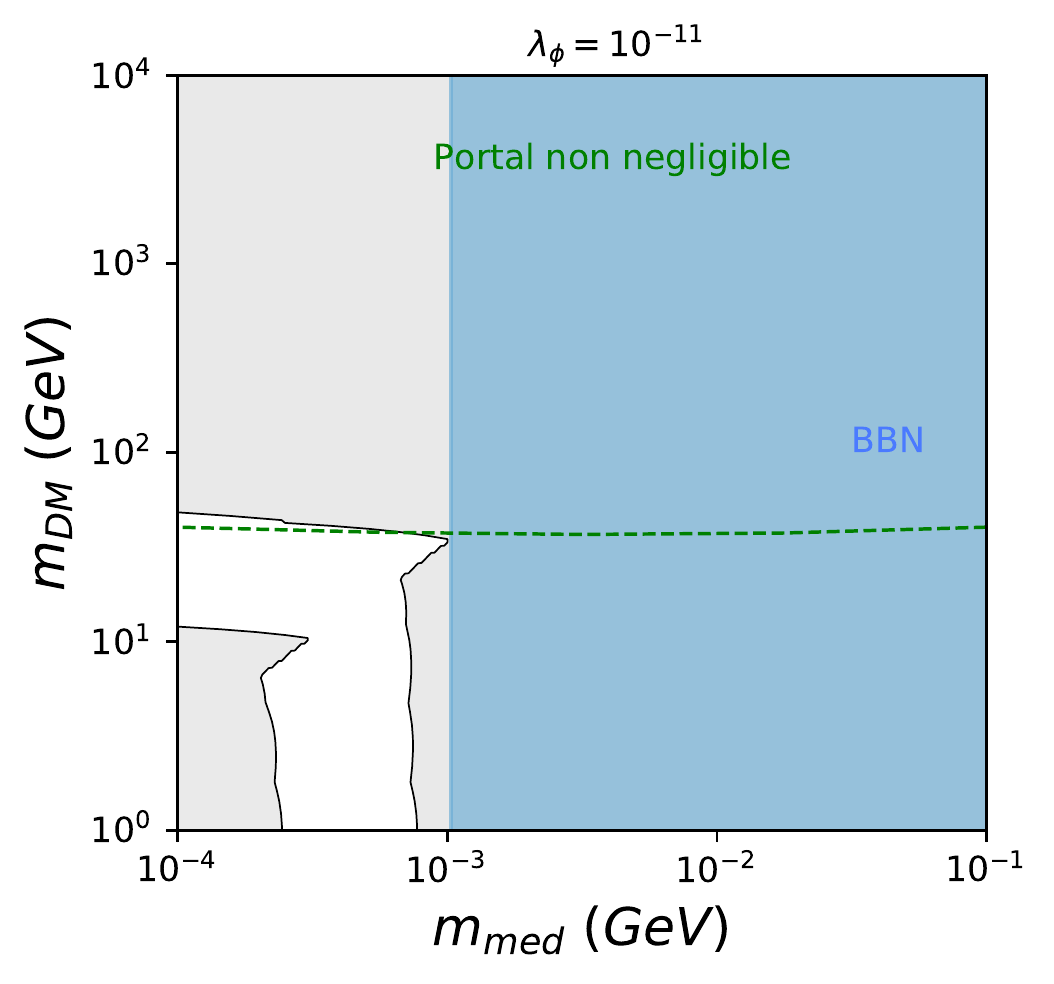}
\end{minipage}\\
\begin{minipage}{0.24\textwidth}
  \centering
\includegraphics[scale=0.42]{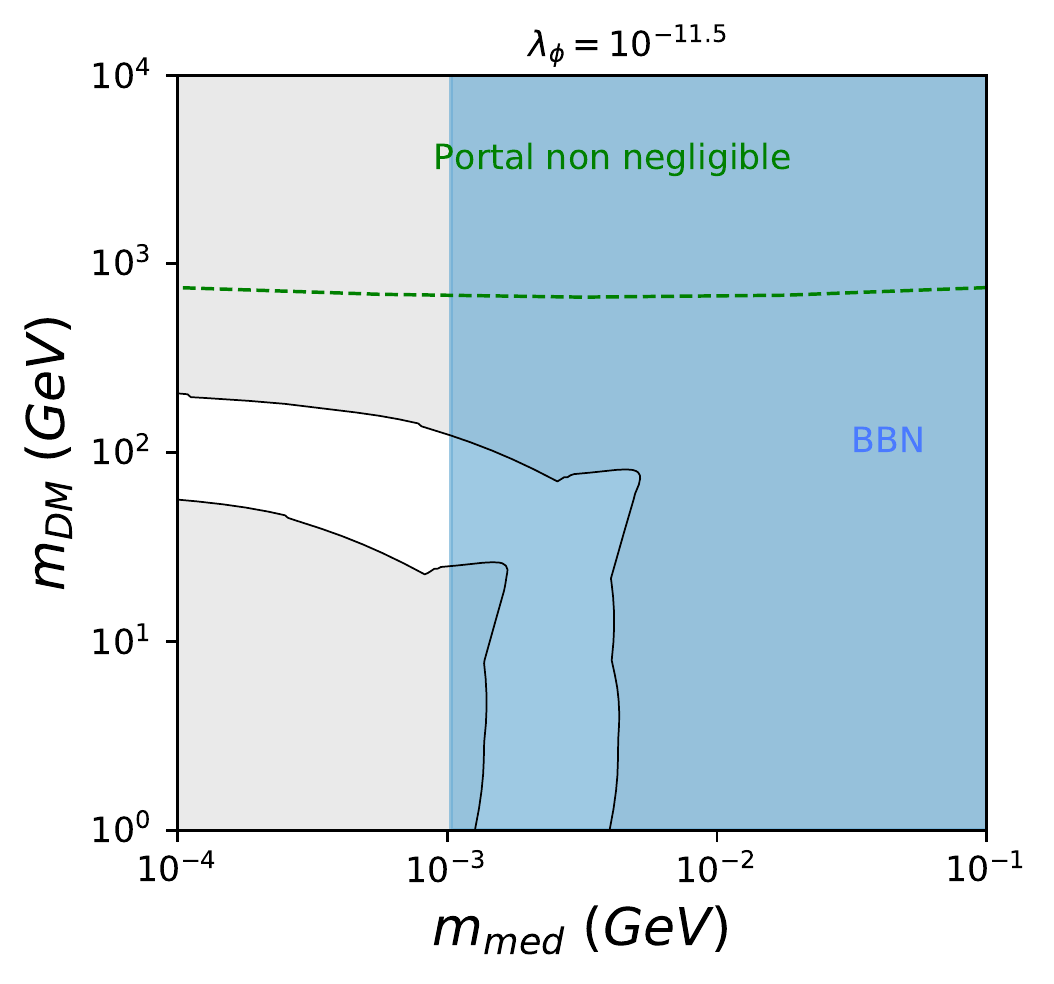}
\end{minipage}
\begin{minipage}{0.23\textwidth}
  \centering
\includegraphics[scale=0.42]{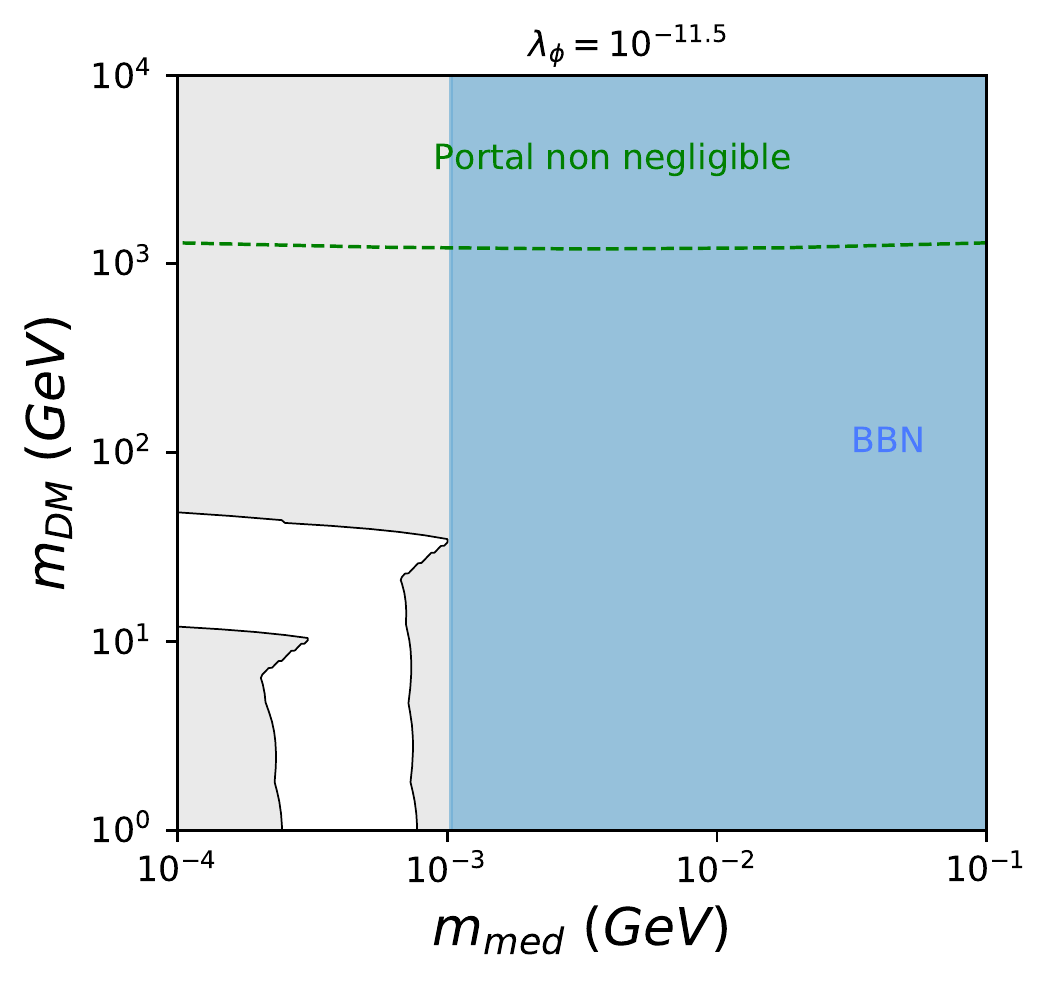}
\end{minipage}\\
\begin{minipage}{0.24\textwidth}
  \centering
\includegraphics[scale=0.42]{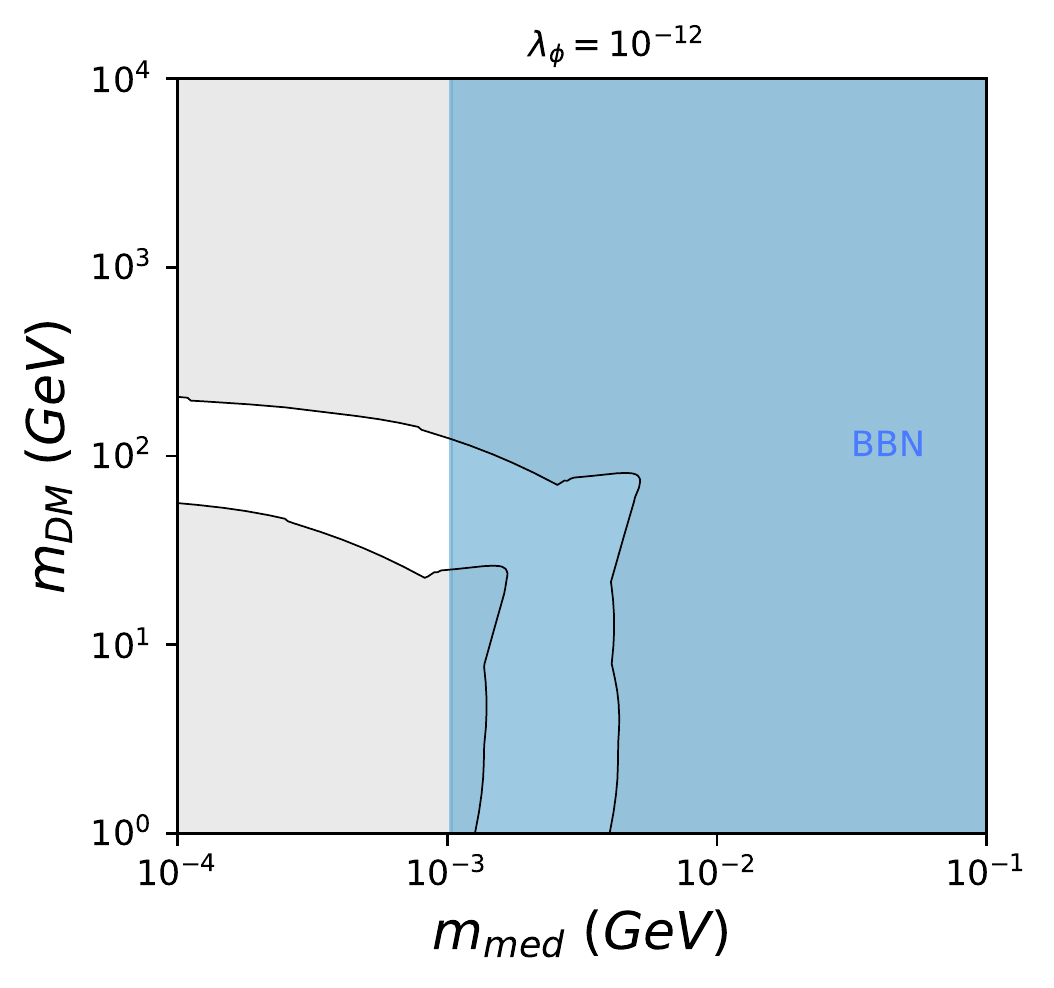}
\end{minipage}
\begin{minipage}{0.23\textwidth}
  \centering
\includegraphics[scale=0.42]{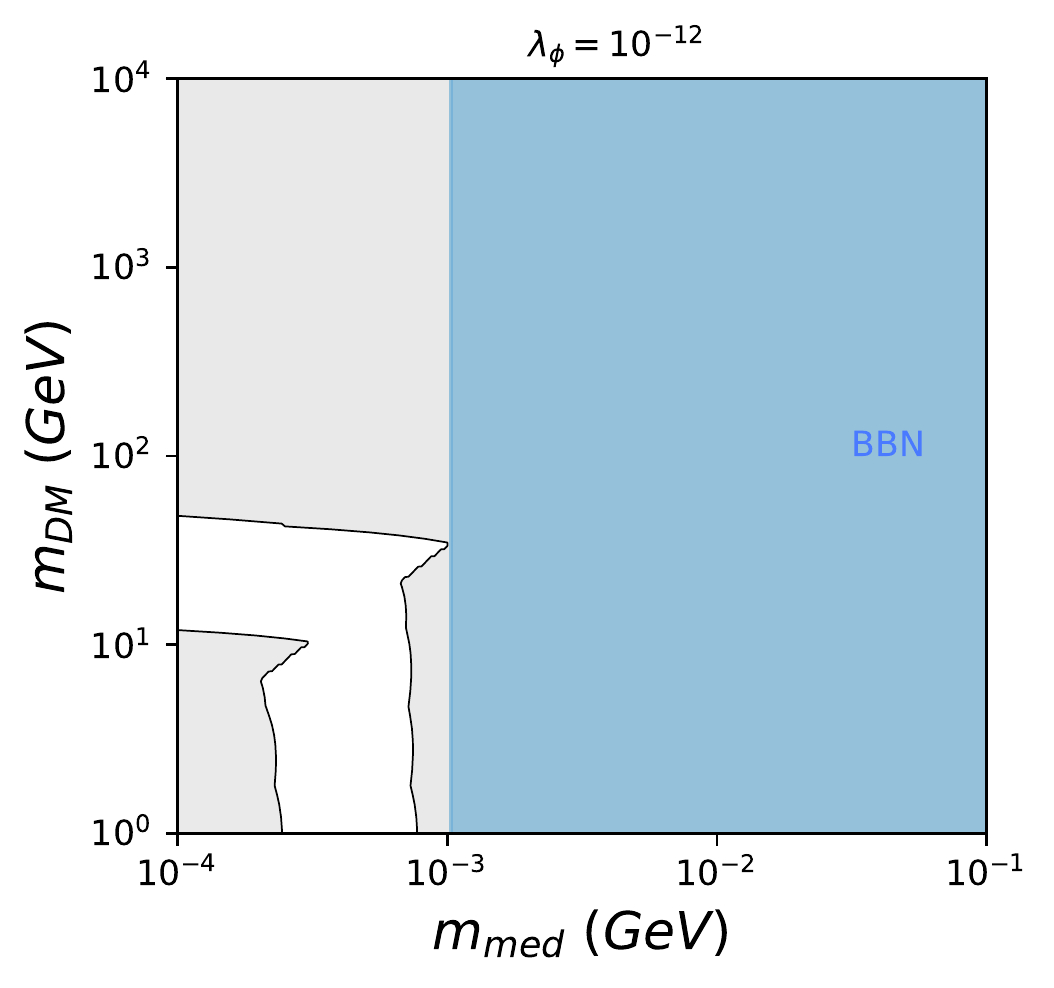}
\end{minipage}\\
\caption{Constraints from CMB, BBN, self-interaction, indirect and direct detection all together for the vector $A_{\phi}$ model with $\alpha _{\phi}= 10^{-4}$  (left) and $\alpha _{\phi}= 10^{-5}$ (right), for three different values of $\lambda _{\phi}$.}\label{fig:CMB_SI_alphapVphi}
\end{figure}

Fig.~\ref{fig:CMB_SI_alphapVphi} applies in a similar way to the scalar $A_{\phi}$ model, for $\alpha_\phi=10^{-4}$ (left) and $\alpha_\phi=10^{-5}$ (right), and 3 values of $\lambda_\phi$. As noted above, for $m_{med}>2 m_e$ the decay width is suppressed not only by the scalar mixing angle, $\tan 2\theta=v_\phi v_{H} \lambda_\phi/(m_{H}^{2}-m_{\phi}^{2})$ but also by the electron Yukawa coupling. As a result, the BBN upper bound on the light mediator lifetime requires relatively large values of $\sin \theta$ and $\lambda_\phi$. However the production of hidden sector particles from the SM thermal bath have processes involving only $\lambda_\phi$, for example $H H\rightarrow \phi\phi$. These processes can easily thermalize the hidden sector with the visible one at high temperature. Thus, the BBN constraints are in tension with the assumption of a low $T'/T$ ratio. This is shown in Figs.\ref{fig:CMB_SI_alphapVphi} where BBN constraints exclude the all parameter space above the electron threshold $m_{med}>2m_{e}$. In particular the narrow region which in the case $T'/T=1$ is still not excluded for this model (for $m_{DM}\sim 0.5$~GeV and $m_{med}\sim 1.1$~MeV \cite{Hufnagel:2018bjp}) disappears for $T'/T<1$ because this region is viable thanks to a large value of the Higgs portal interaction (so that the light mediator decays just fast enough to avoid the BBN constraints) but this large value of the Higgs portal deeply thermalize both sectors, so that $T'/T=1$. However,  as Fig.~\ref{fig:CMB_SI_alphapVphi}  shows, the scalar option is now widely open for $m_{med}<2 m_e$, unlike for the $T'/T=1$ case. Below the $e^+e^-$ threshold, the decay width is loop suppressed and leads to large lifetime which are forbidden in the $T'/T=1$ case but not anymore in the $T'/T<1$ case. For $\alpha'=10^{-5}$ this even allows values of $m_{DM}$ below the GeV scale.


Finally let us stress that, from the discussion of each constraint above, it appears that both model $A_{\gamma '}$ and $A_\phi$ are perfectly viable without any portal at all, i.e.~with a stable mediator. The non overclosure constraint, 
i.e~$\Omega_{med}h^2<0.1188$, requires
\begin{equation}
\frac{T'}{T}\leq 1.14\times 10^{-2}\times \left(\frac{\text{MeV}}{m_{med}}\right)^{1/3}\times  \left( \frac{g^{S}_{\star}(T_{dec})}{g^{eff}_{med}(T'_{dec})} \right)^{1/3},
\label{lowerTprimestable}
\end{equation}
where all quantities are taken at the DM freeze-out time. This upper bound lies well above the lower bound of Eq.~(\ref{eq:TpTlowerbound}), as a result of the fact that $m_{med}\ll m_{DM}$.
As already said above, the Hubble constant modification constraint, Eq.~(\ref{DeltaNeff}), is irrelevant since the number 
of mediator is suppressed by a $(T'/T)^3$ factor, Eq.~(\ref{lowerTprimestable}).


\subsection{The other simplest option: subleading DM annihilation into light mediators \label{subdomannihwayout}}

We move forward with an option which is also quite simple. This option does require at least one extra new particle beyond the DM and light mediator, but not necessarily a light one,
which allow to fulfill all constraints in a much easier way than if this extra particle was light (in particular lighter than the light mediator).
If the DM relic density doesn't result from the freezeout of its annihilation into light mediators, but from another annihilation, say $DM\,DM\rightarrow X\,X$, the annihilation rate into these light mediators could be much smaller than the thermal value. As stressed above, Fig.~\ref{fig:SI_swave_Oh2}, the self-interaction constraints can be fulfilled with couplings to the light mediator much smaller than the ones needed for freezeout. In this case one avoids easily the CMB constraint on the DM annihilation rate into light mediators even if it proceeds in a s-wave way, see Fig.~\ref{fig:CMBsvratio}. On the other hand, the leading annihilation channel, $DM\,DM\rightarrow X\,X$, could be p-wave to satisfy this CMB constraint. The direct detection constraint is relaxed by the fact that the mediator to DM coupling is reduced.
Probably the simplest option of this kind turns out to have as DM a (Dirac) fermion, as light mediator a gauge boson $\gamma '$ (or a scalar $\phi$) and for the extra particle $X$ a scalar $S$ (which could be real or complex, for definiteness we take it to be real and consider only interactions which contain an even number of them), i.e.~to simply add a Yukawa interaction between the DM particle and an extra scalar $S$  to the $A_{\gamma '}$ and $A_\phi$ models above,
\begin{eqnarray}
&&\text{Model}\,\, B_{\gamma '}:\,\,\,{\cal L}\owns -g \gamma '_\mu J^\mu_{DM}-\frac{\epsilon}{2} F^Y_{\mu\nu}F'^{\mu\nu}\nonumber\\
&&\hspace{2.0cm}-y_S S \bar{\psi} \psi +\lambda_{H S} H^{\dagger}H S^{2}\\
&&\text{Model}\,\, B_{\phi}:\,\,\,\owns  - y_\phi \phi\overline{\chi} \chi +h.c. -\lambda \phi^\dagger \phi H^\dagger H\nonumber\\
&&\hspace{2cm}-y_S S \bar{\psi} \psi +\lambda_{H S} H^{\dagger}HS^2 +\lambda_{\phi S} \phi^\dagger \phi S^2\nonumber\\
\end{eqnarray}
with $J_\text{DM}^\mu = \bar \psi \gamma^\mu \psi $.
Here we have also indicated  the possible quartic interactions that can be written down. 
Let us discuss in more details how we can avoid each of the constraint separately along this scenario:
\begin{itemize}
\item {\it CMB} : the DM annihilation into a pair of heavy $S$ scalars doesn't cause any large CMB distortion because this is a p-wave annihilation (for Eq.~(\ref{cmbconstraint})). For the scalar $B_{\phi}$ model, the DM annihilation into light mediators is also p-wave. For the vector $B_{\gamma '}$ model, on the other hand, the DM annihilation into light mediators is of the s-wave type but, since it is subleading, it has a cross section smaller than the thermal value, so that it can be easily slow enough to satisfy the bounds of Fig.~\ref{fig:CMBsvratio}. The decays of the heavy ($S$) and light bosons ($\gamma '$ or $\phi$) do not cause any problem either for CMB because they can easily be fast enough to have any effect on the CMB or $N_{eff}$.\footnote{The vector $\gamma '$ and the scalars $S$ and $\phi$ can decay into a pair of charged fermions through the kinetic mixing or through the mixing with the Higgs scalar respectively.}

\item {\it BBN} : being much beyond the MeV scale, the $S$ scalar can easily have a large enough decay width to decay before it would cause any problems for BBN. For the vector $B_{\gamma '}$ model, the $\gamma '$ can easily have a lifetime short enough to avoid the BBN.  The BBN constraints related to the $\phi$ mediator in the $B_\phi$ model  can also be fulfilled in an easier way because direct detection constraints are relaxed, and thus allows this light mediator to decay faster.

\item {\it Direct detection} : the direct detection constraint are relaxed because the DM to light mediator coupling is reduced. As a result, for the $B_{\gamma '}$ model, the region which is allowed by BBN and direct detection 
is much larger  than for the $A_{\gamma '}$ model (with $T'/T=1$). 
This model illustrates well how simple it can be to fulfill all constraints when one stops to assume that the light mediator is playing the key role in the DM freezeout. 
For the scalar $B_{\phi}$ model, the tension between these constraints, which left only a tiny allowed region for the $A_\phi$ model (with $T'/T=1$), is now reduced, and consequently this region is now enlarged.
The allowed region, nevertheless, still remains relatively small, even if much larger.

\item {\it Indirect detection} : the indirect detection signal from DM annihilation into a pair of light mediators is reduced because the corresponding cross section is now below the thermal value. As for indirect detection from annihilation into a pair of $S$ it is p-wave and for the parameter space which is allowed by the other constraints we found that it is suppressed.

\end{itemize}

Table~\ref{examplesZprimeS} presents numerical examples of sets of parameters which satisfy all constraints for the $B_{\gamma'}$ model, as well as for the $B_\phi$ model.

\begin{table*}[t]
\begin{center}
\small{
\begin{tabular}{|c|c|c|c|c|c|c|c|c|c|c|}
\hline
$m_{DM}$                  & $m_{\gamma '}$                  & $m_S$                     & \multirow{2}{*}{$\alpha '$} & \multirow{2}{*}{$y_S$} & $\sigma _T/m_{DM}$                  & \multirow{2}{*}{$\frac{\sigma_{DM DM \rightarrow \gamma ' \gamma '}}{\sigma_{thermal}}$} & \multirow{2}{*}{$\kappa_{\gamma '}$ $\left(\frac{\kappa_{\gamma '}}{\kappa_{KM}^{DD}}\right)$} & \multirow{2}{*}{$\kappa_{S}$ $\left(\frac{\kappa_{S}}{\kappa_{HP}^{DD}}\right)$} & $\tau_{\gamma '}$               & $\tau_{S}$                \\
(GeV) & (MeV) & (GeV) &                             &                        & $(\text{cm}^2/\text{g})$ &                                                                              &                                                 &                                                 & (sec) & (sec) \\ \hline
83                        & 18                        & 31                        & $1.7\times 10^{-4}$         & 0.25                   & 0.18                                & $1.2\times 10^{-2}$                                                          & $1.8\times 10^{-11}$ (0.55)                                            & $1.1\times 10^{-10}$ ($\ll 1$)                                            & 0.30                      & 0.089                     \\ \hline
326                       & 12                        & 62                        & $6.5\times 10^{-5}$         & 0.51                   & 0.35                                & $1.2\times 10^{-4}$                                                          & $2.3\times 10^{-11}$ (0.35)                                            & $1.8\times 10^{-10}$ ($\ll 1$)                                            & 0.12                      & 0.006                     \\ \hline
617                       & 11                        & 12                        & $3.8\times 10^{-4}$         & 0.70                   & 0.13                                & $1.0\times 10^{-3}$                                                          & $4.4\times 10^{-11}$ (0.47)                                            & $4.4\times 10^{-10}$ ($\ll 1$)                                            & 0.22                      & 0.020                     \\ \hline
\hline
$m_{DM}$                  & $m_{\phi}$                  & $m_S$                     & \multirow{2}{*}{$\alpha_{\phi}$} & \multirow{2}{*}{$y_S$} & $\sigma _T/m_{DM}$                  & \multirow{2}{*}{$\frac{\sigma_{DM DM \rightarrow \phi\phi}}{\sigma_{thermal}}$} & \multirow{2}{*}{$\kappa_{\phi}$ $\left(\frac{\kappa_{\phi}}{\kappa_{HP}^{DD}}\right)$} & \multirow{2}{*}{$\kappa_{S}$ $\left(\frac{\kappa_{S}}{\kappa_{HP}^{DD}}\right)$} & $\tau_{\phi}$               & $\tau_{S}$                \\
(GeV) & (MeV) & (GeV) &                             &                        & $(\text{cm}^2/\text{g})$ &                                                                              &                                                 &                                                 & (sec) & (sec) \\ \hline
0.5                       & 1.1                        & 0.01                        & $1.5\times 10^{-5}$         & 0.02                   & 0.19 & 0.23                                                          & $6.8\times 10^{-7}$ (0.60)                                            & $9.4\times 10^{-8}$ (0.09)                                            & 27                      & 1                     \\ \hline
2                       & 3                       & 0.01                        & $4.5\times 10^{-5}$         & 0.04                   & 0.13 & 0.13                                                          & $1.9\times 10^{-7}$ (0.99)                                            & $1.9\times 10^{-7}$ (0.99)                                            & 24                      & 1                     \\ \hline
326                       & 12                        & 62                        & $6.5\times 10^{-5}$         & 0.51                   & 0.35                                & $1.0\times 10^{-5}$                                                          & $1.4\times 10^{-8}$ (6.17)                                            & $1.8\times 10^{-10}$ ($\ll 1$)                                            & 38                    & 0.006                     \\ \hline
\end{tabular}}
\end{center}
\caption{Examples of parameters values which satisfy the various constraints for model $B_{\gamma '}$ (top) and $B_\phi$ (bottom). $\kappa_{HP}^{DD}$ and $\kappa_{KM}^{DD}$ stand for the current experimental upper limit on $\kappa_{HP}$ and $\kappa_{KM}$ respectively, for the masses considered, see Fig.~\ref{fig:DD_Constraints}. That these sets of couplings satisfy the CMB constraint of Eq.~(\ref{cmbconstraint}) can be seen straightforwardelly from comparing the value of $\sigma_{DM DM \rightarrow \gamma ' \gamma '}/\sigma_{thermal}$ (or $\sigma_{DM DM \rightarrow \phi\phi}/\sigma_{thermal}$) above with Fig.~\ref{fig:CMBsvratio}. For all of these examples, the indirect detection signal is at least two orders of magnitudes below current experimental sensitivities (see Fig. \ref{fig:ID_Constraints}).}
\label{examplesZprimeS}
\end{table*}


\subsection{The p-wave option with reduction of the light mediator number density from decay into extra hidden sector particles\label{sec:pwavemeddecay}}

A next possibility arises assuming that the annihilation into the light mediator is of the p-wave type and dominates the freezeout, and that the light mediator number density is reduced after relativistic decoupling 
from decay of this light mediator into an extra particle.\footnote{A decay exclusively into SM particles through the Higgs portal is forbidden by the combination of  direct detection  and BBN constraints (for $m_{DM}\gtrsim 1$~GeV).}
To have a p-wave annihilation, the light mediator must necessarily be a scalar, $\phi$.
The simplest model along these hypothesis is probably to assume that the extra particle is a scalar particle $S$ (which we take here too to be real for definiteness, with even number of them in all interactions), as following,
\begin{eqnarray}
\hbox{\underline{Model C} :}\quad
{\cal L}&=&- (y_\phi \phi\overline{\chi^c} \chi +h.c.) -\lambda_{\phi H} \phi^\dagger \phi H^\dagger H\nonumber\\
&&-\lambda_{\phi S} \phi^\dagger \phi S^2-\lambda_{H S} H^\dagger H S^2
\label{LagrmodelC}
\end{eqnarray}
with $\phi$ the light mediator and $S$ the extra light scalar into which $\phi$ decays, $\phi\rightarrow S S$, and/or annihilate.
Here we assume for simplicity that there is no $S\overline{\chi} \chi$ Yukawa coupling (which can be justified on the basis of a symmetry), or that it has a negligible effect.  
Note that, in order to decay, the light mediator $\phi$ must have a vev.
The various constraints are fulfilled in the following way:
\begin{itemize}
\item {\it CMB} : the DM annihilation into a pair of scalar light mediators doesn't cause any large CMB distortion because this is a p-wave annihilation (for Eq.~(\ref{cmbconstraint})). 
The light scalars can decay fast enough to have any sizable effect on the CMB. The $N_{eff}$ constraint is satisfied from fast enough decay of the scalars.
\item {\it BBN} : the BBN constraints are solved in a similar way than the $N_{eff}$ constraint.
On the one hand the $\phi\rightarrow S S$ decay can be fast enough if $\lambda_{\phi S}$ is not tiny. On the other hand, the $S\rightarrow SM SM$ decay can also be fast enough,  through $S$-$h$ mixing if the  $\lambda_{H S}$ coupling is large enough. The later feature can be realized without inducing a too large Higgs boson invisible decay width. 
For instance, if the S scalar has a mass above the $e^+e^-$ threshold one gets
\small{\begin{eqnarray}
&&\hspace{-0.6cm}\tau_{\phi\rightarrow SS}\simeq  1\text{s}\cdot \left(\frac{m_{\phi}}{20\text{~MeV}}\right)\left(\frac{2.3\times 10^{-12}}{\lambda _{\phi S}}\right)^{2}\left(\frac{500\text{~MeV}}{v_{\phi}}\right)^{2}\label{phidecay}\\
&&\hspace{-0.6cm}\tau_{S\rightarrow e^+e^-}\simeq  1\text{s}\cdot \left(\frac{2\text{~MeV}}{m_{S}}\right)\left(\frac{7.8\times 10^{-3}}{\lambda _{H S}}\right)^{2}\left(\frac{500\text{~MeV}}{v_{S}}\right)^{2}\label{Sdecay}
\end{eqnarray}}

\normalsize 
\item {\it Direct detection} :  self-interaction constraints requires a sizable DM to $\phi$ coupling but doesn't require any DM to $S$ coupling. Thus, direct detection through DM to S coupling is small if the corresponding coupling is small (or if it doesn't exist as here). Direct detection through DM to $\phi$ interaction can easily be suppressed too, even if $\lambda_{\phi S}$ and $\lambda_{HS}$ couplings are sizable to account for BBN. It just requires that the transition $\phi$ to $H$ is suppressed enough, i.e.~that the $\lambda_{\phi H}$ interaction is small enough, as well as the product of $\lambda_{\phi S}$ and $\lambda_{H S}$ .
\item {\it Indirect detection} : even if the annihilation is of the p-wave type, the Sommerfeld effect can lead in this model to a large indirect detection rate which can be tested (and actually rules out large part of the parameter space). 
A couple of numerical examples which satisfy all constraints are given in the two first lines of Table \ref{examplesphiSmodel}. 
For the second example the annihilation cross section of DM into $\phi$'s is of order $\left\langle\sigma v\right\rangle_{DM DM \rightarrow \phi\phi}\simeq 10^{-25} $ cm$^{3}$s$^{-1}$ in dwarf galaxies\footnote{This cross section has to be multiplied by four in order to get the DM annihilation cross section into SM charged fermions since each of the $\phi$'s will decay into two charged fermions.}. As can be seen from Fig. \ref{fig:ID_Constraints}, this can be constrained by indirect detection experiments. This is due to the fact, discussed above, that the Sommerfeld effect brings a $\propto 1/v^3$ enhancement effect which compensate for the $v^2$ suppression of the p-wave cross section.
\end{itemize}

To sum up, all constraints are satisfied provided $\lambda_{\phi S}$ and $\lambda _{H S}$ are large enough (but not too large so that their product is small enough) and $\lambda_{\phi H}$ is small enough.\footnote{For further details 
on this setup, note that with for example the first set of couplings in Table \ref{examplesphiSmodel}, the $S$ scalar thermalizes with the SM thermal bath through $\lambda_{HS}$, whereas DM and $\phi$ thermalize through $y_\phi$. Both visible and hidden sector never thermalize through the $\lambda_{\phi S}$ interaction but does so through $\lambda_{\phi H}$ until $T\simeq 35$~MeV. Later on, the $T'/T$ ratio doesn't remain equal to unity but remains close to it.  For instance, at $t\simeq 1$~sec,  one has  $T'/T\simeq 0.95$. 
}

\begin{table*}[t]
\begin{center}
\small{
\begin{tabular}{|c|c|c|c|c|c|c|c|c|c|c|c|c|c|c|}
\hline
$m_{DM}$ & $m_{\phi}$ & $m_S$ & $v_\phi$ & $v_S$ & \multirow{2}{*}{$\lambda_{\phi H}$} & \multirow{2}{*}{$\lambda_{\phi S}$} & \multirow{2}{*}{$\lambda_{H S}$} & $\sigma_T/m_{DM}$        & \multirow{2}{*}{$\frac{\kappa_{HP}}{\kappa_{HP}^{DD}}$} & $\tau_{\phi}$ & $\tau_{S}$ & $\Gamma_h^{inv}$ & $\langle \sigma_{\phi\phi\rightarrow S S} v \rangle$ & \multirow{2}{*}{$\Omega_{med}^{0} h^2$} \\
(GeV)    & (MeV)      & (MeV) & (MeV)    & (MeV) &                                     &                                     &                                  & $(\text{cm}^2/\text{g})$ &                                                 & (sec)         & (sec)      & (MeV)            & ($\text{GeV}^{-2}$)                                  &                                     \\ \hline
126      & 20         & 2     & 500      & 500   & $5.4\times 10^{-8}$                 & $6.2\times 10^{-12}$                & 0.008              & 0.28                     & 0.05                                            & 0.14          & 3.0       & 0.31             & $\ll \sigma_{thermal}$ & 0 \\ \hline
382      & 71         & 7     & 436      & 83    & $6.3\times 10^{-7}$                 & $3.2\times 10^{-11}$                & 0.007 & 0.21                     & 0.49                                            & 0.03          & 2.7       & 0.24             & $\ll \sigma_{thermal}$                                                    & 0                                   \\ \hline\hline
83        & 50          & 2     & $\ll 1$        & 500     & $6.3\times 10^{-8}$                                   & 0.010                                   & 0.008                                & 0.11                        & $\ll 1$                                               & $\gg 1$             & 3.0 & 0.31                & $2.5\times 10^{-5}$                                                    & $2.5\times 10^{-5}$                                   \\ \hline
173        & 300          & 10     & $\ll 1$        & 50     & $1.0\times 10^{-6}$                                   & 0.015 & 0.010 & 0.21 & $\ll 1$ & $\gg 1$             & 1.44          & 0.48                & $1.5\times 10^{-6}$                                                    & $4.0\times 10^{-4}$                                   \\ \hline
\end{tabular}}
\end{center}
\caption{Examples of parameter values which satisfy the various constraints for model C, from fast enough mediator decay (first 2 examples) or from efficient enough (stable) mediator annihilation (last 2 examples) . $\kappa_{HP}^{DD}$ stands for the current experimental upper limit on $\kappa_{HP}$ for the masses considered, see Fig.~\ref{fig:DD_Constraints}. $\Omega_{med}^{0} h^2$ refers to the relic density value of the mediator today.}
\label{examplesphiSmodel}
\end{table*}

\subsection{Stable mediator option with annihilation of it into hidden sector particles\label{sec:stablewayout}}

As discussed above, if the mediator is absolutely stable, the main constraints are the non overclosure and modification of the Hubble constant ones.
A simple way to avoid them is to consider a scenario where  $T'/T$ is sizably smaller than one, see Eq.~(\ref{eq:TpTlowerbound}). If  $T'/T\simeq 1$ instead, a more complicated possibility arises if the stable light mediator number density
is reduced after this decoupling from an annihilation into extra lighter particles (implying new constraints related to the existence of this extra light particle).
One finds 2 minimal models realizing this scenario. The first one has already been proposed in Ref.~\cite{Duerr:2018mbd} (see also \cite{Ma:2017ucp}). It involves a s-wave annihilation of DM into a $\gamma '$ light mediator, followed by an annihilation of this $\gamma '$ into a lighter scalar S, followed by decay of this scalar $S$ into SM particles. Omitting less relevant scalar interactions, the Lagrangian of this model is:
\begin{eqnarray}
\hbox{Model} \hbox{ D:}\quad
{\cal L}&=&-g \gamma '_\mu J^\mu_{DM}-g \gamma '_\mu J^\mu_S\nonumber\\
&&-\frac{\epsilon}{2} F^Y_{\mu\nu}F'^{\mu\nu} -\lambda_{H S} S^\dagger S H^\dagger H
\end{eqnarray}
with $J_{S}^\mu = i Q'_S (S^* \partial_\mu S - S \partial_\mu S^*)$ and $J_\text{DM}^\mu$ as for the $A_{\gamma '}$ model of Eq.~(\ref{medcouplV}). The mass of the $\gamma '$ comes from spontaneous breaking of the $U(1)'$ gauge symmetry once the $S$ scalar acquires a vev. The stability of the $\gamma '$ light mediator requires here an extra (charged conjugation) symmetry, in order that a kinetic mixing between the $\gamma '$ and the hypercharge gauge boson is forbidden. A nice feature of this model is that the extra scalar S is the one one has anyway to introduce in the model if one assumes that the $U(1)'$ symmetry is spontaneously broken through the Brout-Englert-Higgs mechanism.
The other minimal model where this turns out to be possible is nothing but the C model of the previous section, in other regions of the parameter space, assuming in particular now that the scalar mediator $\phi$ has no vev, so that it doesn't decay.

The phenomenology of Model D has been analysed in detail in Ref.~\cite{Duerr:2018mbd}. 
Once the non overclosure constraint is satisfied for the $\gamma '$ light mediator, from $\gamma ' \gamma '\rightarrow S S$ annihilations, one is  essentially left with 2 types of constraints. First, one has to care about the modification of the Hubble constant the stable $\gamma'$ implies 
at BBN time, Eq.~(\ref{DeltaNeff}). Second, there are constraints related to the existence of the extra particle $S$.
On the one hand this extra particle must lies above the $2 m_e$ threshold, otherwise its decay is loop suppressed and therefore too slow for CMB or BBN constraints. On the other hand its mass must be below 4.4 MeV to avoid the BBN photodisintegration constraint.
The $N_{eff}$ and Hubble constant/entropy injection constraints are just enough satisfied from the fact that when the $S$ and $\phi$ scalars decouple, $T'/T$ (which is equal to unity at DM decoupling) is already not anymore equal to unity, but slightly smaller (from decoupling of SM particles in between). The CMB constraint on DM annihilation, Eq.~(\ref{cmbconstraint}), doesn't apply for $DM DM\rightarrow \gamma' \gamma'$ because the $\gamma'$ is stable but applies for the $DM DM\rightarrow \gamma' S$ scattering, as well as for $\gamma' \gamma'\rightarrow S S$ scattering. For the last scattering it is easily satisfied since this scattering is not Sommerfeld enhanced. For the former scattering, this constraint still leaves an allowed parameter space.
Direct detection is satisfied from the fact that DM communicates with the SM particles only through a DM-$\gamma '$-$S$-SM chain.

\begin{figure}[h!]
\centering
\begin{minipage}{0.23\textwidth}
  \centering
\includegraphics[scale=0.40]{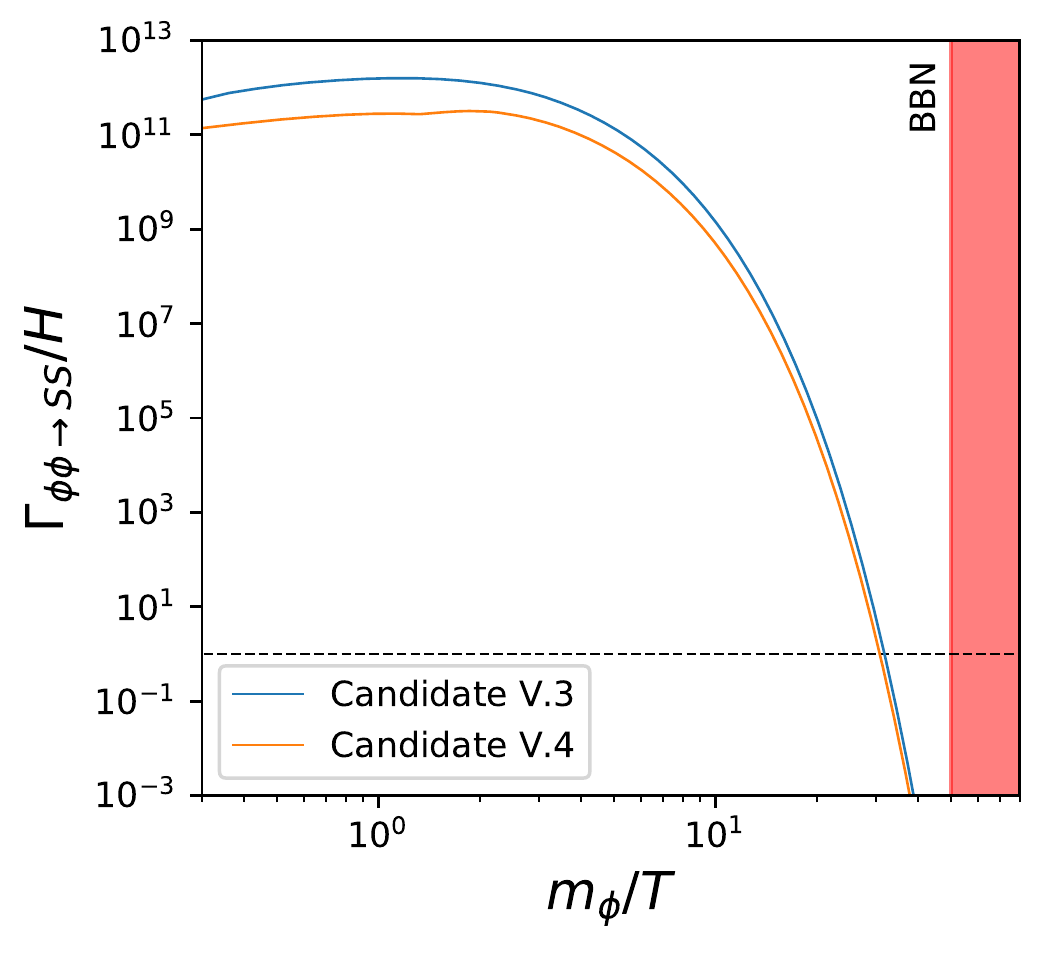}
\end{minipage}
\begin{minipage}{0.23\textwidth}
  \centering
\includegraphics[scale=0.40]{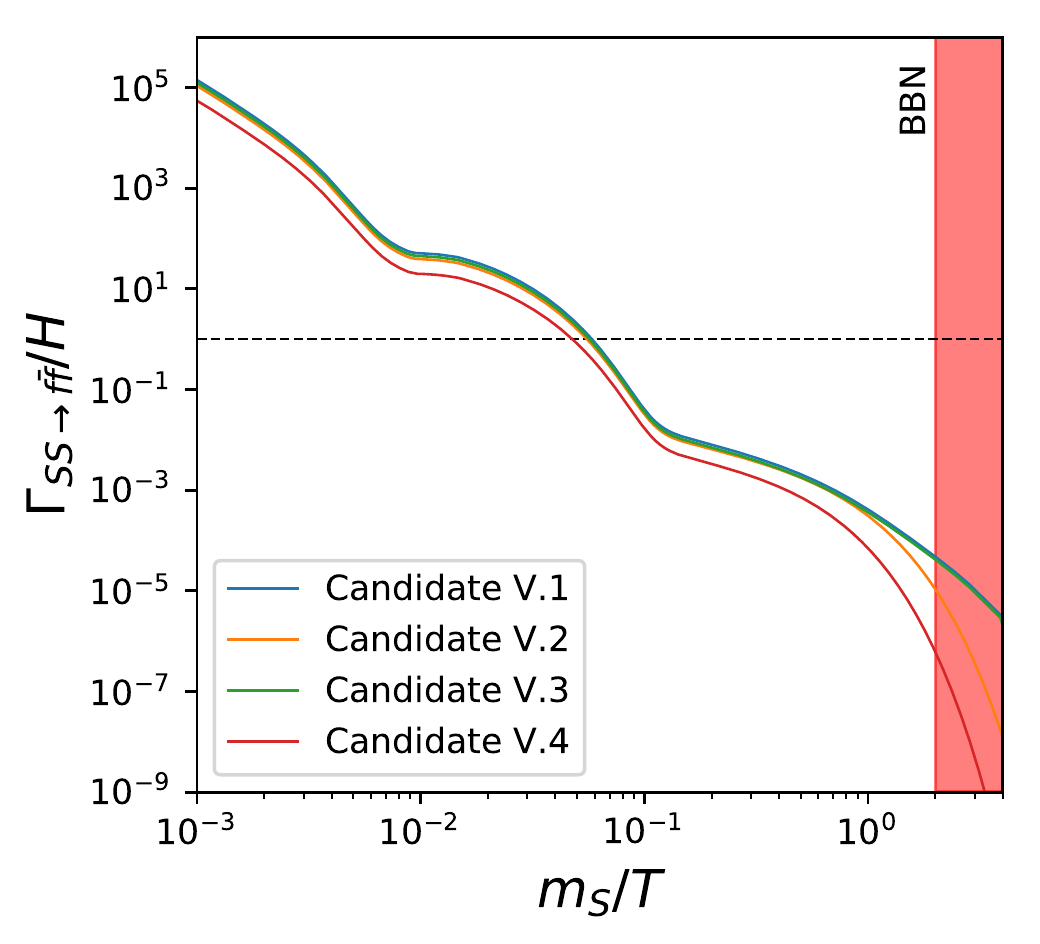}
\end{minipage}
\caption{Left: Evolution of the annihilation rate of a pair of $\phi$'s into a pair of $S$'s, normalized to the Hubble rate for the last two candidates of Table \ref{examplesphiSmodel}. The black dashed line represents the thermalization line, ie: when $\Gamma _{\phi\phi\rightarrow SS}=H$. Right: Evolution of the annihilation rate of a pair of $S$'s into a pair of SM fermions normalized to the Hubble rate for all candidates of Table \ref{examplesphiSmodel}. The black dashed line represents the thermalization line, ie: when $\Gamma _{SS\rightarrow f\bar{f}}=H$. One can also see when BBN starts in red in both plots.}\label{fig:C_Rate}
\end{figure}

As for model C with a stable scalar mediator $\phi$, it can be successful in a perhaps less squeezed way.
The first constraint it has to fulfill is that the $\phi\phi\rightarrow S S$ annihilation process must be in thermal equilibrium at temperature of order $m_\phi$, so that, later on, it Boltzmann suppresses the $\phi$ number density.
This will be the case for model $C$ in another parameter regime than for the decay option of Section~\ref{sec:pwavemeddecay}.
One has to  assume much larger values of $\lambda_{\phi S}$  than above (i.e.~for instance $\lambda_{\phi S}> \lambda _{\phi S}^{th}\simeq 6\times 10^{-9}$ for $m_{med}=30$ MeV). Note that $S$ can annihilate into a pair of SM particles through the Higgs portal, but these $S S\rightarrow f \bar{f}$ processes decouple before the $\phi \phi\rightarrow S S$ process. The later process decouples when $m_\phi/T\sim 30$. 
This is shown in Fig.~\ref{fig:C_Rate} for two examples sets of parameters which satisfy all constraints and are given in Table \ref{examplesphiSmodel}.
Thus, for these two examples (where $m_\phi/m_S\lesssim 30$), the $S$ scalar is a hot relic and its number density is not Boltzmann suppressed.
The way the various constraints are fulfilled is as follows:
 \begin{itemize}
 \item{\it Non overclosure}: the $\phi\phi \rightarrow S S$ scattering can reduce enough the population of stable light mediator $\phi$ to non overclose the Universe.
\item {\it CMB} : the DM annihilation into a pair of scalar light mediators doesn't cause any large CMB distortion because this is a p-wave annihilation (for Eq.~(\ref{cmbconstraint})). The 
decay of $S$ is fast enough to satisfy the $N_{eff}$ constraint associated to $S$ decay. 
\item {\it BBN} : the $S$ decay is also fast enough to satisfy BBN Hubble constant/entropy injection and photodisintegration constraints
 (by taking a large enough $\lambda_{H S}$ coupling, see Eq.~(\ref{Sdecay})).
All this can be realized without inducing a too large Higgs boson invisible decay width. 

\item {\it Direct detection} :  as in Section \ref{sec:pwavemeddecay}, the self-interaction constraints require a sizable DM to $\phi$ coupling but doesn't require any DM to $S$ coupling. Thus, direct detection through DM to S coupling is small if the corresponding coupling is small. Direct detection through DM to $\phi$ doesn't occur at tree level because $\phi$ has no vev.

\item {\it Indirect detection}: if the $\phi$ has no vev and is stable, the annihilation of DM into $\phi$ doesn't produce any SM particles and DM annihilation produce SM particles only at the loop level.

\end{itemize}
All in all, these features make this stable mediator scenario based on model C easily viable. 
\subsection{The neutrino option \label{sec:neutrinowayout}}

If we do not assume any hidden sector extra states, the light mediator can only decay into photons, neutrinos or $e^+e^-$. For the decay to proceed into neutrinos, without producing $e^+ e^-$ pairs, there are 2 simple possibilities. One needs either that the decay proceeds only into muon or tau neutrinos, or that the mass of the mediator lies below $2 m_e$.
With a light scalar mediator it is not easy to have  a dominant and fast enough decay into neutrinos.
As for the $\gamma '$ light mediator option, the possible portals which can make the $\gamma '$ to decay into neutrinos are the kinetic mixing portal or mass mixing portal. As already mentioned above, through kinetic mixing, and below the $e^+e^-$ threshold, the $\gamma '$ decays dominantly to $3\gamma$ at a rate far too slow to proceed before BBN, and even slower to neutrinos.
Through the more involved mass mixing scenario, the decay can dominantly proceed into neutrinos at a rate fast enough to decay before BBN.
The relevant interactions are therefore
\begin{eqnarray}
\hbox{\underline{Model E$_m$} : }
{\cal L} &\owns& -g \gamma '_\mu J^\mu_{DM} -\delta m^2 \gamma '_\mu Z_\mu
\end{eqnarray}
where $J_\text{DM}^\mu = \bar \psi \gamma^\mu \psi $, $J_\text{DM}^\mu = \bar \chi \gamma^\mu \gamma^5 \chi $ and $J_\text{DM}^\mu = i (S^* \partial_\mu S - S \partial_\mu S^*)$
for a Dirac, Majorana and scalar DM particle respectively.
This model (for the Dirac DM case) has already been considered in Appendix B of \cite{Bringmann:2016din}, as a way out to the CMB constraints above, Eq.~(\ref{cmbconstraint}), see the various constraints holding on it in Fig.~4 of this reference.

The kinetic mixing option doesn't allow to have a decay into muon and tau neutrinos without having production of electron neutrinos as well.
However, electron neutrinos production can be avoided in models with an extra $U(1)'$ gauge symmetry, along which the corresponding $Z'$ couples only to $\mu$ and $\tau$ flavors, as well as to DM.
The most straightforward possibility is to assume a $L_\mu-L_\tau$ flavor symmetry so that gauge anomalies cancel, 
\begin{eqnarray}
\hbox{\underline{Model E$_{\mu-\tau}$} : }{\cal L} &\owns& -\frac{1}{4} F'_{\mu\nu} F'^{\mu\nu}- i g' \sum_{i} Q'_{i} \overline{\psi}_i \gamma^\mu Z'_\mu \psi_i  \nonumber\\
&&+ i g'  Q'_{DM} \overline{\chi} \gamma^\mu \chi Z'_\mu\chi\quad
\label{mutaumodel}
\end{eqnarray}
where the sum is over the muon and tau left-handed doublets, as well as over the muon and tau right-handed singlets (with $Q'=1$ for muon spinors and $Q'=-1$ for tau spinors). Here, for definiteness, $\chi$ is taken to be a Dirac fermion. DM could also be a scalar boson charged under the $U(1)'$.
The gauge part of this model \cite{He:1990pn,Foot:1990mn,He:1991qd,Heeck:2011wj} has been considered in many different contexts, including as a possibility of explanation \cite{Gninenko:2001hx,Baek:2001kca,Carone:2013uh,Altmannshofer:2014pba}  for the $(g-2)_\mu$ anomaly. This requires
a $Z'$ with mass 10-100~MeV and $g'\sim 5\times 10^{-4}$.
With the adjunction of a DM particle, as in Eq.~(\ref{mutaumodel}), it has also been considered for various purposes \cite{Cirelli:2008pk,Baek:2008nz,Garani:2019fpa}.
In Ref.~\cite{Garani:2019fpa} it has been noted that the values of $m_{Z'}$ and $g'$ which fit well the $(g-2)_\mu$ anomaly, see above, can also lead to DM self-interactions with $\sigma_T/m_{DM}\sim 1 \hbox{cm}^2/\hbox{g}$ (once $Q'_{DM}$ has been fixed for the annihilation of DM into leptons to have the thermal value).
Here we would like to point out that this model turns out to be good also to avoid the CMB and BBN constraints above. In particular, as already said, a decay of the light mediator into 
muon and/or tau neutrinos allow to avoid the CMB constraint of Eq.~(\ref{cmbconstraint}). The other CMB and the BBN constraints are avoided in a way similar to the way
the $A_{\gamma'}$ model avoids these contraints above (for $T'/T=1$). Note that all this has been analyzed in \cite{Kamada:2018zxi} in the framework of a model where, on top of the interactions of Eq.~(\ref{mutaumodel}), the scalar whose vev breaks spontaneously the $U(1)_{L_{\mu}-L_{\tau}}$ gauge symmetry is considered explicitely, assuming in addition that this scalar has Yukawa interactions with DM, so that for self-interactions the light mediator is this extra scalar, rather than the $Z'$. See also \cite{Kamada:2018kmi} for a model gauging the $U(1)_{(B-L)_{e,\mu,\tau}}$ symmetries with several extra particles.


\section{The asymmetric dark matter option\label{sec:asym}}

So far we have assumed everywhere that there is an equal number of DM particles and antiparticles.
If instead we assume that DM is asymmetric, some of the constraints will change drastically. This possibility has been considered  in Ref.~\cite{Baldes:2017gzu} in the context of a model where a dark proton and a dark electron couple to a dark photon.
On the one hand, all constraints related to DM annihilation are trivially removed, since DM doesn't annihilate anymore. This concerns in particular the CMB constraint of Eq.~(\ref{cmbconstraint}) and the indirect detection constraints.
On the other hand, the constraints on the number of mediators  and on its decay remain.
In an asymmetric setup, still, DM thermalizes with the light mediator at $T\gtrsim m_{DM}$, and the resulting important symmetric DM component must be suppressed afterwards through an annihilation catastrophy. This annihilation catastrophy leaves the mediator as a hot relic, so one is left with as many light mediator as in the symmetric case. 
Thus, if the mediator is stable, the non overclosure constraint remains fully relevant. Similarly, for an unstable mediator, remain relevant the $N_{eff}$ and mediator decay CMB constraints above, as well as the Hubble constant/entropy injection and photodisintegration BBN constraints, as well as direct detection constraints.
In Ref.~\cite{Baldes:2017gzu} the $N_{eff}$ CMB constraint, as well as Hubble constant/entropy injection BBN constraints, have not been considered in details as in the subsequent 
Ref.~\cite{Hufnagel:2018bjp}. However, as already said above, if for the minimal  vector model $A_\gamma'$ of Eq.~(\ref{medcouplV}) (and similarly for the more involved vector model of Ref.~\cite{Baldes:2017gzu}),
one removes the CMB constraint of Eq.~(\ref{cmbconstraint}), a proper incorporation of these constraints leaves allowed a relatively wide region of parameter space. For the scalar mediator model instead, to assume an asymmetric setup instead of a symmetric one doesn't change much the picture because in this model this CMB constraint was already avoided (from the fact that the annihilation is p-wave). In particular the strong tension between direct detection and BBN constraints remains.


\section{Summary}

Very minimal light mediator setups, that could lead to DM self-interactions in agreement with what simulations of formation of small scale structure tend to imply, are suffering from a plethora of constraints.
To assume a light mediator is a particularly simple solution to account for the large disparity of cross sections needed  (large self-interaction vs small annihilation cross sections). It has in particular the virtue to induce a velocity dependence of the self-interaction cross section, which helps accommodating all self-interaction constraints. However,
in minimal setups the light mediator typically decouples relativistically when DM freezes-out, which leads to an overclosure of the Universe if the light mediator is stable.
To assume a portal between the DM/light mediator hidden sector and the SM visible sector allows the light mediator to disappear from the thermal bath into SM particles. 
Such a portal also offers possibilities of tests, beyond purely gravitational ones. A portal, nevertheless, is constrained in many ways.
There is a tension (particularly for a scalar light mediator) between constraints (such as BBN), which tend to require large portal interactions in order that the light mediator decays fast enough, and other constraints which require a tiny portal interaction (such as direct and indirect detection, which are both boosted by the fact that the mediator is light). On top of that, if the light mediator is a vector boson which decays into SM particle through the portal, the s-wave annihilation of DM into a pair of light mediators (which is also boosted by the Sommerfeld enhancement) distorts the CMB beyond what observations allow.
A summary of the many relevant constraints can be found in Section~\ref{sec:constraints}.

In this work we have proposed various simple ways out to these tensions. The simplest one works in the framework of the two very minimal setups generally assumed, Eqs.~(\ref{medcouplV}) and (\ref{medcouplphi}), but relaxing one of the assumption made so far, i.e.~that the hidden sector has same temperature as the visible sector at DM freeze-out. If one assumes instead that both sectors have never thermalized, so that $T'/T$ is sizably smaller than unity, the DM to light mediator interactions needed to account for the observed relic density is smaller, so that all constraints above requiring a tiny portal interaction are relaxed. 
In particular, Fig.~(\ref{fig:CMBsvratio}) shows that as soon as one reduces this DM to light mediator interaction, the CMB upper bound on the s-wave DM annihilation cross section can be easily accommodated.  To have such a reduced DM to light mediator interaction is perfectly compatible with the self-interaction constraints, as Fig.~(\ref{fig:SI_swave_Oh2}) shows. Moreover, all constraints on the decay of the light mediator, such as BBN ones, largely relax because the number density of light mediator is reduced by a factor of $(T'/T)^3$. 
The typical $T'/T$ ratio needed to accommodate all constraints lies from about a few $10^{-2}$ down to the ``$T'/T$ floor'' we have determined in Eq.~(\ref{eq:TpTlowerbound}), typically of order $10^{-4}$.  This floor is due to the fact that if $T'/T$ is too small, the hidden sector will simply not contain enough DM particles to account for the observed relic density. Independently of self-interactions, it corresponds to the value of $T'/T$ for which any DM particle, which has thermalized in a hidden sector, can account for the observed relic density, even though it decouples while still relativistic (i.e.~without Boltzmann suppression) and has a mass
way beyond the usual eV mass scale of hot relics. To assume that both sectors have never thermalized makes full sense for self-interacting dark matter, since there are already strong constraints on the size of the portal, in particular on the kinetic mixing portal in the vector light mediator case. Thus, unless one assumes that there would be an UV interaction to thermalize both sectors, to assume
$T'/T<1$ is quite natural and generic for self-interacting DM. For the vector model we find a rather large parameter space satisfying all constraints with the mediator decaying with a lifetime much smaller than the age of the Universe. 
As for the scalar mediator case with Higgs portal, to reduce $T'/T$ doesn't help with respect to the $T'/T=1$ case for a mediator mass above the $2 m_e$ mass threshold, but
it widely opens the region below this threshold. For both models direct detection should allow to probe some of the allowed parameter space. Note that everywhere we have assumed that the portal interactions do not change sizably the $T'/T$ temperature so that the portal doesn't heat the hidden sector to a $T'$ temperature beyond the one assumed. It would be useful to determine what happen's when this is not the case (leading typically to a reannihilation scenario for the DM relic density). About this $T'/T<1$ scenario, finally, interestingly enough too, note that it also allows to have no portal at all (i.e.~an absolutely stable mediator) as a result of the fact that in this case the light mediator number density is suppressed and consequently doesn't overclose the Universe, see Eq.~(\ref{lowerTprimestable}).

Another simple way out is to assume that DM doesn't annihilate dominantly into a pair of light mediators, but to an extra particle.
In this case the DM to light mediator interaction is also reduced and the various constraints also relax accordingly, in particular direct detection, indirect detection and CMB upper bound on s-wave DM annihilation.
Note importantly that the extra particle, to which DM annihilates dominantly, doesn't need to be as light as the light mediator. This allows to avoid easily the constraints which hold on the extra light particle.
Thus, the model remains truly minimal at a low scale. Here we have considered 2 models which are made of the minimal models of Eqs.~(\ref{medcouplV}) and (\ref{medcouplphi}) to which we have added an extra scalar particle. Annihilation of DM into this new particle is p-wave, whereas the one into light mediators can be s-wave or p-wave. 

Next, we have discussed a more involved possibility where the light mediator would decay into an extra lighter particle, which itself decays subsequently into SM particles. This allows to reduce the number density of light mediator at BBN and CMB time.  The minimal model of this kind we considered, has, on top of a light scalar mediator,  an extra scalar particle which couples to the light mediator but not to DM. All constraints can be satisfied in this case, typically if we assume a hierarchy of couplings between the interactions which, again, must be small (in particular due to direct detection), and the ones which must be larger (such as for inducing fast enough decays for BBN of both the mediator and the extra light particle). 

Next, we considered another more involved possibility, where now the light mediator is stable but with still a portal interaction between the hidden sector and SM visible sector. A model of this kind has already been proposed \cite{Duerr:2018mbd}. 
It considers an extra scalar particle to which a light vector boson mediator can annihilate in order to reduce its number density, with subsequent decay of the extra scalar particle into SM particles through the Higgs portal. Here, we summarized how can work this scenario in this case, together with introducing and discussing another stable mediator setup which, with a light scalar mediator and an extra lighter scalar, can easily work.
 
Next, we discussed the option where the light mediator would decay exclusively to neutrinos. Here we essentially mentioned that  a model of this kind was already proposed \cite{Bringmann:2016din}, through an involved mass mixing of a light mediator vector boson with the Z boson. We also stressed that if the light mediator decays exclusively to muon and tau neutrinos, one could get an easily viable model for a light mediator below the $2m_\mu$ threshold, a possibility which requires a flavor structure. We considered the simple example of a $Z'$ model with global $L_\mu-L_\tau$ symmetry.

Finally, we recalled that if DM is not symmetric but asymmetric, some of the constraints get relaxed and a viable model of this type has been proposed in \cite{Baldes:2017gzu}. Here we are somehow going away from the minimality criteria, which we have taken as basic criteria for this work (since an asymmetric setup requires, on top of an efficient DM annihilation, a mechanism generating the DM asymmetry in the UV) but this is certainly a possibility too.

\section*{Acknowledgments}
We thank Camilo Garcia-Cely and Karsten Jedamzik for several useful discussions and Marco Hufnagel for having transmitted to us the output of his code which we have used for the BBN and $N_{eff}$ constraints discussed above. This work is supported by the F.R.S./FNRS under the Excellence of Science (EoS) project No. 30820817 - be.h ``The H boson gateway to physics beyond the Standard Model'', by the FRIA, by the ``Probing dark matter with neutrinos" ULB-ARC convention and by the IISN convention 4.4503.15. We thank the Erwin Schr\"odinger International Institute (Vienna), the Galileo Galilei Insitute (Florence), the Blaise Pascal Institute (Paris-Saclay astroparticle Symposium) and the TIFR (Mumbai) for hospitality while this work was done.

\newpage
\bibliographystyle{apsrev}
\bibliography{biblio}

\begin{thebibliography}{100}
\expandafter\ifx\csname natexlab\endcsname\relax\def\natexlab#1{#1}\fi
\expandafter\ifx\csname bibnamefont\endcsname\relax
  \def\bibnamefont#1{#1}\fi
\expandafter\ifx\csname bibfnamefont\endcsname\relax
  \def\bibfnamefont#1{#1}\fi
\expandafter\ifx\csname citenamefont\endcsname\relax
  \def\citenamefont#1{#1}\fi
\expandafter\ifx\csname url\endcsname\relax
  \def\url#1{\texttt{#1}}\fi
\expandafter\ifx\csname urlprefix\endcsname\relax\def\urlprefix{URL }\fi
\providecommand{\bibinfo}[2]{#2}
\providecommand{\eprint}[2][]{\url{#2}}

\bibitem[{\citenamefont{Boylan-Kolchin
  et~al.}(2011)\citenamefont{Boylan-Kolchin, Bullock, and
  Kaplinghat}}]{BoylanKolchin:2011de}
\bibinfo{author}{\bibfnamefont{M.}~\bibnamefont{Boylan-Kolchin}},
  \bibinfo{author}{\bibfnamefont{J.~S.} \bibnamefont{Bullock}},
  \bibnamefont{and}
  \bibinfo{author}{\bibfnamefont{M.}~\bibnamefont{Kaplinghat}},
  \bibinfo{journal}{Mon. Not. Roy. Astron. Soc.}
  \textbf{\bibinfo{volume}{415}}, \bibinfo{pages}{L40} (\bibinfo{year}{2011}),
  \eprint{1103.0007}.

\bibitem[{\citenamefont{Boylan-Kolchin
  et~al.}(2012)\citenamefont{Boylan-Kolchin, Bullock, and
  Kaplinghat}}]{BoylanKolchin:2011dk}
\bibinfo{author}{\bibfnamefont{M.}~\bibnamefont{Boylan-Kolchin}},
  \bibinfo{author}{\bibfnamefont{J.~S.} \bibnamefont{Bullock}},
  \bibnamefont{and}
  \bibinfo{author}{\bibfnamefont{M.}~\bibnamefont{Kaplinghat}},
  \bibinfo{journal}{Mon. Not. Roy. Astron. Soc.}
  \textbf{\bibinfo{volume}{422}}, \bibinfo{pages}{1203} (\bibinfo{year}{2012}),
  \eprint{1111.2048}.

\bibitem[{\citenamefont{Spergel and Steinhardt}(2000)}]{Spergel:1999mh}
\bibinfo{author}{\bibfnamefont{D.~N.} \bibnamefont{Spergel}} \bibnamefont{and}
  \bibinfo{author}{\bibfnamefont{P.~J.} \bibnamefont{Steinhardt}},
  \bibinfo{journal}{Phys. Rev. Lett.} \textbf{\bibinfo{volume}{84}},
  \bibinfo{pages}{3760} (\bibinfo{year}{2000}), \eprint{astro-ph/9909386}.

\bibitem[{\citenamefont{Walker and Penarrubia}(2011)}]{Walker:2011zu}
\bibinfo{author}{\bibfnamefont{M.~G.} \bibnamefont{Walker}} \bibnamefont{and}
  \bibinfo{author}{\bibfnamefont{J.}~\bibnamefont{Penarrubia}},
  \bibinfo{journal}{Astrophys. J.} \textbf{\bibinfo{volume}{742}},
  \bibinfo{pages}{20} (\bibinfo{year}{2011}), \eprint{1108.2404}.

\bibitem[{\citenamefont{Kuzio~de Naray and Spekkens}(2011)}]{deNaray:2011hy}
\bibinfo{author}{\bibfnamefont{R.}~\bibnamefont{Kuzio~de Naray}}
  \bibnamefont{and} \bibinfo{author}{\bibfnamefont{K.}~\bibnamefont{Spekkens}},
  \bibinfo{journal}{Astrophys. J.} \textbf{\bibinfo{volume}{741}},
  \bibinfo{pages}{L29} (\bibinfo{year}{2011}), \eprint{1109.1288}.

\bibitem[{\citenamefont{Oman et~al.}(2015)}]{Oman:2015xda}
\bibinfo{author}{\bibfnamefont{K.~A.} \bibnamefont{Oman}} \bibnamefont{et~al.},
  \bibinfo{journal}{Mon. Not. Roy. Astron. Soc.}
  \textbf{\bibinfo{volume}{452}}, \bibinfo{pages}{3650} (\bibinfo{year}{2015}),
  \eprint{1504.01437}.

\bibitem[{\citenamefont{Weinberg et~al.}(2015)\citenamefont{Weinberg, Bullock,
  Governato, Kuzio~de Naray, and Peter}}]{Weinberg:2013aya}
\bibinfo{author}{\bibfnamefont{D.~H.} \bibnamefont{Weinberg}},
  \bibinfo{author}{\bibfnamefont{J.~S.} \bibnamefont{Bullock}},
  \bibinfo{author}{\bibfnamefont{F.}~\bibnamefont{Governato}},
  \bibinfo{author}{\bibfnamefont{R.}~\bibnamefont{Kuzio~de Naray}},
  \bibnamefont{and} \bibinfo{author}{\bibfnamefont{A.~H.~G.}
  \bibnamefont{Peter}}, \bibinfo{journal}{Proc. Nat. Acad. Sci.}
  \textbf{\bibinfo{volume}{112}}, \bibinfo{pages}{12249}
  (\bibinfo{year}{2015}), \eprint{1306.0913}.

\bibitem[{\citenamefont{Wandelt et~al.}(2000)\citenamefont{Wandelt, Dave,
  Farrar, McGuire, Spergel, and Steinhardt}}]{Wandelt:2000ad}
\bibinfo{author}{\bibfnamefont{B.~D.} \bibnamefont{Wandelt}},
  \bibinfo{author}{\bibfnamefont{R.}~\bibnamefont{Dave}},
  \bibinfo{author}{\bibfnamefont{G.~R.} \bibnamefont{Farrar}},
  \bibinfo{author}{\bibfnamefont{P.~C.} \bibnamefont{McGuire}},
  \bibinfo{author}{\bibfnamefont{D.~N.} \bibnamefont{Spergel}},
  \bibnamefont{and} \bibinfo{author}{\bibfnamefont{P.~J.}
  \bibnamefont{Steinhardt}}, in \emph{\bibinfo{booktitle}{{Sources and
  detection of dark matter and dark energy in the universe. Proceedings, 4th
  International Symposium, DM 2000, Marina del Rey, USA, February 23-25,
  2000}}} (\bibinfo{year}{2000}), pp. \bibinfo{pages}{263--274},
  \eprint{astro-ph/0006344},
  \urlprefix\url{http://www.slac.stanford.edu/spires/find/books/www?cl=QB461:I57:2000}.

\bibitem[{\citenamefont{Vogelsberger et~al.}(2012)\citenamefont{Vogelsberger,
  Zavala, and Loeb}}]{Vogelsberger:2012ku}
\bibinfo{author}{\bibfnamefont{M.}~\bibnamefont{Vogelsberger}},
  \bibinfo{author}{\bibfnamefont{J.}~\bibnamefont{Zavala}}, \bibnamefont{and}
  \bibinfo{author}{\bibfnamefont{A.}~\bibnamefont{Loeb}},
  \bibinfo{journal}{Mon. Not. Roy. Astron. Soc.}
  \textbf{\bibinfo{volume}{423}}, \bibinfo{pages}{3740} (\bibinfo{year}{2012}),
  \eprint{1201.5892}.

\bibitem[{\citenamefont{Rocha et~al.}(2013)\citenamefont{Rocha, Peter, Bullock,
  Kaplinghat, Garrison-Kimmel, Onorbe, and Moustakas}}]{Rocha:2012jg}
\bibinfo{author}{\bibfnamefont{M.}~\bibnamefont{Rocha}},
  \bibinfo{author}{\bibfnamefont{A.~H.~G.} \bibnamefont{Peter}},
  \bibinfo{author}{\bibfnamefont{J.~S.} \bibnamefont{Bullock}},
  \bibinfo{author}{\bibfnamefont{M.}~\bibnamefont{Kaplinghat}},
  \bibinfo{author}{\bibfnamefont{S.}~\bibnamefont{Garrison-Kimmel}},
  \bibinfo{author}{\bibfnamefont{J.}~\bibnamefont{Onorbe}}, \bibnamefont{and}
  \bibinfo{author}{\bibfnamefont{L.~A.} \bibnamefont{Moustakas}},
  \bibinfo{journal}{Mon. Not. Roy. Astron. Soc.}
  \textbf{\bibinfo{volume}{430}}, \bibinfo{pages}{81} (\bibinfo{year}{2013}),
  \eprint{1208.3025}.

\bibitem[{\citenamefont{Peter et~al.}(2013)\citenamefont{Peter, Rocha, Bullock,
  and Kaplinghat}}]{Peter:2012jh}
\bibinfo{author}{\bibfnamefont{A.~H.~G.} \bibnamefont{Peter}},
  \bibinfo{author}{\bibfnamefont{M.}~\bibnamefont{Rocha}},
  \bibinfo{author}{\bibfnamefont{J.~S.} \bibnamefont{Bullock}},
  \bibnamefont{and}
  \bibinfo{author}{\bibfnamefont{M.}~\bibnamefont{Kaplinghat}},
  \bibinfo{journal}{Mon. Not. Roy. Astron. Soc.}
  \textbf{\bibinfo{volume}{430}}, \bibinfo{pages}{105} (\bibinfo{year}{2013}),
  \eprint{1208.3026}.

\bibitem[{\citenamefont{Zavala et~al.}(2013)\citenamefont{Zavala, Vogelsberger,
  and Walker}}]{Zavala:2012us}
\bibinfo{author}{\bibfnamefont{J.}~\bibnamefont{Zavala}},
  \bibinfo{author}{\bibfnamefont{M.}~\bibnamefont{Vogelsberger}},
  \bibnamefont{and} \bibinfo{author}{\bibfnamefont{M.~G.}
  \bibnamefont{Walker}}, \bibinfo{journal}{Mon. Not. Roy. Astron. Soc.}
  \textbf{\bibinfo{volume}{431}}, \bibinfo{pages}{L20} (\bibinfo{year}{2013}),
  \eprint{1211.6426}.

\bibitem[{\citenamefont{Vogelsberger et~al.}(2014)\citenamefont{Vogelsberger,
  Zavala, Simpson, and Jenkins}}]{Vogelsberger:2014pda}
\bibinfo{author}{\bibfnamefont{M.}~\bibnamefont{Vogelsberger}},
  \bibinfo{author}{\bibfnamefont{J.}~\bibnamefont{Zavala}},
  \bibinfo{author}{\bibfnamefont{C.}~\bibnamefont{Simpson}}, \bibnamefont{and}
  \bibinfo{author}{\bibfnamefont{A.}~\bibnamefont{Jenkins}},
  \bibinfo{journal}{Mon. Not. Roy. Astron. Soc.}
  \textbf{\bibinfo{volume}{444}}, \bibinfo{pages}{3684} (\bibinfo{year}{2014}),
  \eprint{1405.5216}.

\bibitem[{\citenamefont{Elbert et~al.}(2015)\citenamefont{Elbert, Bullock,
  Garrison-Kimmel, Rocha, Oñorbe, and Peter}}]{Elbert:2014bma}
\bibinfo{author}{\bibfnamefont{O.~D.} \bibnamefont{Elbert}},
  \bibinfo{author}{\bibfnamefont{J.~S.} \bibnamefont{Bullock}},
  \bibinfo{author}{\bibfnamefont{S.}~\bibnamefont{Garrison-Kimmel}},
  \bibinfo{author}{\bibfnamefont{M.}~\bibnamefont{Rocha}},
  \bibinfo{author}{\bibfnamefont{J.}~\bibnamefont{Oñorbe}}, \bibnamefont{and}
  \bibinfo{author}{\bibfnamefont{A.~H.~G.} \bibnamefont{Peter}},
  \bibinfo{journal}{Mon. Not. Roy. Astron. Soc.}
  \textbf{\bibinfo{volume}{453}}, \bibinfo{pages}{29} (\bibinfo{year}{2015}),
  \eprint{1412.1477}.

\bibitem[{\citenamefont{Kaplinghat et~al.}(2016)\citenamefont{Kaplinghat,
  Tulin, and Yu}}]{Kaplinghat:2015aga}
\bibinfo{author}{\bibfnamefont{M.}~\bibnamefont{Kaplinghat}},
  \bibinfo{author}{\bibfnamefont{S.}~\bibnamefont{Tulin}}, \bibnamefont{and}
  \bibinfo{author}{\bibfnamefont{H.-B.} \bibnamefont{Yu}},
  \bibinfo{journal}{Phys. Rev. Lett.} \textbf{\bibinfo{volume}{116}},
  \bibinfo{pages}{041302} (\bibinfo{year}{2016}), \eprint{1508.03339}.

\bibitem[{\citenamefont{Kamada et~al.}(2017)\citenamefont{Kamada, Kaplinghat,
  Pace, and Yu}}]{Kamada:2016euw}
\bibinfo{author}{\bibfnamefont{A.}~\bibnamefont{Kamada}},
  \bibinfo{author}{\bibfnamefont{M.}~\bibnamefont{Kaplinghat}},
  \bibinfo{author}{\bibfnamefont{A.~B.} \bibnamefont{Pace}}, \bibnamefont{and}
  \bibinfo{author}{\bibfnamefont{H.-B.} \bibnamefont{Yu}},
  \bibinfo{journal}{Phys. Rev. Lett.} \textbf{\bibinfo{volume}{119}},
  \bibinfo{pages}{111102} (\bibinfo{year}{2017}), \eprint{1611.02716}.

\bibitem[{\citenamefont{Tulin and Yu}(2018)}]{Tulin:2017ara}
\bibinfo{author}{\bibfnamefont{S.}~\bibnamefont{Tulin}} \bibnamefont{and}
  \bibinfo{author}{\bibfnamefont{H.-B.} \bibnamefont{Yu}},
  \bibinfo{journal}{Phys. Rept.} \textbf{\bibinfo{volume}{730}},
  \bibinfo{pages}{1} (\bibinfo{year}{2018}), \eprint{1705.02358}.

\bibitem[{\citenamefont{Feng et~al.}(2010)\citenamefont{Feng, Kaplinghat, and
  Yu}}]{Feng:2009hw}
\bibinfo{author}{\bibfnamefont{J.~L.} \bibnamefont{Feng}},
  \bibinfo{author}{\bibfnamefont{M.}~\bibnamefont{Kaplinghat}},
  \bibnamefont{and} \bibinfo{author}{\bibfnamefont{H.-B.} \bibnamefont{Yu}},
  \bibinfo{journal}{Phys. Rev. Lett.} \textbf{\bibinfo{volume}{104}},
  \bibinfo{pages}{151301} (\bibinfo{year}{2010}), \eprint{0911.0422}.

\bibitem[{\citenamefont{Buckley and Fox}(2010)}]{Buckley:2009in}
\bibinfo{author}{\bibfnamefont{M.~R.} \bibnamefont{Buckley}} \bibnamefont{and}
  \bibinfo{author}{\bibfnamefont{P.~J.} \bibnamefont{Fox}},
  \bibinfo{journal}{Phys. Rev.} \textbf{\bibinfo{volume}{D81}},
  \bibinfo{pages}{083522} (\bibinfo{year}{2010}), \eprint{0911.3898}.

\bibitem[{\citenamefont{Loeb and Weiner}(2011)}]{Loeb:2010gj}
\bibinfo{author}{\bibfnamefont{A.}~\bibnamefont{Loeb}} \bibnamefont{and}
  \bibinfo{author}{\bibfnamefont{N.}~\bibnamefont{Weiner}},
  \bibinfo{journal}{Phys. Rev. Lett.} \textbf{\bibinfo{volume}{106}},
  \bibinfo{pages}{171302} (\bibinfo{year}{2011}), \eprint{1011.6374}.

\bibitem[{\citenamefont{Tulin et~al.}(2013)\citenamefont{Tulin, Yu, and
  Zurek}}]{Tulin:2013teo}
\bibinfo{author}{\bibfnamefont{S.}~\bibnamefont{Tulin}},
  \bibinfo{author}{\bibfnamefont{H.-B.} \bibnamefont{Yu}}, \bibnamefont{and}
  \bibinfo{author}{\bibfnamefont{K.~M.} \bibnamefont{Zurek}},
  \bibinfo{journal}{Phys. Rev.} \textbf{\bibinfo{volume}{D87}},
  \bibinfo{pages}{115007} (\bibinfo{year}{2013}), \eprint{1302.3898}.

\bibitem[{\citenamefont{Harvey et~al.}(2015)\citenamefont{Harvey, Massey,
  Kitching, Taylor, and Tittley}}]{Harvey:2015hha}
\bibinfo{author}{\bibfnamefont{D.}~\bibnamefont{Harvey}},
  \bibinfo{author}{\bibfnamefont{R.}~\bibnamefont{Massey}},
  \bibinfo{author}{\bibfnamefont{T.}~\bibnamefont{Kitching}},
  \bibinfo{author}{\bibfnamefont{A.}~\bibnamefont{Taylor}}, \bibnamefont{and}
  \bibinfo{author}{\bibfnamefont{E.}~\bibnamefont{Tittley}},
  \bibinfo{journal}{Science} \textbf{\bibinfo{volume}{347}},
  \bibinfo{pages}{1462} (\bibinfo{year}{2015}), \eprint{1503.07675}.

\bibitem[{\citenamefont{Bondarenko et~al.}(2018)\citenamefont{Bondarenko,
  Boyarsky, Bringmann, and Sokolenko}}]{Bondarenko_2018}
\bibinfo{author}{\bibfnamefont{K.}~\bibnamefont{Bondarenko}},
  \bibinfo{author}{\bibfnamefont{A.}~\bibnamefont{Boyarsky}},
  \bibinfo{author}{\bibfnamefont{T.}~\bibnamefont{Bringmann}},
  \bibnamefont{and}
  \bibinfo{author}{\bibfnamefont{A.}~\bibnamefont{Sokolenko}},
  \bibinfo{journal}{Journal of Cosmology and Astroparticle Physics}
  \textbf{\bibinfo{volume}{2018}}, \bibinfo{pages}{049} (\bibinfo{year}{2018}),
  \urlprefix\url{https://doi.org/10.1088%2F1475-7516%2F2018%2F04%2F049}.

\bibitem[{\citenamefont{Clowe et~al.}(2004)\citenamefont{Clowe, Gonzalez, and
  Markevitch}}]{Clowe:2003tk}
\bibinfo{author}{\bibfnamefont{D.}~\bibnamefont{Clowe}},
  \bibinfo{author}{\bibfnamefont{A.}~\bibnamefont{Gonzalez}}, \bibnamefont{and}
  \bibinfo{author}{\bibfnamefont{M.}~\bibnamefont{Markevitch}},
  \bibinfo{journal}{Astrophys. J.} \textbf{\bibinfo{volume}{604}},
  \bibinfo{pages}{596} (\bibinfo{year}{2004}), \eprint{astro-ph/0312273}.

\bibitem[{\citenamefont{Markevitch et~al.}(2004)\citenamefont{Markevitch,
  Gonzalez, Clowe, Vikhlinin, Forman, Jones, Murray, and
  Tucker}}]{Markevitch_2004}
\bibinfo{author}{\bibfnamefont{M.}~\bibnamefont{Markevitch}},
  \bibinfo{author}{\bibfnamefont{A.~H.} \bibnamefont{Gonzalez}},
  \bibinfo{author}{\bibfnamefont{D.}~\bibnamefont{Clowe}},
  \bibinfo{author}{\bibfnamefont{A.}~\bibnamefont{Vikhlinin}},
  \bibinfo{author}{\bibfnamefont{W.}~\bibnamefont{Forman}},
  \bibinfo{author}{\bibfnamefont{C.}~\bibnamefont{Jones}},
  \bibinfo{author}{\bibfnamefont{S.}~\bibnamefont{Murray}}, \bibnamefont{and}
  \bibinfo{author}{\bibfnamefont{W.}~\bibnamefont{Tucker}},
  \bibinfo{journal}{The Astrophysical Journal} \textbf{\bibinfo{volume}{606}},
  \bibinfo{pages}{819} (\bibinfo{year}{2004}),
  \urlprefix\url{https://doi.org/10.1086%2F383178}.

\bibitem[{\citenamefont{Randall et~al.}(2008)\citenamefont{Randall, Markevitch,
  Clowe, Gonzalez, and Brada{\v{c}}}}]{Randall_2008}
\bibinfo{author}{\bibfnamefont{S.~W.} \bibnamefont{Randall}},
  \bibinfo{author}{\bibfnamefont{M.}~\bibnamefont{Markevitch}},
  \bibinfo{author}{\bibfnamefont{D.}~\bibnamefont{Clowe}},
  \bibinfo{author}{\bibfnamefont{A.~H.} \bibnamefont{Gonzalez}},
  \bibnamefont{and}
  \bibinfo{author}{\bibfnamefont{M.}~\bibnamefont{Brada{\v{c}}}},
  \bibinfo{journal}{The Astrophysical Journal} \textbf{\bibinfo{volume}{679}},
  \bibinfo{pages}{1173} (\bibinfo{year}{2008}),
  \urlprefix\url{https://doi.org/10.1086%2F587859}.

\bibitem[{\citenamefont{Kaplinghat et~al.}(2014)\citenamefont{Kaplinghat,
  Tulin, and Yu}}]{PhysRevD.89.035009}
\bibinfo{author}{\bibfnamefont{M.}~\bibnamefont{Kaplinghat}},
  \bibinfo{author}{\bibfnamefont{S.}~\bibnamefont{Tulin}}, \bibnamefont{and}
  \bibinfo{author}{\bibfnamefont{H.-B.} \bibnamefont{Yu}},
  \bibinfo{journal}{Phys. Rev. D} \textbf{\bibinfo{volume}{89}},
  \bibinfo{pages}{035009} (\bibinfo{year}{2014}),
  \urlprefix\url{https://link.aps.org/doi/10.1103/PhysRevD.89.035009}.

\bibitem[{\citenamefont{Nobile et~al.}(2015)\citenamefont{Nobile, Kaplinghat,
  and Yu}}]{Nobile_2015}
\bibinfo{author}{\bibfnamefont{E.~D.} \bibnamefont{Nobile}},
  \bibinfo{author}{\bibfnamefont{M.}~\bibnamefont{Kaplinghat}},
  \bibnamefont{and} \bibinfo{author}{\bibfnamefont{H.-B.} \bibnamefont{Yu}},
  \bibinfo{journal}{Journal of Cosmology and Astroparticle Physics}
  \textbf{\bibinfo{volume}{2015}}, \bibinfo{pages}{055} (\bibinfo{year}{2015}),
  \urlprefix\url{https://doi.org/10.1088%2F1475-7516%2F2015%2F10%2F055}.

\bibitem[{\citenamefont{Bernal et~al.}(2016)\citenamefont{Bernal, Chu,
  Garcia-Cely, Hambye, and Zaldivar}}]{Bernal:2015ova}
\bibinfo{author}{\bibfnamefont{N.}~\bibnamefont{Bernal}},
  \bibinfo{author}{\bibfnamefont{X.}~\bibnamefont{Chu}},
  \bibinfo{author}{\bibfnamefont{C.}~\bibnamefont{Garcia-Cely}},
  \bibinfo{author}{\bibfnamefont{T.}~\bibnamefont{Hambye}}, \bibnamefont{and}
  \bibinfo{author}{\bibfnamefont{B.}~\bibnamefont{Zaldivar}},
  \bibinfo{journal}{JCAP} \textbf{\bibinfo{volume}{1603}}, \bibinfo{pages}{018}
  (\bibinfo{year}{2016}), \eprint{1510.08063}.

\bibitem[{\citenamefont{Bringmann et~al.}(2017)\citenamefont{Bringmann,
  Kahlhoefer, Schmidt-Hoberg, and Walia}}]{Bringmann:2016din}
\bibinfo{author}{\bibfnamefont{T.}~\bibnamefont{Bringmann}},
  \bibinfo{author}{\bibfnamefont{F.}~\bibnamefont{Kahlhoefer}},
  \bibinfo{author}{\bibfnamefont{K.}~\bibnamefont{Schmidt-Hoberg}},
  \bibnamefont{and} \bibinfo{author}{\bibfnamefont{P.}~\bibnamefont{Walia}},
  \bibinfo{journal}{Phys. Rev. Lett.} \textbf{\bibinfo{volume}{118}},
  \bibinfo{pages}{141802} (\bibinfo{year}{2017}), \eprint{1612.00845}.

\bibitem[{\citenamefont{Cirelli et~al.}(2017)\citenamefont{Cirelli, Panci,
  Petraki, Sala, and Taoso}}]{Cirelli:2016rnw}
\bibinfo{author}{\bibfnamefont{M.}~\bibnamefont{Cirelli}},
  \bibinfo{author}{\bibfnamefont{P.}~\bibnamefont{Panci}},
  \bibinfo{author}{\bibfnamefont{K.}~\bibnamefont{Petraki}},
  \bibinfo{author}{\bibfnamefont{F.}~\bibnamefont{Sala}}, \bibnamefont{and}
  \bibinfo{author}{\bibfnamefont{M.}~\bibnamefont{Taoso}},
  \bibinfo{journal}{JCAP} \textbf{\bibinfo{volume}{1705}}, \bibinfo{pages}{036}
  (\bibinfo{year}{2017}), \eprint{1612.07295}.

\bibitem[{\citenamefont{Kahlhoefer et~al.}(2017)\citenamefont{Kahlhoefer,
  Schmidt-Hoberg, and Wild}}]{Kahlhoefer:2017umn}
\bibinfo{author}{\bibfnamefont{F.}~\bibnamefont{Kahlhoefer}},
  \bibinfo{author}{\bibfnamefont{K.}~\bibnamefont{Schmidt-Hoberg}},
  \bibnamefont{and} \bibinfo{author}{\bibfnamefont{S.}~\bibnamefont{Wild}},
  \bibinfo{journal}{JCAP} \textbf{\bibinfo{volume}{1708}}, \bibinfo{pages}{003}
  (\bibinfo{year}{2017}), \eprint{1704.02149}.

\bibitem[{\citenamefont{Hufnagel
  et~al.}(2018{\natexlab{a}})\citenamefont{Hufnagel, Schmidt-Hoberg, and
  Wild}}]{Hufnagel:2017dgo}
\bibinfo{author}{\bibfnamefont{M.}~\bibnamefont{Hufnagel}},
  \bibinfo{author}{\bibfnamefont{K.}~\bibnamefont{Schmidt-Hoberg}},
  \bibnamefont{and} \bibinfo{author}{\bibfnamefont{S.}~\bibnamefont{Wild}},
  \bibinfo{journal}{JCAP} \textbf{\bibinfo{volume}{1802}}, \bibinfo{pages}{044}
  (\bibinfo{year}{2018}{\natexlab{a}}), \eprint{1712.03972}.

\bibitem[{\citenamefont{Hufnagel
  et~al.}(2018{\natexlab{b}})\citenamefont{Hufnagel, Schmidt-Hoberg, and
  Wild}}]{Hufnagel:2018bjp}
\bibinfo{author}{\bibfnamefont{M.}~\bibnamefont{Hufnagel}},
  \bibinfo{author}{\bibfnamefont{K.}~\bibnamefont{Schmidt-Hoberg}},
  \bibnamefont{and} \bibinfo{author}{\bibfnamefont{S.}~\bibnamefont{Wild}},
  \bibinfo{journal}{JCAP} \textbf{\bibinfo{volume}{1811}}, \bibinfo{pages}{032}
  (\bibinfo{year}{2018}{\natexlab{b}}), \eprint{1808.09324}.

\bibitem[{\citenamefont{Bernal et~al.}(2019)\citenamefont{Bernal, Chu,
  Kulkarni, and Pradler}}]{Bernal:2019uqr}
\bibinfo{author}{\bibfnamefont{N.}~\bibnamefont{Bernal}},
  \bibinfo{author}{\bibfnamefont{X.}~\bibnamefont{Chu}},
  \bibinfo{author}{\bibfnamefont{S.}~\bibnamefont{Kulkarni}}, \bibnamefont{and}
  \bibinfo{author}{\bibfnamefont{J.}~\bibnamefont{Pradler}}
  (\bibinfo{year}{2019}), \eprint{1912.06681}.

\bibitem[{\citenamefont{Hochberg et~al.}(2014)\citenamefont{Hochberg, Kuflik,
  Volansky, and Wacker}}]{Hochberg:2014dra}
\bibinfo{author}{\bibfnamefont{Y.}~\bibnamefont{Hochberg}},
  \bibinfo{author}{\bibfnamefont{E.}~\bibnamefont{Kuflik}},
  \bibinfo{author}{\bibfnamefont{T.}~\bibnamefont{Volansky}}, \bibnamefont{and}
  \bibinfo{author}{\bibfnamefont{J.~G.} \bibnamefont{Wacker}},
  \bibinfo{journal}{Phys. Rev. Lett.} \textbf{\bibinfo{volume}{113}},
  \bibinfo{pages}{171301} (\bibinfo{year}{2014}), \eprint{1402.5143}.

\bibitem[{\citenamefont{{Carlson} et~al.}(1992)\citenamefont{{Carlson},
  {Machacek}, and {Hall}}}]{1992ApJ...398...43C}
\bibinfo{author}{\bibfnamefont{E.~D.} \bibnamefont{{Carlson}}},
  \bibinfo{author}{\bibfnamefont{M.~E.} \bibnamefont{{Machacek}}},
  \bibnamefont{and} \bibinfo{author}{\bibfnamefont{L.~J.}
  \bibnamefont{{Hall}}}, \bibinfo{journal}{\apj}
  \textbf{\bibinfo{volume}{398}}, \bibinfo{pages}{43} (\bibinfo{year}{1992}).

\bibitem[{\citenamefont{Farina et~al.}(2016)\citenamefont{Farina, Pappadopulo,
  Ruderman, and Trevisan}}]{Farina:2016llk}
\bibinfo{author}{\bibfnamefont{M.}~\bibnamefont{Farina}},
  \bibinfo{author}{\bibfnamefont{D.}~\bibnamefont{Pappadopulo}},
  \bibinfo{author}{\bibfnamefont{J.~T.} \bibnamefont{Ruderman}},
  \bibnamefont{and} \bibinfo{author}{\bibfnamefont{G.}~\bibnamefont{Trevisan}},
  \bibinfo{journal}{JHEP} \textbf{\bibinfo{volume}{12}}, \bibinfo{pages}{039}
  (\bibinfo{year}{2016}), \eprint{1607.03108}.

\bibitem[{\citenamefont{Pappadopulo et~al.}(2016)\citenamefont{Pappadopulo,
  Ruderman, and Trevisan}}]{Pappadopulo:2016pkp}
\bibinfo{author}{\bibfnamefont{D.}~\bibnamefont{Pappadopulo}},
  \bibinfo{author}{\bibfnamefont{J.~T.} \bibnamefont{Ruderman}},
  \bibnamefont{and} \bibinfo{author}{\bibfnamefont{G.}~\bibnamefont{Trevisan}},
  \bibinfo{journal}{Phys. Rev.} \textbf{\bibinfo{volume}{D94}},
  \bibinfo{pages}{035005} (\bibinfo{year}{2016}), \eprint{1602.04219}.

\bibitem[{\citenamefont{Hambye et~al.}(2018)\citenamefont{Hambye, Tytgat,
  Vandecasteele, and Vanderheyden}}]{Hambye:2018dpi}
\bibinfo{author}{\bibfnamefont{T.}~\bibnamefont{Hambye}},
  \bibinfo{author}{\bibfnamefont{M.~H.~G.} \bibnamefont{Tytgat}},
  \bibinfo{author}{\bibfnamefont{J.}~\bibnamefont{Vandecasteele}},
  \bibnamefont{and}
  \bibinfo{author}{\bibfnamefont{L.}~\bibnamefont{Vanderheyden}},
  \bibinfo{journal}{Phys. Rev.} \textbf{\bibinfo{volume}{D98}},
  \bibinfo{pages}{075017} (\bibinfo{year}{2018}), \eprint{1807.05022}.

\bibitem[{\citenamefont{Chu et~al.}(2012)\citenamefont{Chu, Hambye, and
  Tytgat}}]{Chu:2011be}
\bibinfo{author}{\bibfnamefont{X.}~\bibnamefont{Chu}},
  \bibinfo{author}{\bibfnamefont{T.}~\bibnamefont{Hambye}}, \bibnamefont{and}
  \bibinfo{author}{\bibfnamefont{M.~H.~G.} \bibnamefont{Tytgat}},
  \bibinfo{journal}{JCAP} \textbf{\bibinfo{volume}{1205}}, \bibinfo{pages}{034}
  (\bibinfo{year}{2012}), \eprint{1112.0493}.

\bibitem[{\citenamefont{Slatyer}(2016)}]{Slatyer:2015jla}
\bibinfo{author}{\bibfnamefont{T.~R.} \bibnamefont{Slatyer}},
  \bibinfo{journal}{Phys. Rev.} \textbf{\bibinfo{volume}{D93}},
  \bibinfo{pages}{023527} (\bibinfo{year}{2016}), \eprint{1506.03811}.

\bibitem[{\citenamefont{Pospelov et~al.}(2008)\citenamefont{Pospelov, Ritz, and
  Voloshin}}]{Pospelov:2008jk}
\bibinfo{author}{\bibfnamefont{M.}~\bibnamefont{Pospelov}},
  \bibinfo{author}{\bibfnamefont{A.}~\bibnamefont{Ritz}}, \bibnamefont{and}
  \bibinfo{author}{\bibfnamefont{M.~B.} \bibnamefont{Voloshin}},
  \bibinfo{journal}{Phys. Rev.} \textbf{\bibinfo{volume}{D78}},
  \bibinfo{pages}{115012} (\bibinfo{year}{2008}), \eprint{0807.3279}.

\bibitem[{\citenamefont{Slatyer}(2013)}]{Slatyer:2012yq}
\bibinfo{author}{\bibfnamefont{T.~R.} \bibnamefont{Slatyer}},
  \bibinfo{journal}{Phys. Rev.} \textbf{\bibinfo{volume}{D87}},
  \bibinfo{pages}{123513} (\bibinfo{year}{2013}), \eprint{1211.0283}.

\bibitem[{\citenamefont{Poulin et~al.}(2017)\citenamefont{Poulin, Lesgourgues,
  and Serpico}}]{Poulin:2016anj}
\bibinfo{author}{\bibfnamefont{V.}~\bibnamefont{Poulin}},
  \bibinfo{author}{\bibfnamefont{J.}~\bibnamefont{Lesgourgues}},
  \bibnamefont{and} \bibinfo{author}{\bibfnamefont{P.~D.}
  \bibnamefont{Serpico}}, \bibinfo{journal}{JCAP}
  \textbf{\bibinfo{volume}{1703}}, \bibinfo{pages}{043} (\bibinfo{year}{2017}),
  \eprint{1610.10051}.

\bibitem[{\citenamefont{Essig et~al.}(2013)\citenamefont{Essig, Kuflik,
  McDermott, Volansky, and Zurek}}]{Essig:2013goa}
\bibinfo{author}{\bibfnamefont{R.}~\bibnamefont{Essig}},
  \bibinfo{author}{\bibfnamefont{E.}~\bibnamefont{Kuflik}},
  \bibinfo{author}{\bibfnamefont{S.~D.} \bibnamefont{McDermott}},
  \bibinfo{author}{\bibfnamefont{T.}~\bibnamefont{Volansky}}, \bibnamefont{and}
  \bibinfo{author}{\bibfnamefont{K.~M.} \bibnamefont{Zurek}},
  \bibinfo{journal}{JHEP} \textbf{\bibinfo{volume}{11}}, \bibinfo{pages}{193}
  (\bibinfo{year}{2013}), \eprint{1309.4091}.

\bibitem[{\citenamefont{Boddy and Kumar}(2015)}]{Boddy:2015efa}
\bibinfo{author}{\bibfnamefont{K.~K.} \bibnamefont{Boddy}} \bibnamefont{and}
  \bibinfo{author}{\bibfnamefont{J.}~\bibnamefont{Kumar}},
  \bibinfo{journal}{Phys. Rev.} \textbf{\bibinfo{volume}{D92}},
  \bibinfo{pages}{023533} (\bibinfo{year}{2015}), \eprint{1504.04024}.

\bibitem[{\citenamefont{Riemer-Sørensen
  et~al.}(2015)}]{Riemer-Sorensen:2015kqa}
\bibinfo{author}{\bibfnamefont{S.}~\bibnamefont{Riemer-Sørensen}}
  \bibnamefont{et~al.}, \bibinfo{journal}{Astrophys. J.}
  \textbf{\bibinfo{volume}{810}}, \bibinfo{pages}{48} (\bibinfo{year}{2015}),
  \eprint{1507.01378}.

\bibitem[{\citenamefont{Aghanim et~al.}(2018)}]{Aghanim:2018eyx}
\bibinfo{author}{\bibfnamefont{N.}~\bibnamefont{Aghanim}} \bibnamefont{et~al.}
  (\bibinfo{collaboration}{Planck}) (\bibinfo{year}{2018}),
  \eprint{1807.06209}.

\bibitem[{\citenamefont{Jedamzik}(2006)}]{Jedamzik:2006xz}
\bibinfo{author}{\bibfnamefont{K.}~\bibnamefont{Jedamzik}},
  \bibinfo{journal}{Phys. Rev.} \textbf{\bibinfo{volume}{D74}},
  \bibinfo{pages}{103509} (\bibinfo{year}{2006}), \eprint{hep-ph/0604251}.

\bibitem[{\citenamefont{Forestell et~al.}(2019)\citenamefont{Forestell,
  Morrissey, and White}}]{Forestell:2018txr}
\bibinfo{author}{\bibfnamefont{L.}~\bibnamefont{Forestell}},
  \bibinfo{author}{\bibfnamefont{D.~E.} \bibnamefont{Morrissey}},
  \bibnamefont{and} \bibinfo{author}{\bibfnamefont{G.}~\bibnamefont{White}},
  \bibinfo{journal}{JHEP} \textbf{\bibinfo{volume}{01}}, \bibinfo{pages}{074}
  (\bibinfo{year}{2019}), \eprint{1809.01179}.

\bibitem[{\citenamefont{Scherrer and Turner}(1988)}]{Scherrer:1987rr}
\bibinfo{author}{\bibfnamefont{R.~J.} \bibnamefont{Scherrer}} \bibnamefont{and}
  \bibinfo{author}{\bibfnamefont{M.~S.} \bibnamefont{Turner}},
  \bibinfo{journal}{Astrophys. J.} \textbf{\bibinfo{volume}{331}},
  \bibinfo{pages}{19} (\bibinfo{year}{1988}), \bibinfo{note}{[Astrophys.
  J.331,33(1988)]}.

\bibitem[{\citenamefont{Menestrina and Scherrer}(2012)}]{Menestrina:2011mz}
\bibinfo{author}{\bibfnamefont{J.~L.} \bibnamefont{Menestrina}}
  \bibnamefont{and} \bibinfo{author}{\bibfnamefont{R.~J.}
  \bibnamefont{Scherrer}}, \bibinfo{journal}{Phys. Rev.}
  \textbf{\bibinfo{volume}{D85}}, \bibinfo{pages}{047301}
  (\bibinfo{year}{2012}), \eprint{1111.0605}.

\bibitem[{\citenamefont{Duerr et~al.}(2018)\citenamefont{Duerr, Schmidt-Hoberg,
  and Wild}}]{Duerr:2018mbd}
\bibinfo{author}{\bibfnamefont{M.}~\bibnamefont{Duerr}},
  \bibinfo{author}{\bibfnamefont{K.}~\bibnamefont{Schmidt-Hoberg}},
  \bibnamefont{and} \bibinfo{author}{\bibfnamefont{S.}~\bibnamefont{Wild}},
  \bibinfo{journal}{JCAP} \textbf{\bibinfo{volume}{1809}}, \bibinfo{pages}{033}
  (\bibinfo{year}{2018}), \eprint{1804.10385}.

\bibitem[{\citenamefont{Patrignani et~al.}(2016)}]{Patrignani:2016xqp}
\bibinfo{author}{\bibfnamefont{C.}~\bibnamefont{Patrignani}}
  \bibnamefont{et~al.} (\bibinfo{collaboration}{Particle Data Group}),
  \bibinfo{journal}{Chin.\ Phys.\ C} \textbf{\bibinfo{volume}{40}},
  \bibinfo{pages}{100001} (\bibinfo{year}{2016}).

\bibitem[{\citenamefont{Kazanas et~al.}(2014)\citenamefont{Kazanas, Mohapatra,
  Nussinov, Teplitz, and Zhang}}]{Kazanas:2014mca}
\bibinfo{author}{\bibfnamefont{D.}~\bibnamefont{Kazanas}},
  \bibinfo{author}{\bibfnamefont{R.~N.} \bibnamefont{Mohapatra}},
  \bibinfo{author}{\bibfnamefont{S.}~\bibnamefont{Nussinov}},
  \bibinfo{author}{\bibfnamefont{V.~L.} \bibnamefont{Teplitz}},
  \bibnamefont{and} \bibinfo{author}{\bibfnamefont{Y.}~\bibnamefont{Zhang}},
  \bibinfo{journal}{Nucl. Phys.} \textbf{\bibinfo{volume}{B890}},
  \bibinfo{pages}{17} (\bibinfo{year}{2014}), \eprint{1410.0221}.

\bibitem[{\citenamefont{Chang et~al.}(2017)\citenamefont{Chang, Essig, and
  McDermott}}]{Chang:2016ntp}
\bibinfo{author}{\bibfnamefont{J.~H.} \bibnamefont{Chang}},
  \bibinfo{author}{\bibfnamefont{R.}~\bibnamefont{Essig}}, \bibnamefont{and}
  \bibinfo{author}{\bibfnamefont{S.~D.} \bibnamefont{McDermott}},
  \bibinfo{journal}{JHEP} \textbf{\bibinfo{volume}{01}}, \bibinfo{pages}{107}
  (\bibinfo{year}{2017}), \eprint{1611.03864}.

\bibitem[{\citenamefont{Mahoney et~al.}(2017)\citenamefont{Mahoney, Leibovich,
  and Zentner}}]{Mahoney:2017jqk}
\bibinfo{author}{\bibfnamefont{C.}~\bibnamefont{Mahoney}},
  \bibinfo{author}{\bibfnamefont{A.~K.} \bibnamefont{Leibovich}},
  \bibnamefont{and} \bibinfo{author}{\bibfnamefont{A.~R.}
  \bibnamefont{Zentner}}, \bibinfo{journal}{Phys. Rev.}
  \textbf{\bibinfo{volume}{D96}}, \bibinfo{pages}{043018}
  (\bibinfo{year}{2017}), \eprint{1706.08871}.

\bibitem[{\citenamefont{Chang et~al.}(2018{\natexlab{a}})\citenamefont{Chang,
  Essig, and McDermott}}]{Chang:2018rso}
\bibinfo{author}{\bibfnamefont{J.~H.} \bibnamefont{Chang}},
  \bibinfo{author}{\bibfnamefont{R.}~\bibnamefont{Essig}}, \bibnamefont{and}
  \bibinfo{author}{\bibfnamefont{S.~D.} \bibnamefont{McDermott}},
  \bibinfo{journal}{JHEP} \textbf{\bibinfo{volume}{09}}, \bibinfo{pages}{051}
  (\bibinfo{year}{2018}{\natexlab{a}}), \eprint{1803.00993}.

\bibitem[{\citenamefont{Cline et~al.}(2013)\citenamefont{Cline, Kainulainen,
  Scott, and Weniger}}]{Cline:2013gha}
\bibinfo{author}{\bibfnamefont{J.~M.} \bibnamefont{Cline}},
  \bibinfo{author}{\bibfnamefont{K.}~\bibnamefont{Kainulainen}},
  \bibinfo{author}{\bibfnamefont{P.}~\bibnamefont{Scott}}, \bibnamefont{and}
  \bibinfo{author}{\bibfnamefont{C.}~\bibnamefont{Weniger}},
  \bibinfo{journal}{Phys. Rev.} \textbf{\bibinfo{volume}{D88}},
  \bibinfo{pages}{055025} (\bibinfo{year}{2013}), \bibinfo{note}{[Erratum:
  Phys. Rev.D92,no.3,039906(2015)]}, \eprint{1306.4710}.

\bibitem[{\citenamefont{Krnjaic}(2016)}]{Krnjaic:2015mbs}
\bibinfo{author}{\bibfnamefont{G.}~\bibnamefont{Krnjaic}},
  \bibinfo{journal}{Phys. Rev.} \textbf{\bibinfo{volume}{D94}},
  \bibinfo{pages}{073009} (\bibinfo{year}{2016}), \eprint{1512.04119}.

\bibitem[{\citenamefont{Chang et~al.}(2018{\natexlab{b}})\citenamefont{Chang,
  Modak, and Ng}}]{Chang:2017ynj}
\bibinfo{author}{\bibfnamefont{W.-F.} \bibnamefont{Chang}},
  \bibinfo{author}{\bibfnamefont{T.}~\bibnamefont{Modak}}, \bibnamefont{and}
  \bibinfo{author}{\bibfnamefont{J.~N.} \bibnamefont{Ng}},
  \bibinfo{journal}{Phys. Rev.} \textbf{\bibinfo{volume}{D97}},
  \bibinfo{pages}{055020} (\bibinfo{year}{2018}{\natexlab{b}}),
  \eprint{1711.05722}.

\bibitem[{\citenamefont{Aprile et~al.}(2018)}]{Aprile:2018dbl}
\bibinfo{author}{\bibfnamefont{E.}~\bibnamefont{Aprile}} \bibnamefont{et~al.}
  (\bibinfo{collaboration}{XENON}), \bibinfo{journal}{Phys. Rev. Lett.}
  \textbf{\bibinfo{volume}{121}}, \bibinfo{pages}{111302}
  (\bibinfo{year}{2018}), \eprint{1805.12562}.

\bibitem[{\citenamefont{Agrawal et~al.}(2020)\citenamefont{Agrawal, Parikh, and
  Reece}}]{Agrawal:2020lea}
\bibinfo{author}{\bibfnamefont{P.}~\bibnamefont{Agrawal}},
  \bibinfo{author}{\bibfnamefont{A.}~\bibnamefont{Parikh}}, \bibnamefont{and}
  \bibinfo{author}{\bibfnamefont{M.}~\bibnamefont{Reece}}
  (\bibinfo{year}{2020}), \eprint{2003.00021}.

\bibitem[{\citenamefont{De~Simone et~al.}(2014)\citenamefont{De~Simone,
  Giudice, and Strumia}}]{deSimone:2014pda}
\bibinfo{author}{\bibfnamefont{A.}~\bibnamefont{De~Simone}},
  \bibinfo{author}{\bibfnamefont{G.~F.} \bibnamefont{Giudice}},
  \bibnamefont{and} \bibinfo{author}{\bibfnamefont{A.}~\bibnamefont{Strumia}},
  \bibinfo{journal}{JHEP} \textbf{\bibinfo{volume}{06}}, \bibinfo{pages}{081}
  (\bibinfo{year}{2014}), \eprint{1402.6287}.

\bibitem[{\citenamefont{Chu et~al.}(2016)\citenamefont{Chu, Garcia-Cely, and
  Hambye}}]{Chu:2016pew}
\bibinfo{author}{\bibfnamefont{X.}~\bibnamefont{Chu}},
  \bibinfo{author}{\bibfnamefont{C.}~\bibnamefont{Garcia-Cely}},
  \bibnamefont{and} \bibinfo{author}{\bibfnamefont{T.}~\bibnamefont{Hambye}},
  \bibinfo{journal}{JHEP} \textbf{\bibinfo{volume}{11}}, \bibinfo{pages}{048}
  (\bibinfo{year}{2016}), \eprint{1609.00399}.

\bibitem[{\citenamefont{Blennow et~al.}(2017)\citenamefont{Blennow, Clementz,
  and Herrero-Garcia}}]{Blennow:2016gde}
\bibinfo{author}{\bibfnamefont{M.}~\bibnamefont{Blennow}},
  \bibinfo{author}{\bibfnamefont{S.}~\bibnamefont{Clementz}}, \bibnamefont{and}
  \bibinfo{author}{\bibfnamefont{J.}~\bibnamefont{Herrero-Garcia}},
  \bibinfo{journal}{JCAP} \textbf{\bibinfo{volume}{1703}}, \bibinfo{pages}{048}
  (\bibinfo{year}{2017}), \eprint{1612.06681}.

\bibitem[{\citenamefont{Das and Dasgupta}(2017)}]{Das:2016ced}
\bibinfo{author}{\bibfnamefont{A.}~\bibnamefont{Das}} \bibnamefont{and}
  \bibinfo{author}{\bibfnamefont{B.}~\bibnamefont{Dasgupta}},
  \bibinfo{journal}{Phys. Rev. Lett.} \textbf{\bibinfo{volume}{118}},
  \bibinfo{pages}{251101} (\bibinfo{year}{2017}), \eprint{1611.04606}.

\bibitem[{\citenamefont{Ackermann et~al.}(2012)\citenamefont{Ackermann, Ajello,
  Albert, Baldini, Barbiellini, Bechtol, Bellazzini, Berenji, Blandford, Bloom
  et~al.}}]{Ackermann_2012}
\bibinfo{author}{\bibfnamefont{M.}~\bibnamefont{Ackermann}},
  \bibinfo{author}{\bibfnamefont{M.}~\bibnamefont{Ajello}},
  \bibinfo{author}{\bibfnamefont{A.}~\bibnamefont{Albert}},
  \bibinfo{author}{\bibfnamefont{L.}~\bibnamefont{Baldini}},
  \bibinfo{author}{\bibfnamefont{G.}~\bibnamefont{Barbiellini}},
  \bibinfo{author}{\bibfnamefont{K.}~\bibnamefont{Bechtol}},
  \bibinfo{author}{\bibfnamefont{R.}~\bibnamefont{Bellazzini}},
  \bibinfo{author}{\bibfnamefont{B.}~\bibnamefont{Berenji}},
  \bibinfo{author}{\bibfnamefont{R.~D.} \bibnamefont{Blandford}},
  \bibinfo{author}{\bibfnamefont{E.~D.} \bibnamefont{Bloom}},
  \bibnamefont{et~al.}, \bibinfo{journal}{Physical Review D}
  \textbf{\bibinfo{volume}{86}} (\bibinfo{year}{2012}), ISSN
  \bibinfo{issn}{1550-2368},
  \urlprefix\url{http://dx.doi.org/10.1103/PhysRevD.86.022002}.

\bibitem[{\citenamefont{Abramowski et~al.}(2013)\citenamefont{Abramowski,
  Acero, Aharonian, Akhperjanian, Anton, Balenderan, Balzer, Barnacka,
  Becherini, Becker~Tjus et~al.}}]{Abramowski_2013}
\bibinfo{author}{\bibfnamefont{A.}~\bibnamefont{Abramowski}},
  \bibinfo{author}{\bibfnamefont{F.}~\bibnamefont{Acero}},
  \bibinfo{author}{\bibfnamefont{F.}~\bibnamefont{Aharonian}},
  \bibinfo{author}{\bibfnamefont{A.~G.} \bibnamefont{Akhperjanian}},
  \bibinfo{author}{\bibfnamefont{G.}~\bibnamefont{Anton}},
  \bibinfo{author}{\bibfnamefont{S.}~\bibnamefont{Balenderan}},
  \bibinfo{author}{\bibfnamefont{A.}~\bibnamefont{Balzer}},
  \bibinfo{author}{\bibfnamefont{A.}~\bibnamefont{Barnacka}},
  \bibinfo{author}{\bibfnamefont{Y.}~\bibnamefont{Becherini}},
  \bibinfo{author}{\bibfnamefont{J.}~\bibnamefont{Becker~Tjus}},
  \bibnamefont{et~al.}, \bibinfo{journal}{Physical Review Letters}
  \textbf{\bibinfo{volume}{110}} (\bibinfo{year}{2013}), ISSN
  \bibinfo{issn}{1079-7114},
  \urlprefix\url{http://dx.doi.org/10.1103/PhysRevLett.110.041301}.

\bibitem[{\citenamefont{Ackermann et~al.}(2015)\citenamefont{Ackermann, Ajello,
  Albert, Anderson, Atwood, Baldini, Barbiellini, Bastieri, Bellazzini,
  Bissaldi et~al.}}]{Ackermann_2015}
\bibinfo{author}{\bibfnamefont{M.}~\bibnamefont{Ackermann}},
  \bibinfo{author}{\bibfnamefont{M.}~\bibnamefont{Ajello}},
  \bibinfo{author}{\bibfnamefont{A.}~\bibnamefont{Albert}},
  \bibinfo{author}{\bibfnamefont{B.}~\bibnamefont{Anderson}},
  \bibinfo{author}{\bibfnamefont{W.}~\bibnamefont{Atwood}},
  \bibinfo{author}{\bibfnamefont{L.}~\bibnamefont{Baldini}},
  \bibinfo{author}{\bibfnamefont{G.}~\bibnamefont{Barbiellini}},
  \bibinfo{author}{\bibfnamefont{D.}~\bibnamefont{Bastieri}},
  \bibinfo{author}{\bibfnamefont{R.}~\bibnamefont{Bellazzini}},
  \bibinfo{author}{\bibfnamefont{E.}~\bibnamefont{Bissaldi}},
  \bibnamefont{et~al.}, \bibinfo{journal}{Physical Review D}
  \textbf{\bibinfo{volume}{91}} (\bibinfo{year}{2015}), ISSN
  \bibinfo{issn}{1550-2368},
  \urlprefix\url{http://dx.doi.org/10.1103/PhysRevD.91.122002}.

\bibitem[{\citenamefont{Abdallah et~al.}(2016)\citenamefont{Abdallah,
  Abramowski, Aharonian, Ait~Benkhali, Akhperjanian, Angüner, Arrieta, Aubert,
  Backes, Balzer et~al.}}]{Abdallah_2016}
\bibinfo{author}{\bibfnamefont{H.}~\bibnamefont{Abdallah}},
  \bibinfo{author}{\bibfnamefont{A.}~\bibnamefont{Abramowski}},
  \bibinfo{author}{\bibfnamefont{F.}~\bibnamefont{Aharonian}},
  \bibinfo{author}{\bibfnamefont{F.}~\bibnamefont{Ait~Benkhali}},
  \bibinfo{author}{\bibfnamefont{A.}~\bibnamefont{Akhperjanian}},
  \bibinfo{author}{\bibfnamefont{E.}~\bibnamefont{Angüner}},
  \bibinfo{author}{\bibfnamefont{M.}~\bibnamefont{Arrieta}},
  \bibinfo{author}{\bibfnamefont{P.}~\bibnamefont{Aubert}},
  \bibinfo{author}{\bibfnamefont{M.}~\bibnamefont{Backes}},
  \bibinfo{author}{\bibfnamefont{A.}~\bibnamefont{Balzer}},
  \bibnamefont{et~al.}, \bibinfo{journal}{Physical Review Letters}
  \textbf{\bibinfo{volume}{117}} (\bibinfo{year}{2016}), ISSN
  \bibinfo{issn}{1079-7114},
  \urlprefix\url{http://dx.doi.org/10.1103/PhysRevLett.117.111301}.

\bibitem[{\citenamefont{Albert et~al.}(2017)}]{Albert:2016emp}
\bibinfo{author}{\bibfnamefont{A.}~\bibnamefont{Albert}} \bibnamefont{et~al.},
  \bibinfo{journal}{Phys. Lett.} \textbf{\bibinfo{volume}{B769}},
  \bibinfo{pages}{249} (\bibinfo{year}{2017}), \bibinfo{note}{[Erratum: Phys.
  Lett.B796,253(2019)]}, \eprint{1612.04595}.

\bibitem[{201(2016)}]{2016}
\bibinfo{journal}{Journal of Cosmology and Astroparticle Physics}
  \textbf{\bibinfo{volume}{2016}}, \bibinfo{pages}{039–039}
  (\bibinfo{year}{2016}), ISSN \bibinfo{issn}{1475-7516},
  \urlprefix\url{http://dx.doi.org/10.1088/1475-7516/2016/02/039}.

\bibitem[{\citenamefont{Sirunyan et~al.}(2019)}]{Sirunyan:2018owy}
\bibinfo{author}{\bibfnamefont{A.~M.} \bibnamefont{Sirunyan}}
  \bibnamefont{et~al.} (\bibinfo{collaboration}{CMS}), \bibinfo{journal}{Phys.
  Lett.} \textbf{\bibinfo{volume}{B793}}, \bibinfo{pages}{520}
  (\bibinfo{year}{2019}), \eprint{1809.05937}.

\bibitem[{\citenamefont{Aad et~al.}(2016)}]{Khachatryan:2016vau}
\bibinfo{author}{\bibfnamefont{G.}~\bibnamefont{Aad}} \bibnamefont{et~al.}
  (\bibinfo{collaboration}{ATLAS, CMS}), \bibinfo{journal}{JHEP}
  \textbf{\bibinfo{volume}{08}}, \bibinfo{pages}{045} (\bibinfo{year}{2016}),
  \eprint{1606.02266}.

\bibitem[{\citenamefont{Bondarenko et~al.}(2020)\citenamefont{Bondarenko,
  Boyarsky, Bringmann, Hufnagel, Schmidt-Hoberg, and
  Sokolenko}}]{Bondarenko:2019vrb}
\bibinfo{author}{\bibfnamefont{K.}~\bibnamefont{Bondarenko}},
  \bibinfo{author}{\bibfnamefont{A.}~\bibnamefont{Boyarsky}},
  \bibinfo{author}{\bibfnamefont{T.}~\bibnamefont{Bringmann}},
  \bibinfo{author}{\bibfnamefont{M.}~\bibnamefont{Hufnagel}},
  \bibinfo{author}{\bibfnamefont{K.}~\bibnamefont{Schmidt-Hoberg}},
  \bibnamefont{and}
  \bibinfo{author}{\bibfnamefont{A.}~\bibnamefont{Sokolenko}},
  \bibinfo{journal}{JHEP} \textbf{\bibinfo{volume}{03}}, \bibinfo{pages}{118}
  (\bibinfo{year}{2020}), \bibinfo{note}{[JHEP20,118(2020)]},
  \eprint{1909.08632}.

\bibitem[{\citenamefont{Berezhiani et~al.}(2001)\citenamefont{Berezhiani,
  Comelli, and Villante}}]{Berezhiani:2000gw}
\bibinfo{author}{\bibfnamefont{Z.}~\bibnamefont{Berezhiani}},
  \bibinfo{author}{\bibfnamefont{D.}~\bibnamefont{Comelli}}, \bibnamefont{and}
  \bibinfo{author}{\bibfnamefont{F.~L.} \bibnamefont{Villante}},
  \bibinfo{journal}{Phys. Lett.} \textbf{\bibinfo{volume}{B503}},
  \bibinfo{pages}{362} (\bibinfo{year}{2001}), \eprint{hep-ph/0008105}.

\bibitem[{\citenamefont{Goodsell et~al.}(2009)\citenamefont{Goodsell, Jaeckel,
  Redondo, and Ringwald}}]{Goodsell:2009xc}
\bibinfo{author}{\bibfnamefont{M.}~\bibnamefont{Goodsell}},
  \bibinfo{author}{\bibfnamefont{J.}~\bibnamefont{Jaeckel}},
  \bibinfo{author}{\bibfnamefont{J.}~\bibnamefont{Redondo}}, \bibnamefont{and}
  \bibinfo{author}{\bibfnamefont{A.}~\bibnamefont{Ringwald}},
  \bibinfo{journal}{JHEP} \textbf{\bibinfo{volume}{11}}, \bibinfo{pages}{027}
  (\bibinfo{year}{2009}), \eprint{0909.0515}.

\bibitem[{\citenamefont{Jaeckel and Ringwald}(2010)}]{Jaeckel:2010ni}
\bibinfo{author}{\bibfnamefont{J.}~\bibnamefont{Jaeckel}} \bibnamefont{and}
  \bibinfo{author}{\bibfnamefont{A.}~\bibnamefont{Ringwald}},
  \bibinfo{journal}{Ann. Rev. Nucl. Part. Sci.} \textbf{\bibinfo{volume}{60}},
  \bibinfo{pages}{405} (\bibinfo{year}{2010}), \eprint{1002.0329}.

\bibitem[{\citenamefont{Berezhiani and Lepidi}(2009)}]{Berezhiani:2008gi}
\bibinfo{author}{\bibfnamefont{Z.}~\bibnamefont{Berezhiani}} \bibnamefont{and}
  \bibinfo{author}{\bibfnamefont{A.}~\bibnamefont{Lepidi}},
  \bibinfo{journal}{Phys. Lett.} \textbf{\bibinfo{volume}{B681}},
  \bibinfo{pages}{276} (\bibinfo{year}{2009}), \eprint{0810.1317}.

\bibitem[{\citenamefont{McDermott et~al.}(2011)\citenamefont{McDermott, Yu, and
  Zurek}}]{McDermott:2010pa}
\bibinfo{author}{\bibfnamefont{S.~D.} \bibnamefont{McDermott}},
  \bibinfo{author}{\bibfnamefont{H.-B.} \bibnamefont{Yu}}, \bibnamefont{and}
  \bibinfo{author}{\bibfnamefont{K.~M.} \bibnamefont{Zurek}},
  \bibinfo{journal}{Phys. Rev.} \textbf{\bibinfo{volume}{D83}},
  \bibinfo{pages}{063509} (\bibinfo{year}{2011}), \eprint{1011.2907}.

\bibitem[{\citenamefont{Foot and Vagnozzi}(2015)}]{Foot:2014uba}
\bibinfo{author}{\bibfnamefont{R.}~\bibnamefont{Foot}} \bibnamefont{and}
  \bibinfo{author}{\bibfnamefont{S.}~\bibnamefont{Vagnozzi}},
  \bibinfo{journal}{Phys. Rev.} \textbf{\bibinfo{volume}{D91}},
  \bibinfo{pages}{023512} (\bibinfo{year}{2015}), \eprint{1409.7174}.

\bibitem[{\citenamefont{Foot and Vagnozzi}(2016)}]{Foot:2016wvj}
\bibinfo{author}{\bibfnamefont{R.}~\bibnamefont{Foot}} \bibnamefont{and}
  \bibinfo{author}{\bibfnamefont{S.}~\bibnamefont{Vagnozzi}},
  \bibinfo{journal}{JCAP} \textbf{\bibinfo{volume}{1607}}, \bibinfo{pages}{013}
  (\bibinfo{year}{2016}), \eprint{1602.02467}.

\bibitem[{\citenamefont{Hambye et~al.}(2019)\citenamefont{Hambye, Tytgat,
  Vandecasteele, and Vanderheyden}}]{Hambye:2019dwd}
\bibinfo{author}{\bibfnamefont{T.}~\bibnamefont{Hambye}},
  \bibinfo{author}{\bibfnamefont{M.~H.~G.} \bibnamefont{Tytgat}},
  \bibinfo{author}{\bibfnamefont{J.}~\bibnamefont{Vandecasteele}},
  \bibnamefont{and}
  \bibinfo{author}{\bibfnamefont{L.}~\bibnamefont{Vanderheyden}},
  \bibinfo{journal}{Phys. Rev.} \textbf{\bibinfo{volume}{D100}},
  \bibinfo{pages}{095018} (\bibinfo{year}{2019}), \eprint{1908.09864}.

\bibitem[{\citenamefont{Ma}(2017)}]{Ma:2017ucp}
\bibinfo{author}{\bibfnamefont{E.}~\bibnamefont{Ma}}, \bibinfo{journal}{Phys.
  Lett.} \textbf{\bibinfo{volume}{B772}}, \bibinfo{pages}{442}
  (\bibinfo{year}{2017}), \eprint{1704.04666}.

\bibitem[{\citenamefont{He et~al.}(1991{\natexlab{a}})\citenamefont{He, Joshi,
  Lew, and Volkas}}]{He:1990pn}
\bibinfo{author}{\bibfnamefont{X.~G.} \bibnamefont{He}},
  \bibinfo{author}{\bibfnamefont{G.~C.} \bibnamefont{Joshi}},
  \bibinfo{author}{\bibfnamefont{H.}~\bibnamefont{Lew}}, \bibnamefont{and}
  \bibinfo{author}{\bibfnamefont{R.~R.} \bibnamefont{Volkas}},
  \bibinfo{journal}{Phys. Rev.} \textbf{\bibinfo{volume}{D43}},
  \bibinfo{pages}{22} (\bibinfo{year}{1991}{\natexlab{a}}).

\bibitem[{\citenamefont{Foot}(1991)}]{Foot:1990mn}
\bibinfo{author}{\bibfnamefont{R.}~\bibnamefont{Foot}}, \bibinfo{journal}{Mod.
  Phys. Lett.} \textbf{\bibinfo{volume}{A6}}, \bibinfo{pages}{527}
  (\bibinfo{year}{1991}).

\bibitem[{\citenamefont{He et~al.}(1991{\natexlab{b}})\citenamefont{He, Joshi,
  Lew, and Volkas}}]{He:1991qd}
\bibinfo{author}{\bibfnamefont{X.-G.} \bibnamefont{He}},
  \bibinfo{author}{\bibfnamefont{G.~C.} \bibnamefont{Joshi}},
  \bibinfo{author}{\bibfnamefont{H.}~\bibnamefont{Lew}}, \bibnamefont{and}
  \bibinfo{author}{\bibfnamefont{R.~R.} \bibnamefont{Volkas}},
  \bibinfo{journal}{Phys. Rev.} \textbf{\bibinfo{volume}{D44}},
  \bibinfo{pages}{2118} (\bibinfo{year}{1991}{\natexlab{b}}).

\bibitem[{\citenamefont{Heeck and Rodejohann}(2011)}]{Heeck:2011wj}
\bibinfo{author}{\bibfnamefont{J.}~\bibnamefont{Heeck}} \bibnamefont{and}
  \bibinfo{author}{\bibfnamefont{W.}~\bibnamefont{Rodejohann}},
  \bibinfo{journal}{Phys. Rev.} \textbf{\bibinfo{volume}{D84}},
  \bibinfo{pages}{075007} (\bibinfo{year}{2011}), \eprint{1107.5238}.

\bibitem[{\citenamefont{Gninenko and Krasnikov}(2001)}]{Gninenko:2001hx}
\bibinfo{author}{\bibfnamefont{S.~N.} \bibnamefont{Gninenko}} \bibnamefont{and}
  \bibinfo{author}{\bibfnamefont{N.~V.} \bibnamefont{Krasnikov}},
  \bibinfo{journal}{Phys. Lett.} \textbf{\bibinfo{volume}{B513}},
  \bibinfo{pages}{119} (\bibinfo{year}{2001}), \eprint{hep-ph/0102222}.

\bibitem[{\citenamefont{Baek et~al.}(2001)\citenamefont{Baek, Deshpande, He,
  and Ko}}]{Baek:2001kca}
\bibinfo{author}{\bibfnamefont{S.}~\bibnamefont{Baek}},
  \bibinfo{author}{\bibfnamefont{N.~G.} \bibnamefont{Deshpande}},
  \bibinfo{author}{\bibfnamefont{X.~G.} \bibnamefont{He}}, \bibnamefont{and}
  \bibinfo{author}{\bibfnamefont{P.}~\bibnamefont{Ko}}, \bibinfo{journal}{Phys.
  Rev.} \textbf{\bibinfo{volume}{D64}}, \bibinfo{pages}{055006}
  (\bibinfo{year}{2001}), \eprint{hep-ph/0104141}.

\bibitem[{\citenamefont{Carone}(2013)}]{Carone:2013uh}
\bibinfo{author}{\bibfnamefont{C.~D.} \bibnamefont{Carone}},
  \bibinfo{journal}{Phys. Lett.} \textbf{\bibinfo{volume}{B721}},
  \bibinfo{pages}{118} (\bibinfo{year}{2013}), \eprint{1301.2027}.

\bibitem[{\citenamefont{Altmannshofer et~al.}(2014)\citenamefont{Altmannshofer,
  Gori, Pospelov, and Yavin}}]{Altmannshofer:2014pba}
\bibinfo{author}{\bibfnamefont{W.}~\bibnamefont{Altmannshofer}},
  \bibinfo{author}{\bibfnamefont{S.}~\bibnamefont{Gori}},
  \bibinfo{author}{\bibfnamefont{M.}~\bibnamefont{Pospelov}}, \bibnamefont{and}
  \bibinfo{author}{\bibfnamefont{I.}~\bibnamefont{Yavin}},
  \bibinfo{journal}{Phys. Rev. Lett.} \textbf{\bibinfo{volume}{113}},
  \bibinfo{pages}{091801} (\bibinfo{year}{2014}), \eprint{1406.2332}.

\bibitem[{\citenamefont{Cirelli et~al.}(2009)\citenamefont{Cirelli, Kadastik,
  Raidal, and Strumia}}]{Cirelli:2008pk}
\bibinfo{author}{\bibfnamefont{M.}~\bibnamefont{Cirelli}},
  \bibinfo{author}{\bibfnamefont{M.}~\bibnamefont{Kadastik}},
  \bibinfo{author}{\bibfnamefont{M.}~\bibnamefont{Raidal}}, \bibnamefont{and}
  \bibinfo{author}{\bibfnamefont{A.}~\bibnamefont{Strumia}},
  \bibinfo{journal}{Nucl. Phys.} \textbf{\bibinfo{volume}{B813}},
  \bibinfo{pages}{1} (\bibinfo{year}{2009}), \bibinfo{note}{[Addendum: Nucl.
  Phys.B873,530(2013)]}, \eprint{0809.2409}.

\bibitem[{\citenamefont{Baek and Ko}(2009)}]{Baek:2008nz}
\bibinfo{author}{\bibfnamefont{S.}~\bibnamefont{Baek}} \bibnamefont{and}
  \bibinfo{author}{\bibfnamefont{P.}~\bibnamefont{Ko}}, \bibinfo{journal}{JCAP}
  \textbf{\bibinfo{volume}{0910}}, \bibinfo{pages}{011} (\bibinfo{year}{2009}),
  \eprint{0811.1646}.

\bibitem[{\citenamefont{Garani and Heeck}(2019)}]{Garani:2019fpa}
\bibinfo{author}{\bibfnamefont{R.}~\bibnamefont{Garani}} \bibnamefont{and}
  \bibinfo{author}{\bibfnamefont{J.}~\bibnamefont{Heeck}},
  \bibinfo{journal}{Phys. Rev.} \textbf{\bibinfo{volume}{D100}},
  \bibinfo{pages}{035039} (\bibinfo{year}{2019}), \eprint{1906.10145}.

\bibitem[{\citenamefont{Kamada et~al.}(2018)\citenamefont{Kamada, Kaneta,
  Yanagi, and Yu}}]{Kamada:2018zxi}
\bibinfo{author}{\bibfnamefont{A.}~\bibnamefont{Kamada}},
  \bibinfo{author}{\bibfnamefont{K.}~\bibnamefont{Kaneta}},
  \bibinfo{author}{\bibfnamefont{K.}~\bibnamefont{Yanagi}}, \bibnamefont{and}
  \bibinfo{author}{\bibfnamefont{H.-B.} \bibnamefont{Yu}},
  \bibinfo{journal}{JHEP} \textbf{\bibinfo{volume}{06}}, \bibinfo{pages}{117}
  (\bibinfo{year}{2018}), \eprint{1805.00651}.

\bibitem[{\citenamefont{Kamada et~al.}(2019)\citenamefont{Kamada, Yamada, and
  Yanagida}}]{Kamada:2018kmi}
\bibinfo{author}{\bibfnamefont{A.}~\bibnamefont{Kamada}},
  \bibinfo{author}{\bibfnamefont{M.}~\bibnamefont{Yamada}}, \bibnamefont{and}
  \bibinfo{author}{\bibfnamefont{T.~T.} \bibnamefont{Yanagida}},
  \bibinfo{journal}{JHEP} \textbf{\bibinfo{volume}{03}}, \bibinfo{pages}{021}
  (\bibinfo{year}{2019}), \eprint{1811.02567}.

\bibitem[{\citenamefont{Baldes et~al.}(2018)\citenamefont{Baldes, Cirelli,
  Panci, Petraki, Sala, and Taoso}}]{Baldes:2017gzu}
\bibinfo{author}{\bibfnamefont{I.}~\bibnamefont{Baldes}},
  \bibinfo{author}{\bibfnamefont{M.}~\bibnamefont{Cirelli}},
  \bibinfo{author}{\bibfnamefont{P.}~\bibnamefont{Panci}},
  \bibinfo{author}{\bibfnamefont{K.}~\bibnamefont{Petraki}},
  \bibinfo{author}{\bibfnamefont{F.}~\bibnamefont{Sala}}, \bibnamefont{and}
  \bibinfo{author}{\bibfnamefont{M.}~\bibnamefont{Taoso}},
  \bibinfo{journal}{SciPost Phys.} \textbf{\bibinfo{volume}{4}},
  \bibinfo{pages}{041} (\bibinfo{year}{2018}), \eprint{1712.07489}.

\end{thebibliography}

\end{document}